\documentclass[structabstract]{aa}  

\usepackage{natbib}
\usepackage[applemac]{inputenc}
\usepackage{amsmath}
\usepackage{graphicx}
\usepackage{longtable}
\usepackage{txfonts}
\usepackage{color}
\usepackage{soul}

\newcommand{\cc}{cm$^{-3}$}

\newcommand{\sqc}{cm$^{-2}$}

\newcommand{\mjy}{MJy\,sr$^{-1}$}
\newcommand{\kjy}{kJy\,sr$^{-1}$}

\newcommand{\Av}{A$_\mathrm{V}$}

\newcommand{\pdix}[1]{$\times$\,10$^{#1}$}

\newcommand\mic{$\mu$m}

\begin{document}

   \title{Dust properties inside molecular clouds from coreshine modeling and observations
   }

   \author{C. Lef{\`e}vre            \inst{1}
          \and
          L. Pagani\inst{1}
         \and
         M. Juvela\inst{2}
         \and
         R. Paladini\inst{3}
         \and
         R. Lallement\inst{4}
         \and
         D.J. Marshall\inst{5}
	 \and
	 M. Andersen\inst{6,7}
	 \and
	 A. Bacmann\inst{6,7}
	 \and
	 P. M. McGehee\inst{3}
	 \and
	 L. Montier\inst{8,9}
	 \and
	 A. Noriega-Crespo\inst{3,10}
	 \and
	V.-M. Pelkonen	 \inst{2,11}
	 \and
	 I. Ristorcelli\inst{8,9} 
	 \and
	 J. Steinacker\inst{6,7}}

   \offprints{C.Lef{\`e}vre}

 \institute{ LERMA \& UMR8112 du CNRS, Observatoire de
  Paris, 61, Av. de l'Observatoire, 75014 Paris, France\\
\email{charlene.lefevre@obspm.fr}
\and
Department of Physics, P.O.Box 64, FI-00014, University of Helsinki, Finland
\and
Infrared Processing and Analysis Center, California Institute of Technology, Pasadena, CA 91125, USA       
\and
GEPI Observatoire de Paris, CNRS, Universit{\'e} Paris Diderot, Place Jules Janssen 92190 Meudon, France
\and
Laboratoire d’astrophysique  Instrumentation Modelisation (AIM) Paris-Saclay, CEA Saclay, 91191 Gif-sur-Yvette, France
\and
Universit{\'e} Grenoble Alpes, IPAG, F-38000 Grenoble, France
\and
CNRS, IPAG, F-38000 Grenoble, France
\and
Université de Toulouse, UPS-OMP, IRAP, 31028 Toulouse Cedex 4, France
\and
CNRS, IRAP, 9 Av. colonel Roche, BP 44346, 31028 Toulouse Cedex 4, France
\and
Space Telescope Science Institute, Baltimore, MD, 21218, USA
\and
Finnish Centre for astronomy with ESO, University of Turku, Väisäläntie 20, FI-21500 PIIKKIÖ, Finland
}

   \date{Received 28 April 2014; Accepted 16 July 2014}

%\authorrunning{Lef\`evre, Pagani, Juvela et al.}
%\titlerunning{Coreshine modeling and observations}

  \abstract
  {Using observations to deduce dust properties, grain size distribution, and physical conditions in {molecular} clouds is a highly degenerate problem.} 
   {The coreshine phenomenon, a scattering process at 3.6 and 4.5\,\mic\ that dominates absorption, has revealed its ability to explore the densest parts of clouds. We want to use this effect to constrain the dust parameters. The goal is to investigate to what extent grain growth (at constant dust mass) inside molecular clouds is able to explain 
the coreshine observations. We aim to find dust models that can explain a sample of Spitzer coreshine data. We also look at the consistency with near-infrared data we obtained for a few clouds.}
   {We selected four regions with a very high occurrence of coreshine cases: Taurus--Perseus, Cepheus, Chameleon and L183/L134. We built a grid of dust models and investigated the key parameters to reproduce the general trend of surface brightnesses and {intensity ratios} of both coreshine and near-infrared observations with the help of a 3D Monte-Carlo radiative transfer code. {The grid parameters allow to investigate} the effect of coagulation upon spherical grains up to 5\,\mic\ in size  derived from the DustEm diffuse interstellar medium grains. Fluffiness {(porosity or fractal degree)}, ices, and a handful of {classical grain size distributions} were also tested. We used the  near-- and mostly mid--infrared {intensity ratios} as strong discriminants between dust models.}
  { {The determination of the background field intensity at each wavelength is a key issue. In particular, an especially strong background field explains why we do not see coreshine in the Galactic plane at 3.6 and 4.5\,\mic. 
  %For the same  reason, there is no such effect at 5.8\,\mic\ in any direction which is constraining for the modeling. 
  For starless cores, where detected, the observed 4.5\,\mic\,/\,3.6\,\mic\ coreshine intensity ratio is always lower than $\sim$0.5 which is also what we find in the models for the {Taurus--Perseus} and L183 directions. Embedded sources can lead to higher fluxes (up to four times greater than the strongest starless core fluxes) and higher coreshine ratios (from 0.5 to 1.1 in our selected sample). 
  %This ratio enhancement is readily explained by the reddening of the sources, especially with Class 0 and 0/I objects, that changes the local balance of the radiation field at the mid--infrared wavelengths.
  }\\
    Normal interstellar radiation field conditions are sufficient to find suitable grain models at all wavelengths for starless cores. The standard interstellar grains are not able to reproduce observations and, due to the multi-wavelength approach, only a few grain types meet the criteria set by the data. Porosity does not affect {the flux ratios} while {the fractal dimension} helps to explain {coreshine ratios} but does not seem {able to} reproduce near--infrared observations without a mix of other grain types.}
   {Combined near-- and mid--infrared wavelengths confirm {the potential} to reveal the nature and size distribution of dust grains. Careful assessment of the environmental parameters (interstellar and background fields, embedded or nearby reddened sources) is required to validate this new diagnostic. }   


   \maketitle

%---------------------INTRO--------------------------------------
%________________________________________________________________

\section{Introduction}
The study of low mass star and of planet formation starts by understanding the place where they form and evolve, that is inside dense molecular clouds. {There, the gas and the dust are in constant interaction through collisions that can lead to heat exchange and, in suitable conditions, to the freezing of gas molecules onto dust grains.} Inside the molecular cloud, the dust content is known to evolve mainly via grain growth: by accretion of heavy gas particles on the dust grains that increases the total dust mass \citep{2012MNRAS.422.1263H}, and by the presence of sticky ice mantles (volatile species frozen onto the grains, \citealt{2004A&A...418.1035W}) which favors coagulation \citep{1993A&A...280..617O, 2009A&A...502..845O}. Ice mantle formation beyond A$_\mathrm{v}\sim\,$3~mag (\citealt{2001ApJ...547..872W,2013ApJ...774..102W}) implies a change in the grain properties throughout molecular clouds. In addition, interstellar grains evolve with time during the prestellar phase. They continue to grow or possibly reach a stationary state in the cloud envelope \citep{2009A&A...502..845O} while, in the densest region, the dust evolution becomes complex {in the presence of an embedded object such as Class 0 or Class I protostars} (\citealt{2000prpl.conf...59A}). The thermo-mechanical action of the protostellar object will affect the grains by shattering them in the outflows  \citep{2013A&A...556A..69A} and by evaporating the grain mantles, releasing water (and other species) in the surrounding gas \citep{2001MNRAS.327.1165F}, as seen by Herschel \citep{2012A&A...542A...8K,VanDerTak2013}. In this context, to infer the molecular cloud stage from the dust properties is a complex problem which starts by understanding {the dust grain content}.

Grain properties can be investigated in different manners, for example via the characterization of the extinction curve. In the optical and UV, this curve changes depending on the dominant grain growth mechanism: accretion or coagulation  \citep{2014MNRAS.437.1636H}. In the near-infrared (NIR) and mid-infrared (MIR) ranges, this change is a clue to  coagulation  \citep{2009ApJ...690..496C, 2012arXiv1211.6556A} and to the presence of ice mantles (\citealt{2009ApJ...693L..81M}). Nevertheless, grain growth deduced indirectly from the extinction curve is sensitive to the wavelength normalization  \citep{Fitzpatrick:2007gg}.
{In this context, a very efficent way to learn about the properties of larger grains (micrometer size) is through the recently discovered effect of MIR dust scattering or "coreshine effect" \citep{2010Sci...329.1622P, 2010A&A...511A...9S}.} The coreshine effect is widespread and detected in at least half of the molecular clouds investigated by \cite{2010Sci...329.1622P} and Paladini et al. (in prep.),  and thus can be used as a tool to explore the properties of the dust responsible for such a phenomenon. 

{Coreshine is observed at those MIR wavelengths, 3--5\,\mic, where the scattering by large grains is strong enough to be seen in emission.} The best examples are seen in the 3.6 and 4.5\,\mic\ Spitzer Infrared Array Camera (IRAC) filters \citep{2010Sci...329.1622P}. {When the absence of emission in the  5.8 and 8~\mic\ IRAC filters excludes the presence of Polycyclic Aromatic Hydrocarbons (PAHs), there is no need to consider an active emission process in the modeling and only scattering and absorption have to be treated, as presented in \cite{2010A&A...511A...9S}.} The same restriction to absorption and scattering modeling is also pertinent for shorter wavelengths (optical and NIR) as demonstrated by \citet{1996A&A...309..570L} for a standard InterStellar Radiation Field (ISRF). 

The investigation potential of coreshine, combined with the modeling of other wavelengths provides an opportunity to better constrain both cloud structures and dust properties. Indeed, thanks to the low opacities  at MIR wavelengths and to the anisotropic scattering, coreshine provides access deep inside the clouds and brings information on their 3D structure.
 The feasibility of the method has been shown \citep{2010A&A...511A...9S, 2013A&A...559A..60A} but to quantify our capacity to build a 3D model of a real cloud and deduce grain properties from a combination of wavelengths, we need to evaluate the impact of the free parameters and of the boundary conditions on the modeling. Here, we compare our models to observations and deduce general trends on the grain properties for the {regions where most of the clouds} present coreshine. The set of {dust models} will constitute a future data base to start modeling molecular clouds in absorption and scattering from visible to MIR. In a forthcoming paper, we will model a specific cloud while also including  the far--infrared (FIR) emission to further constrain the dust properties and cloud structure.

In Sect. \ref{sect:obs}, we present a summary of the observations and the strategy adopted to analyze them. In Sect. \ref{sect:modeling}, we describe {the Monte Carlo radiative transfer} code used for the simulations, focus on the importance of constraining the radiation field and present the cloud model and the dust content which are the free parameters to be explored. In Sect. \ref{sect:res}, we describe the results {obtained from the observational data and confront them with our grid of models.} We discuss the coreshine phenomenon and what a multi-wavelength approach can bring to investigate the grain properties. We present our conclusions in Sect. \ref{sect:conclu}.

%--------------OBSERVATIONS AND ANALYSIS---------------------------
%__________________________________________________________________

\section{Observations and Analysis}\label{sect:obs}
\subsection{NIR data}
NIR data have been  obtained with WIRCAM, the CFHT Wide IR CAMera. Its field of view is 20\arcmin$\times$20\arcmin\ with a pixel size of 0.3\arcsec. {It is large enough to cover each of the selected targets including a large area around them. The observations were obtained (and continue currently to be taken) for a total of eight sources (\object{L183} and a set of sources in the Taurus region) in the standard J, H, and Ks (hereafter K) spectral bands.} The observations, data processing and data themselves will be presented in detail in a forthcoming paper. We use some J and K band observations to have a first look at the comparison with models, without performing any exact fits.

\subsection{Coreshine data}
{To investigate the presence of coreshine phenomenon inside a sample of molecular clouds, we used Spitzer data taken by IRAC \citep{2004ApJS..154...10F} at 3.6, 4.5, 5.8 and 8\,\mic\ (Cryogenic mission) complemented by that from the Warm mission (3.6 and 4.5 \mic\ only).}  {When possible, the 3.6 and 4.5 \mic\ data have been compared to Spitzer 8\,\mic\ and/or WISE 12\,\mic\  maps\footnote{http://skyview.gsfc.nasa.gov/current/cgi/titlepage.pl} \citep{2010AJ....140.1868W} to define the size of the core in absorption and optical data (in R or B) to delimit the size of the surrounding cloud.}
{These Spitzer observations were collected at different epochs and show a median frame integration time per pixel from 50 seconds (From molecular cores to Planet forming disks (c2d) -- \citealt{2009ApJS..181..321E}) to 1800 seconds (Hunting Coreshine Survey HCS, Spitzer cycles 8 and 9 -- Paladini et al. in prep.)}. The sensitivity {for a unit exposure time} in the warm mode (HCS) is comparable to the one in the cold mode (P94 program: Search for Low-Luminosity YSOs -  \citealt{2004sptz.prop...94L}, c2d)\footnote{\label{senspet}http://ssc.spitzer.caltech.edu/warmmission/propkit/pet/senspet/}, which allows us to merge the observations. The aim of the HCS proposal was to obtain an unbiased sample, and our analysis confirms the {$\sim$50\% detection rate of coreshine as in}~\citet[ Paladini et al. in prep.]{2010Sci...329.1622P}. 

The coverage is entirely dependent on the program and on the target. The Spitzer field of view size is 5.12\arcmin\ $\times$  5.12\arcmin\ with {a native pixel size of 1.2\arcsec. The Full Width Half Maximum (FWHM) of the Point Spread Function (PSF) varies from 1.66\arcsec\ at 3.6 \mic\ to 1.98\arcsec\ at 8 \mic. For the nearby molecular clouds,} the maps are large enough to include both the cores and their environment. The sensitivity for an extended source at 4.5\,\mic\ is about 90\% of the 3.6\,\mic\ sensitivity for both warm and cold campaigns\footnotemark[2]. However, the data must be corrected for 
 column pulldown and zodiacal emission\footnote{http://irsa.ipac.caltech.edu/data/SPITZER/docs/dataanalysistools /tools/contributed/irac/wcpc/} (Paladini et al., in prep.). {It must be noted {that even if  Spitzer data are absolutely calibrated using point sources, the extended emission calibration can be as uncertain as 10\%. Indeed, while the drift of the electronics zero level is small, the zodiacal light estimate remains slightly uncertain because of its time dependence and the spatial resolution of the model used to derive it.} Though this is of no importance for differential measurements, such as the {coreshine intensity defined as the net signal above the background}, it becomes a problem for evaluating the absolute background intensity behind the clouds. Indeed, {the observational value I$_\mathrm{back,Sp}$} can vary by a factor of two from one campaign to another as the uncertainty on the zodiacal light estimate is of the same order as the background value itself {(and consequently cannot be used as a reference for I$_\mathrm{back}$ in the modeling, Sect.~\ref{sect:back})}.
 
\begin{figure*}
\flushleft
\sidecaption
\includegraphics[trim = 5cm 3cm 5cm 3cm, clip,width = 18cm]{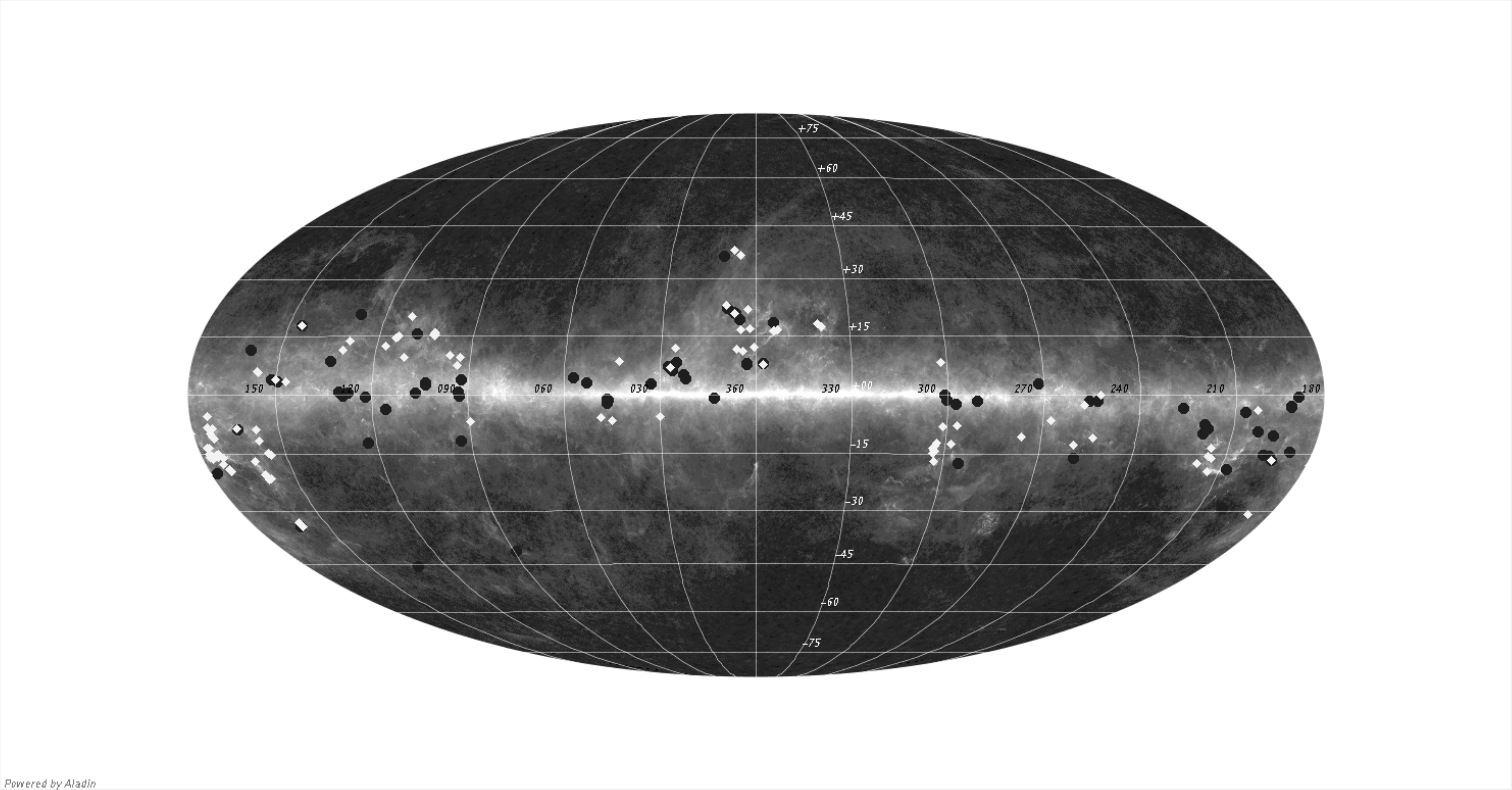}
\caption{Mollweide map of the 3.6 \mic\ coreshine spatial distribution according to the selected sample of 215 sources across the Galaxy. Black filled circles represent negative cases , white diamonds are associated with clouds that show coreshine. The background image is the combined Planck 353-545-857 GHz map. {Coordinate steps in longitude are 30 degrees and 15 degrees in latitude.}}
\label{fig:planck}
\end{figure*}

\begin{figure*}[t!]
\centering
\sidecaption
\includegraphics[width=17cm]{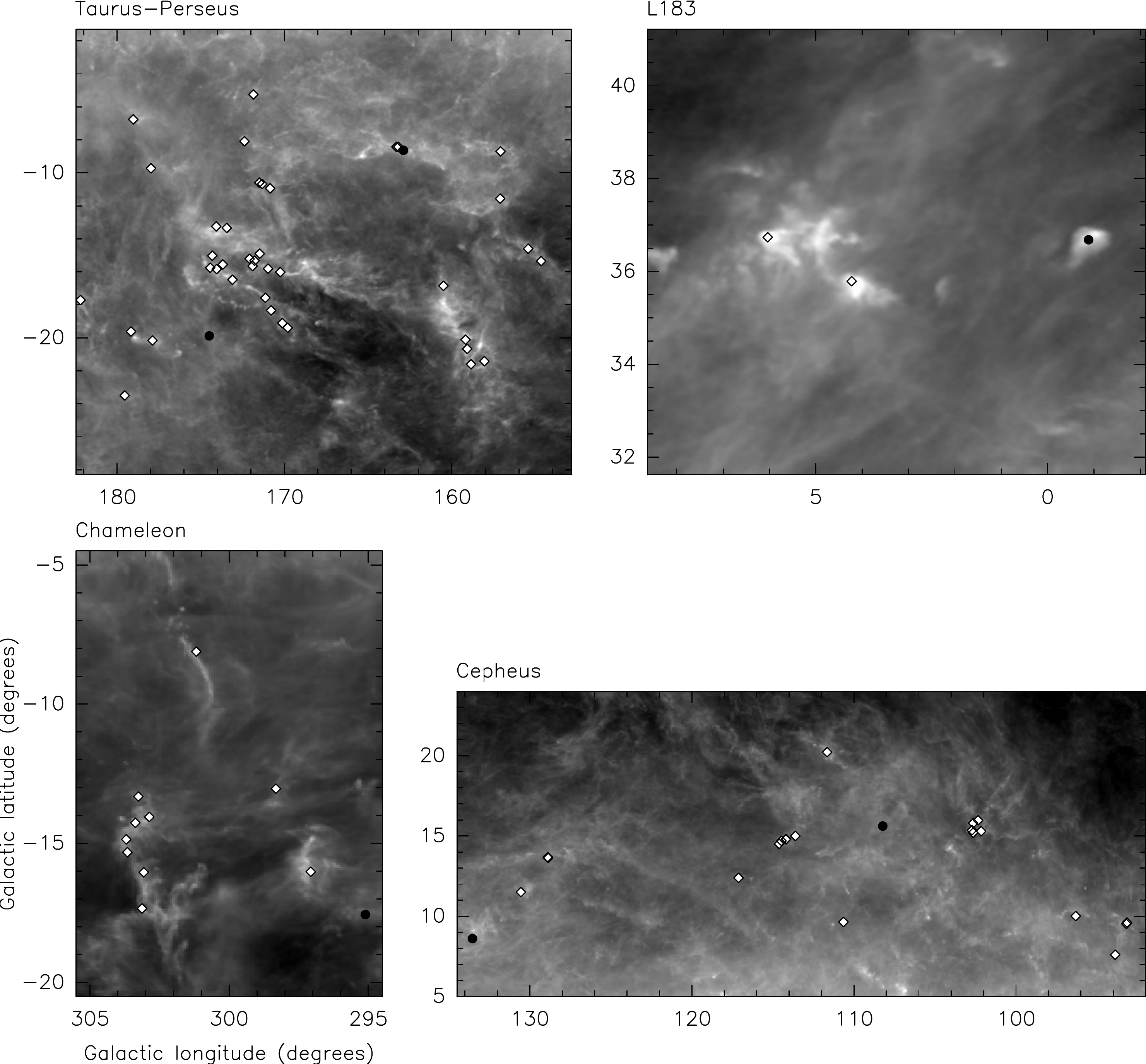}
\caption{Zoom from Fig. \ref{fig:planck} on the four regions of interest.}
\label{fig:4regions}
\end{figure*}

The sample of the {215 sources} presented in Table \ref{tab:sources} and in Fig. \ref{fig:planck} is essentially a compilation of previous surveys including c2d and P94 surveys which were the main archives for the identification of coreshine sources in \citet{2010Sci...329.1622P}, the {new} HCS survey (Paladini et al. in prep), plus a few more targets of interest detected thanks to WISE \citep{2010AJ....140.1868W} or recently identified in the Spitzer Archive (including the Gum/Vela region, \citealt{Pagani2012}). {The total number of clouds which show coreshine is 108 with some preferential directions on the sky. Indeed, the Taurus, Perseus, Aquila and Aries complex ({hereafter Taurus--Perseus}) reveals almost 100~\% coreshine detection (Fig. \ref{fig:4regions}, Table \ref{tab:sources}). Cepheus and Chameleon regions also display a large fraction of coreshine cases. Finally, because of its high galactic latitude and its complexity \object{L183} will be also one of the regions of interest for this study. {Among these 108 positive coreshine detections, we chose to ignore the cores located in Orion/Monoceros and $\rho$ Oph regions where the coreshine occurrence drops to $\sim$50\% (Table \ref{tab:sources}). In these regions the coreshine occurrence is likely dominated by {the effect of local sources which make the radiation field difficult to constrain, adding another degree of freedom. Therefore, they are beyond the goal of our global study of coreshine}. After this selection criterium, we are left with 72 coreshine sources.}}

\begin{figure*}
\flushleft
\sidecaption
\includegraphics[width=13.8cm]{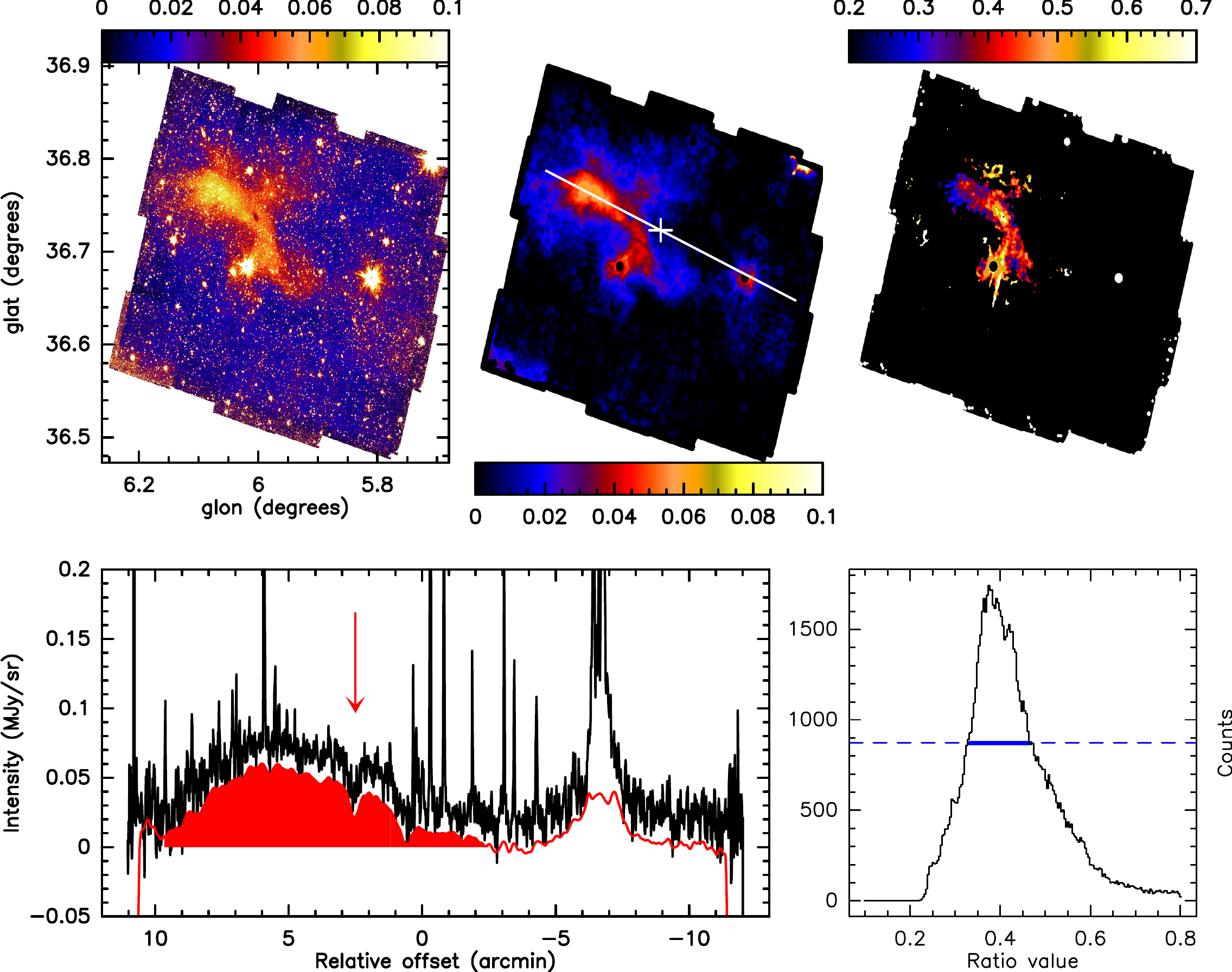}
\caption{Top: From left to right, L183 original image in galactic coordinates from the HCS survey (IRAC 1, 3.6\,\mic), point sources and background subtracted image showing the 3.6\,\mic\ coreshine intensity (with a gaussian smoothing), and coreshine ratio (4.5/3.6) image. The white line shows the position of the cut presented below, {the white cross indicates the zero position reference}. Bottom left: the profile through the cut for the source data (black) and through the cleaned and smoothed image (red); the 3.6 \mic\ coreshine zone is highlighted by the red shading. {The intensities are given in MJy\,sr$^{-1}$. The cut has been chosen to fit both a region with strong coreshine intensity ($\sim$~50 kJy\,sr$^{-1}$ here) and to show the internal depression, which traces self--absorption, $\sim$ 2.5 arcmin from the zero reference, marked by the red arrow on the profile. Bottom right: the coreshine ratio histogram. The blue line displays the FWHM of the histogram.}}
\label{fig:L183_sex}
\end{figure*}

\subsection{Analysis}\label{sect:analysis}

{From the observations, we want to quantify the scattering excess above the surrounding background. To proceed to the measurement two steps are mandatory: i) remove the point sources to keep only the extended emission, ii) subtract the background contribution (I$_\mathrm{back,Sp}$) under the core to deduce the remaining intensity. For the present analysis, the SExtractor software \citep{1996A&AS..117..393B} has been used and its parameters tuned in order to remove point sources and keep only the extended emission.}

First, SExtractor computes the background from a large scale mesh. This coarse background is resampled to the original resolution,  the fluctuations are dampened by applying a median filter to the mesh points. The result is subtracted internally from the original image. {Next, the object detection step works on a smoothed image and requires to set up a detection threshold. {The objects are detected from an image smoothed with a gaussian kernel that has a FWHM equal to 1.5 pixels. Each detection is required to contain at least 3 pixels above 1.5 $\sigma$.} The difficulty here is to adjust the background mesh size so that the  coreshine emission is removed as part of the background while the extended saturated stars remain in order to be detected as objects by SExtractor.} If the mesh size is too small they will appear in the final image. On the other hand, if the mesh size is too large, the extended coreshine emission would not be included in the background estimate but detected instead as a source. The object identification file is checked and for most coreshine cases, we chose a mesh size of {32$\times$32 pixels, but sometimes we adapted the mesh size to 64$\times$64 instead}. Then, the object list can be used to either remove the sources from the image for display or to produce a mask of the point sources for measurements. 
This source subtraction method with the optimized parameters works on almost all Spitzer maps (even in crowded fields) and allows us to mask or to subtract compact objects and to retrieve only extended emission (Fig. \ref{fig:L183_sex}). 

Because coreshine is a phenomenon defined as a signal in excess with respect to the surrounding background emission, the coreshine intensity has to be background subtracted. To perform this subtraction, it is mandatory to interpolate the background value (I$_\mathrm{back,Sp}$) at the core position. Due to the extended nature of the coreshine effect, the process cannot be automatized. After having subtracted the point sources, we mask the coreshine region by hand and subtract either a plane by least square fitting or repeat the background subtraction process with SExtractor, interpolating across the coreshine masked region. This interpolation is usually safe when the masked region remains small but can become less accurate for large sources. However, the interpolation of the background is normally close to a plane. We compared both subtraction methods for the very extended coreshine emission present in L183 and found good agreement between the two methods, with a difference smaller than 10\%.

\addtocounter{table}{1}

The 3.6 and 4.5\,\mic\ cleaned images are convolved with a gaussian kernel of 10\arcsec\ FWHM to reduce the noise by a factor of {$\sim$6}. Then we measure the peak flux for the 3.6\,\mic\ image and we build the 4.5\,\mic\,/\,3.6\,\mic\ ratio image eliminating all pixels with a 4.5\,\mic\ flux below {3$\sigma$ (after smoothing)} which is a good compromise to avoid the large fluctuations due to noise. We compute the histogram of the ratio map and take {the maximum}, and the FWHM to characterize its dispersion (Fig. \ref{fig:L183_sex}). In this paper, we have selected four regions of interest (Fig.2) where most of the clouds present coreshine. We apply the above method to the 72 cores from these regions, which show a positive coreshine signal (Table 1).

%----------------MODELING DESCRIPTION------------------------------
%__________________________________________________________________

\section{3D Radiative transfer modeling}\label{sect:modeling}

{Many 3D dust radiative transfer codes are available today (see review by \citealt{2013ARA&A..51...63S}). In this Section, we first present one of these 3D codes (Continuum Radiative Transfer - CRT)\footnote{https://wiki.helsinki.fi/display/$\sim$mjuvela@helsinki.fi/CRT}. We discuss the main input parameters:  the radiation field, the cloud model and the dust properties. 
While the radiation field parameter is important to constrain, we want to focus on the dust properties' variations to see the influence of grain growth (especially coagulation) in reproducing the observational trends region by region.}

%%%----%%%
\subsection{Monte Carlo radiative transfer code}\label{sect:crt}

The radiative transfer calculations are done with the CRT program
\citep{2003A&A...397..201J, 2005A&A...440..531J}. This implements the basic Monte
Carlo scheme where a number of photon packages is sent out from each
of the radiation sources, the propagation and scattering of the
photons is followed, and the intensity of the scattered radiation
exiting the medium is registered. The density field can be discretized
using spherical or cylinder geometries \citep[see][]{2012A&A...542A..21Y} or, as
in the case of this paper, using full three-dimensional Cartesian
grids. CRT allows the use of multiple dust populations and spatial
variations of their abundance. For scattering calculations, in
addition to the density field, only the dust optical opacity (e.g.,
relative to hydrogen), the albedo, and the scattering phase function need
to be specified. For the scattering phase function, i.e. the probability
distribution of the scattering angles, CRT allows the use of the
Henyey--Greenstein approximation with the asymmetry dust parameter $\mathrm{g}$ = $\langle$$cos\, \theta$$\rangle$, or one can use any scattering function tabulated as a
function of the scattering angle $\theta$. In this paper, we use the usual 
Henyey-Greenstein approximation. Its validity will be discussed in Sect. \ref{sect:dust}.

In our calculations, the main source of radiation is the interstellar
radiation field that is described with all-sky DIRBE maps \citep{1998ApJ...508...25H}
in HEALPix format \citep{2005ApJ...622..759G}, using a separate map to describe the sky
brightness at each of the simulated wavelengths {(Sect. \ref{sect:Illumination}).} 
%We aim to model six individual wavelengths: J (1.25\,\mic), K (2.2\,\mic), IRAC1 (3.6\,\mic), IRAC2 (4.5\,\mic), IRAC3 (5.8\,\mic) and IRAC4 (8\,\mic). 
The influence of using a single wavelength instead of several averaged wavelengths taking into account the filter response has been tested and shows no statistical differences for the coreshine wavelengths. The simulation runs have been done with 100 million photon packets and we estimated the numerical uncertainty on the modeling result to be 1 kJy\,sr$^{-1}$.
The original positions of the emitted photon packages are weighted with the sky intensity, so that
more packages (with correspondingly smaller weight or smaller photon
number) are generated from the Galactic plane and, in particular,
from the direction of the Galactic center. CRT uses the forced first
scattering method \citep{1970A&A.....9...53M} to ensure adequate sampling of
scattered flux {in regions of low opacity}. To improve the
quality of the scattered light images, the peel-off technique
\citep{1984ApJ...278..186Y} is used where, once a photon package is
scattered, CRT always explicitly calculates the fraction of photons
that scatter towards the observer and escape the cloud without further
interactions {(I$_\mathrm{sca}$, Sect. \ref{sect:back})}. 

The images of scattered light are built using the peeled photons and
they represent the surface brightness visible for an observer far
outside the cloud. CRT has the option for calculating peel--off images
for several directions during the same run. In the present case,
images are only calculated in one direction as determined by the relative
locations of the selected clouds and the observer. In practice, the cloud model is
viewed along one coordinate axis and the background DIRBE maps are
rotated so that the illumination geometry is correct.

%%%----%%%
\subsection{Incident and background interstellar radiation fields}\label{sect:ISRF}

%The way the photons are launched in the Monte Carlo modeling can be determined by using an isotropic interstellar radiation field or an anisotropic radiation field. 

To model dust extinction and emission, the ISRF has to be determined. Two different quantities are needed: the sky brightness in all directions, determining the illumination of the cloud, and the surface brightness behind the cloud, determining the net effect of absorption along the line of sight of our observations. The background field determination requires a precise treatment, which we discuss in further detail  below.

\subsubsection{Incident ISRF}\label{sect:Illumination}

The all-sky illumination has to take into account the contribution of stellar sources; O and B stars dominate the UV field (\citealt{1968BAN....19..421H}) and mostly K stars/red giants for the longer wavelengths. The diffuse part due to ambient stellar light scattered from small grains, UV light reprocessed in PDRs, and PAHs emission must also be considered. Different galactocentric distances can lead to different intensity estimates \citep{1983A&A...128..212M}, but taking into account the fact that the molecular clouds we are studying are close to us {(with a distance range  from 100 pc -- L183 -- to 325 pc -- Cepheus)} we made the approximation that the illumination seen by the objects is the same as the one observed from the Earth. Only an anisotropic ISRF is considered in this study, since its presence is essential to be able to see scattered light in excess of the background field (Appendix \ref{sect:aniso}). 

The Galaxy, and especially the Galactic Center,  is {the main} source of the anisotropic ISRF illuminating the clouds.
The zodiacal subtracted mission average (ZSMA) DIRBE survey provides us directly with {the 3D all-sky maps\footnote{http://cade.irap.omp.eu/documents/Ancillary/4Aladin/}$^,$\footnote{http://lambda.gsfc.nasa.gov/product/cobe/dirbe$\_$overview.cfm}} {with an accurate estimate of the sky flux} at J, K, 3.5 $\mu$m, 4.9 $\mu$m and 12 $\mu$m wavelengths (and up to 240\,\mic, out of scope here,  \citealt{1998ApJ...508...25H}). We use HEALPix maps with parameter N$_{\rm SIDE}$ = 256, giving sky pixels with a size of 13.7$\arcmin$, smaller than the DIRBE resolution of $\sim$\ 40\arcmin\  \citep{1998ApJ...508...25H}.
At NIR wavelengths {the} radiation field is directly obtained thanks to filter conversion from DIRBE to 2MASS \citep{Levenson:2007gh}: DIRBE 1 (J) has been divided by 0.97 and DIRBE 2 (K) by 0.88 to obtain the NIR radiation map inputs. The ISRF at the MIR wavelengths is obtained by rescaling DIRBE 3 (3.5 \,\mic), 4 (4.9\,\mic) and 5 (12\,\mic) map fluxes to Spitzer fluxes (at 3.6, 4.5 and 8\,\mic\ respectively) {thanks to filter color corrections and wavelength scaling deduced from the Galactic spectrum} by \citet[their Table 2]{2006A&A...453..969F}.
In order to deduce the illumination at 5.8\,\mic, we took the 0.3 {observed color value} $R_{5.8/8.0}$ for GLIMPSE \citep{2003PASP..115..953B,2006A&A...453..969F},
consistent with \cite{2001ApJ...554..778L} and not dependent on the line of sight, which assumes that the stellar contribution at this wavelength is completely negligible compared to the diffuse contribution. This presumes that we are dominated by PAH emission and there is no offset between bands \citep[Fig. 9 in ][]{2006A&A...453..969F}.

\subsubsection{Cloud background field}\label{sect:back}
 
To illuminate the cloud, it is mandatory to consider all the contributions (both stellar  $\mathrm{I^*(\lambda)}$ and diffuse  $\mathrm{I_{diff} (\lambda)}$). On the other hand, {the observed signal is the combination of radiation coming from behind the cloud, attenuated on its way through the cloud, and of the fraction of the radiation field that it scatters towards the Earth. Therefore we have to be careful on how to evaluate this background value, I$_\mathrm{back}$.} The DIRBE resolution implies that the stellar contribution in a particular beam is always present and wavelength dependent:

\begin{equation} \label{eq:DIRBE_diff}
\mathrm{I_\mathrm{back} (\lambda) = I_\mathrm{diff, (l,b)} (\lambda)*bg+CIRB}
\end{equation}

\begin{equation}\label{eq:eq_bckg}
\mathrm{I_\mathrm{diff, (l,b)} (\lambda) = DIRBE_\mathrm{(l,b)} (\lambda) - I^*(\lambda)}
\end{equation}

\noindent where bg is the fraction of the diffuse light on the line of sight coming from behind the cloud and CIRB is the Cosmic Infra Red Background due to unresolved galaxies from the early Universe \citep[e.g.][]{2000ApJ...536..550G,Levenson:2007gh}. To  evaluate this stellar contribution we consider two different approaches.  In the first method, we make a  sky--direction independent guess on the proportion of the stellar contribution relative to the diffuse contribution in each DIRBE filter. We then use the J/K band ratio to deduce the extinction on the line of sight, and  evaluate the stellar contribution in the 3.5 and 4.9\,\mic\ filters by using stellar color ratios and the extinction previously deduced (see \citealt{1994A&A...291L...5B} and Appendix \ref{sect:JPB} for details). This method is powerful because it gives an all-sky map of the stellar contribution at each wavelength. Nevertheless, it implies the major hypothesis that the signals in DIRBE 1 and DIRBE 2 bands are only due to the stellar contribution. Even though the relative contribution of the diffuse emission is weak in the J and K bands (10\,--\,20\,\%), the previous hypothesis is too strong  since we are interested in this residual value. {Despite this limitation, this method gives a good approximation at 3.6 and 4.5\,\mic\ of the diffuse cloud background field intensity. }
\\
\\

\begin{table*}
\flushleft
\caption{\label{tab:Iback_table}I$_\mathrm{diff}$ (in kJy\,sr$^{-1}$) and bg values for the four lines of sight and the six wavelengths. {  For the four directions, we give the central pixel value, the average value of the nine pixels {in brackets} and the dispersion of the nine pixels as an estimate of the uncertainty. }}
\begin{center}
\begin{tabular}{lcccccccc}
\hline 
\hline 
Line of Sight & I$_\mathrm{diff}(J) $ & I$_\mathrm{diff}(K)$ & I$_\mathrm{diff}(3.6)$ & I$_\mathrm{diff}(4.5)$ & I$_\mathrm{diff}(5.8)$ & I$_\mathrm{diff}(8.0)$ & bg1 & bg2 \tabularnewline
\hline  
L183 & 99 $\langle$88$\rangle$ $\pm$20 & 63 $\langle$52$\rangle$ $\pm$22 & 58 $\langle$51$\rangle$ $\pm$14 & 63 $\langle$53$\rangle$ $\pm$22 & 602 $\langle$604$\rangle$ $\pm$3 &  1826 $\langle$1832$\rangle$ $\pm$9& 0.75 & 0.5 \tabularnewline
Taurus--Perseus & 70 $\langle$73$\rangle$ $\pm$21 & 31 $\langle$34$\rangle$ $\pm$13 &  70 $\langle$72$\rangle$ $\pm$7 & 65 $\langle$75$\rangle$ $\pm$7 &  725 $\langle$678$\rangle$ $\pm$10&  2201 $\langle$2057$\rangle$$\pm$31 & 0.95 & 0.75\tabularnewline
Chameleon & 101 $\langle$96$\rangle$ $\pm$12 & 46 $\langle$48$\rangle$ $\pm$7 & 53 $\langle$56$\rangle$ $\pm$4 & 31 $\langle$33$\rangle$ $\pm$5 &  498 $\langle$500$\rangle$ $\pm$14&  1512$\langle$1517$\rangle$ $\pm$42& 0.7 & *\tabularnewline
Cepheus & 28 $\langle$51$\rangle$ $\pm$23 & 3 $\langle$27$\rangle$ $\pm$22 & 25 $\langle$36$\rangle$ $\pm$10 & 15 $\langle$25$\rangle$ $\pm$10 &  475 $\langle$475$\rangle$ $\pm$5&  1441 $\langle$1442$\rangle$ $\pm$16& 0.45 & *\tabularnewline
CIRB &	8.9$\pm$6.3&14.7$\pm$4.5	&15.6$\pm$3.3	&(14)\tablefootmark{a}	&	(12)\tablefootmark{a}&(11)\tablefootmark{a}	&	&\\
\hline 
\end{tabular}
\end{center}
\tablefoot{
\tablefoottext{a}{estimated values from \cite{Levenson:2007gh} Fig. 10}
}
\end{table*}

\begin{figure}[t]
\centering
\includegraphics[height=8cm]{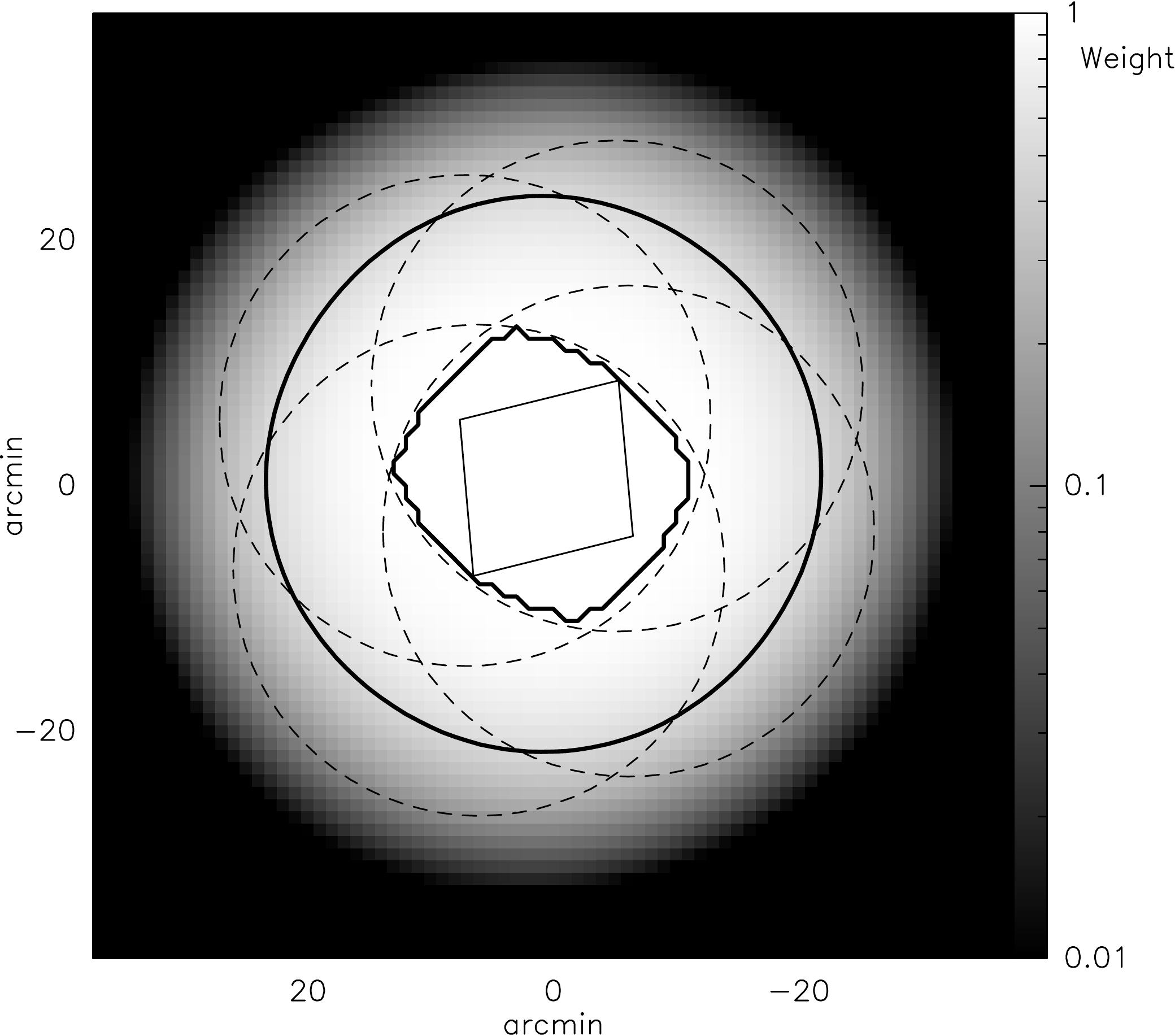}
\caption{{Weight map of a DIRBE healpixel. This represents the probability} for a star to contribute to the healpixel flux as a function of its position in the sky. The diamond shape represents the particular healpixel (here in the Taurus--Perseus region). The thick pillow-like contour delimits the region inside which all stars contribute in all individual observations (probability = 1). The dashed circles represent the probability = 1 for a star to contribute for the given pointing. Each of the four circles is centered at one of the four corners of the diamond shape.  The outer thick circle is the probability = 0.5 for a star to contribute.}
\label{fig:DIRBE_proba}
\end{figure}

\begin{figure*}
\centering
\sidecaption
\includegraphics[width=14cm]{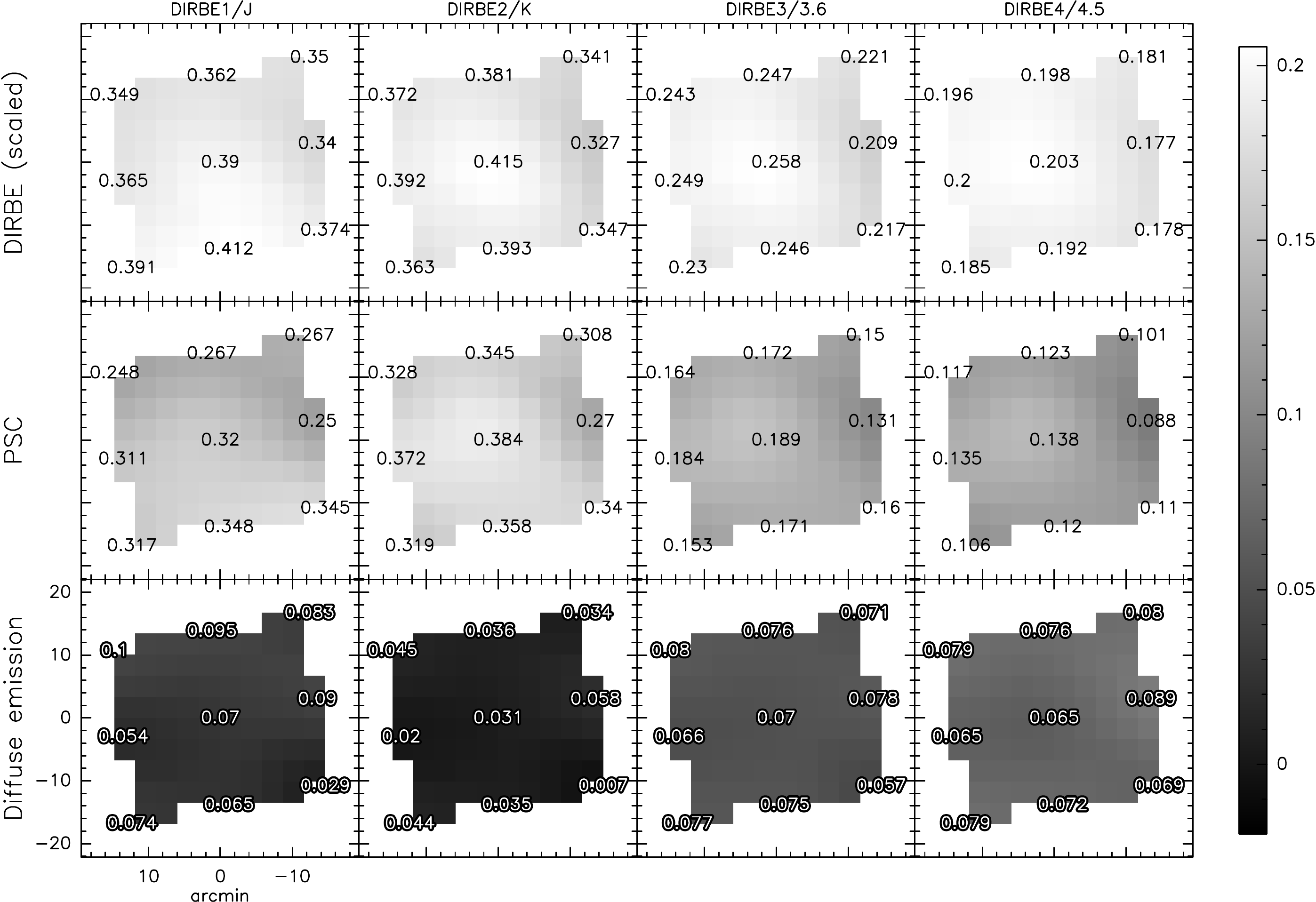}
\caption{The nine nearest healpixels towards a single Taurus--Perseus position (here, \object{L1544}). Each column corresponds to a wavelength. Top row is the DIRBE values (rescaled respectively to 2MASS or Spitzer units), middle row the star flux contribution (PSC) and bottom row the remaining diffuse emission {I$_\mathrm{diff} (\lambda)$}. Grey scale and values are given in MJy\,sr$^{-1}$}
\label{fig:back_diffus}
\end{figure*}

The second method is based on the flux subtraction of the sources by summing them from 2MASS and WISE point source catalogs \citep{2006AJ....131.1163S, 2012yCat.2311....0C,2012wise.rept....1C}. 
This has been done in several studies that tried to estimate the CIRB and we basically apply the same method as described in  \cite{Levenson:2007gh}. The DIRBE HEALPix maps are made of pixels of equal area but of different shape (healpixels, \citealt{2005ApJ...622..759G}). {Each healpixel contains the average of all pointings that fall inside that pixel, for all orientations since DIRBE observed each position in the sky many times with different position angles. The probability that a star contributes to the measured flux of a healpixel is therefore not trivial to evaluate \citep{Levenson:2007gh}. For each cloud direction, the healpixel shape is different and we have to compute the probability map of the stellar contributions for each of them separately. We start from the probability that a star falls inside the DIRBE pixel for any position angle (this probability is 1 inside a 20\arcmin\ radius and 0 outside a 28\arcmin\ radius) and we add together the DIRBE pixel probability map for all positions inside the healpixel shape (discretized to the arcmin level). The resultant {weight} map is renormalized to unity in its center. Such a map is displayed in Fig. \ref{fig:DIRBE_proba} for the Taurus--Perseus region. This {weight mask} is applied to a catalog of point sources to compute their contribution to the healpixel.

For each cloud of interest, we  retrieve the DIRBE pixel value that covers the cloud and its eight neighbors, and the shape of that particular healpixel. We compute the weight map as described above and retrieve the 2MASS and WISE point source catalogues of the corresponding region (from the Vizier database). The point source fluxes are obtained by using the zero flux reference of \cite{2011ApJ...735..112J} for WISE and of \cite{2003AJ....126.1090C} for 2MASS. {Then, for each band, individual stellar contribution is respectively converted to Spitzer fluxes or kept in 2MASS fluxes, summed with the appropriate weight maps and subtracted {from} the DIRBE pixel value (scaled to Spitzer or 2MASS fluxes).} The number of point sources considered varies from 5,000 in the 2MASS catalogue at high galactic latitude {(36\degr, L183) up to 30,000 sources in the WISE catalogue for lower latitudes ($-$14\degr, Chameleon)}. An example measurement using nine pixels along with a point source contribution estimate is shown in Fig. \ref{fig:back_diffus}.

{The results for the four regions discussed in this paper are summarized in Table \ref{tab:Iback_table}. For each line of sight and wavelength, we give the value of the central pixel, the nine pixels average value and the dispersion. The I$_\mathrm{diff}$ values obtained from individual lines of sight have been chosen to be representative of the complete region for the {Taurus--Perseus complex}, the southern part of Chameleon and L183 direction. The Cepheus region is more heterogeneous. It shows a gradient in the Galactic Plane direction and has to be considered more cautiously (Fig. \ref{fig:all_back}).} 

One can note that the CIRB contribution (eq. \ref{eq:DIRBE_diff}) is negligible compared to I$_\mathrm{diff}$ in J but is in the  50-100\% range in K. It represents $\sim$ 25-50\% of the diffuse flux at 3.6\,\mic\ and 4.5\,\mic. At 5.8 and 8\,\mic, the star contribution is considered to be negligible and the background emission is directly measured from the interpolated DIRBE fluxes in these bands as explained in Sect. \ref{sect:Illumination}.

\begin{figure*}
\centering
\includegraphics[width=13cm]{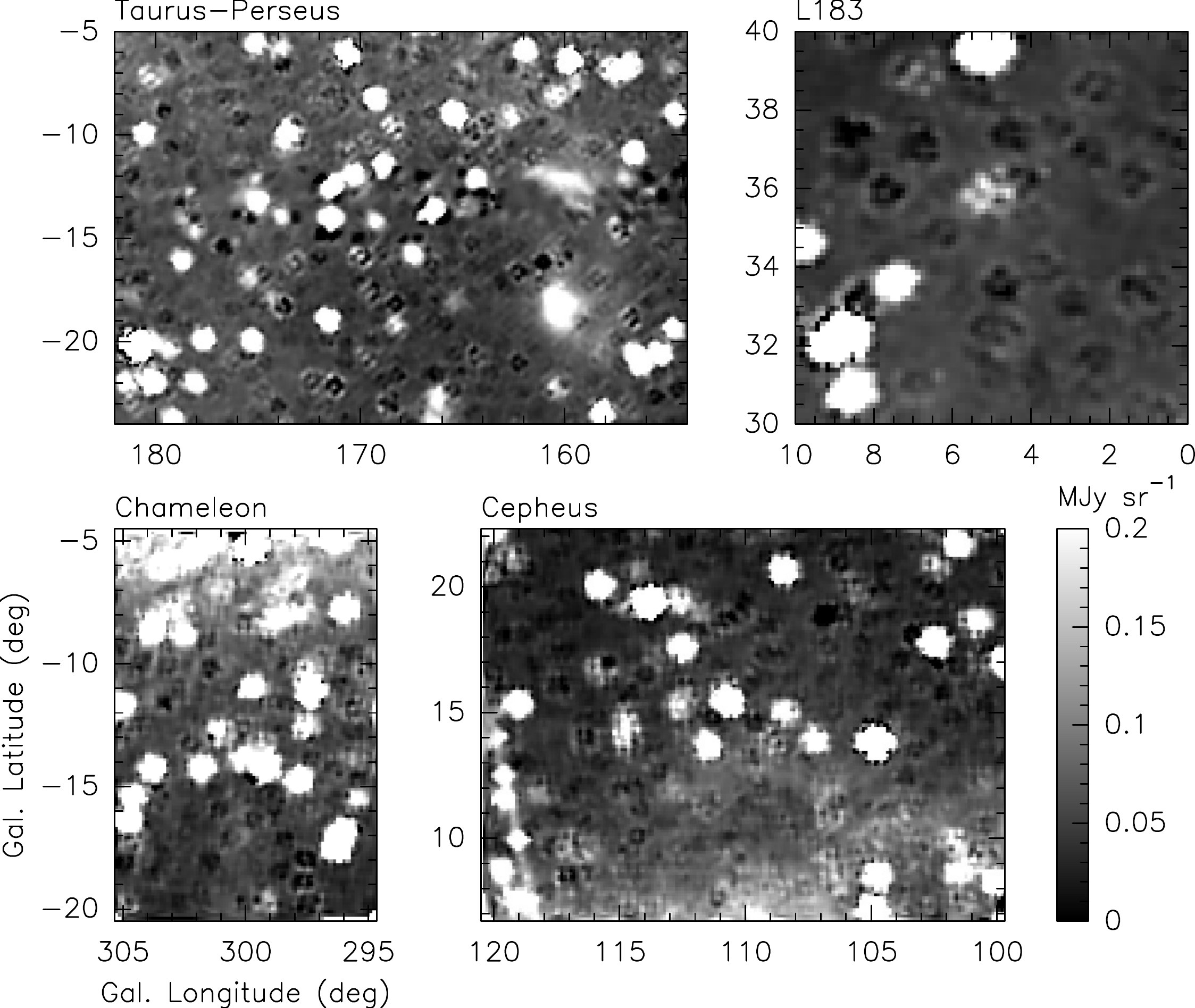}
\caption{Maps of the diffuse emission I$_\mathrm{diff}$ at 3.6\,\mic\ for the four regions obtained with the second method at the DIRBE resolution.}
\label{fig:all_back}
\end{figure*}

The method is efficient but can become inaccurate when very bright stars enter the field since their flux is not always correctly estimated in the WISE or 2MASS catalogs though it represents a major fraction of the total flux in the DIRBE pixel. {For a discussion about non-linearities and saturation due to bright stars in the determination of the photometry, the reader is referred to the explanatory supplements of the WISE and 2MASS missions\footnote{http://www.ipac.caltech.edu/2mass/releases/allsky/doc/sec1$\_$6b.html\\http://wise2.ipac.caltech.edu/docs/release/allsky/expsup/sec1$\_$4b.html$\#$brt}}. {The adopted correction also assumes that the completeness of the input catalogs is similar to the depth to which point sources are removed in our analysis of the diffuse signal. Indeed, the faint end of the source distribution is only a few \kjy\ in J and K and even less at coreshine wavelengths \citep{Levenson:2007gh}. The uncertainty on the method can be estimated from the fluctuations between the nine adjacent pixels and from the fact that no pixel should show a flux lower than the CIRB flux (Table \ref{tab:Iback_table}). The zodiacal light subtraction we use is also challenged by some authors and the result slightly depends on the adopted zodiacal light correction (see Table 5 in \citealt{Levenson:2007gh}).}

The two methods are  qualitatively in agreement at 3.5 and 4.9 $\mu$m but not in J and K bands. 
While the first method supposes that all the J band flux is due to the stars, the second method finds a sizable fraction of the flux to be due to the diffuse light (10 to 25\,\%). This is expected since the standard interstellar grains in the diffuse medium are more efficient at scattering light at 1.2~$\mu$m than at 2.2\,\mic. It is in the K band that the diffuse contribution reaches a minimum.   The diffuse emission fraction is also higher in the 3.5 and 4.9\,\mic\ bands since the stellar contribution decreases with increasing wavelength. When we obtained results below the CIRB intensity we replaced I$_\mathrm{diff}$ by the CIRB intensity in our models. This happened only in the Cepheus direction, for the K, 3.6, and 4.5\,\mic\ bands. The second method is more reliable at getting the measurement of $\mathrm{I_{diff} (\lambda)}$ when we investigate a particular direction (modeling a cloud) while the first is more useful in exploring what happens at Galactic scales in the MIR.

The measurement of $\mathrm{I_\mathrm{diff} (\lambda)}$ is not the only parameter we need to know to reconstruct the absorption part of the final map; the proportion of the diffuse emission which is in front of our cloud (foreground = fg) and the one which is behind (background = bg) is a key element too (see eq. \ref{eq:DIRBE_diff}). These quantities are directly linked to the proportion of dust in the diffuse medium located in front and behind the cloud. Their evaluation is based on a model of  dust in our Galaxy up to 300 pc by {\citet[bg1]{2014A&A...561A..91L}} and has to be taken as an indication rather than a precise value. That is why, for certain lines of sight, and when possible, we also used  an other estimate based on a different method by \cite{2006A&A...453..635M} to evaluate the uncertainty on the fg and bg values.  {This method, known to be unreliable below one kpc, has been refined and give better approximations at smaller heliocentric distances (Marshall et al., in prep, bg2).}

{At the end, it is the product of $\mathrm{I_\mathrm{diff} (\lambda)}$ with the bg value that gives $\mathrm{I_\mathrm{back}}$ (eq. \ref{eq:DIRBE_diff}). The contrast level of the emergent flux (I$_\mathrm{final}$) in final model maps will depend on this I$_\mathrm{back}$ value as follows:
\begin{equation} \label{eq:trans}
\mathrm{I_\mathrm{final} = (I_\mathrm{sca} +I_\mathrm{back}*\mathrm{exp (-\tau)}\-) - I_\mathrm{back}\-.}
\end{equation}}

\noindent with I$_\mathrm{sca}$ the scattered light image, $\tau$ the integrated extinction opacity map, both obtained from the radiative transfer code, and $\mathrm{(I_\mathrm{sca} +I_\mathrm{back}*\mathrm{exp (-\tau)}\-)}$ the transmitted signal from the cloud. The focus on four different regions allows us to explore different ($\mathrm{I_\mathrm{diff}}$,  bg) pair values and see the impact on {the emergent intensity} in the modeling.

%%%----%%%
\subsection{Molecular cloud content}\label{sect:moc}

\subsubsection{Cloud model}\label{sect:cloud}

 Because modeling in 3 dimensions is mandatory (Appendix \ref{sect:aniso}), we chose to use a general 3D shape that represents most of the data clouds. An inclined ellipsoid has been taken as the cloud model. This cloud model was built to correspond to 40 pixels for a cloud radius (core+envelope) equal to 16000 AU (Fig. \ref{fig:cloudprofile}).  {This corresponds to a resolution ranging from 2\arcsec\ to 7\arcsec\ depending on the distance of the region}, a few times the Spitzer resolution. Because the coreshine varies slowly through the cloud ({red area} Fig. \ref{fig:L183_sex}) and because we are interested in the surface brightness (in MJy\,sr$^{-1}$), this resolution is not crucial. {We made a compromise by limiting the total number of cells (with a size of 104$\times$104$\times$104 cells) to reduce the computational time. However, we keep enough cells to dedicate reasonable physical space to the cloud in our modeling cube while keeping some room for the external part. {This also explains why we chose a jump at lower density to the envelope instead of a smooth variation. This has a minor impact on the modeling.}}
 
 The minor axis has been designed to be half the major axis, and the inclination angle is 60° to see a possible gradient effect as a function of the direction of the Galactic Center. The adopted density profile for the core is a Plummer profile adapted to an ellipsoid (from \citealt{2005MNRAS.359..228D} and \citealt{2001ApJ...547..317W})
 
 \begin{equation}
n(x,y,z)=\frac{n_{0}} 
{1+\left[
\frac{r}{a_0}
\right]
^{\alpha}}
,
\end{equation}
\noindent with:
\begin{equation}
r=\sqrt{
\frac{x^{2}}{a^{2}} + \frac{y^{2}}{b^{2}} +
\frac{z^{2}}{c^{2}}}
\end{equation}

where $n$ is the number density, 
$n_0$ is the reference density, $a_0$ the radius associated with $\frac{n_0}{2}$, $(x,y,z)$ defines the position,
$(a,b,c)$ are parameters specifying the shape of the ellipsoid,
with an index of $\alpha$ = 2.5. Generally used on a sphere, this type of profile (Fig. \ref{fig:cloudprofile}) is common and realistic for simple molecular clouds like \object{L1544} in the {Taurus--Perseus region} \citep{2005MNRAS.359..228D}.

\begin{figure}[t]
\centering
\sidecaption
\includegraphics[width=9cm]{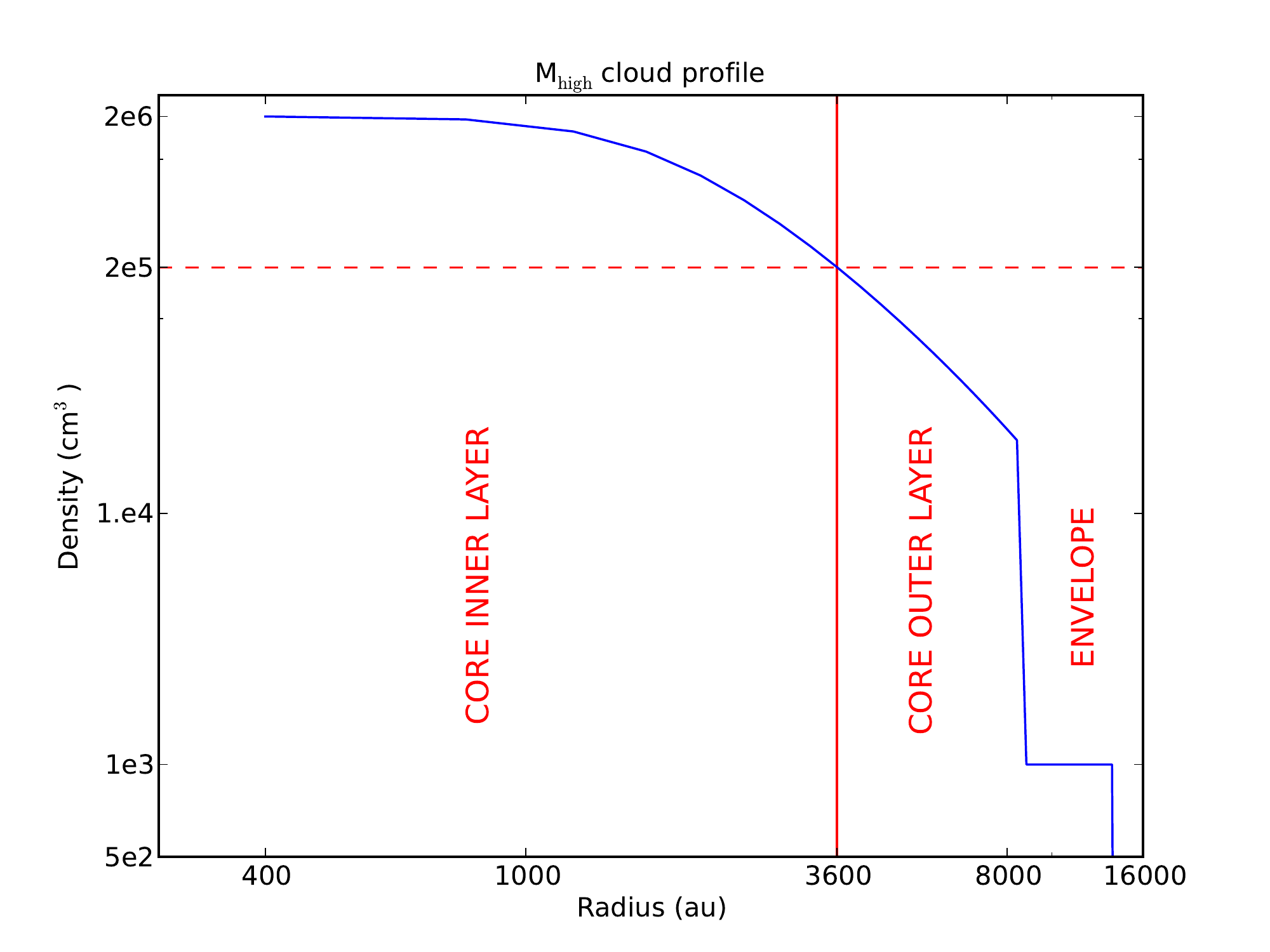}
\caption{Cloud profile for the second cloud model (M$_\mathrm{high}$) with a central density of 2 \pdix6 \cc.}
\label{fig:cloudprofile}
\end{figure}

While the detailed shape does not really matter for our toy cloud model, the column density has to be more representative of the range of our observations. We defined two models, one with a central density of n$_0$(H${_2}$)= 2\,\pdix 6 cm$^{-3}$ (M$_\mathrm{high}$ = 1.5 M$_{\sun}$) and another one with n$_0$(H${_2}$) = 5 \pdix 5 cm$^{-3}$ (M$_\mathrm{low}$ = 0.4 M$_{\sun}$) {which gives a peak column density of 9.2 \pdix{22} \sqc and 2.3~\pdix{22}~\sqc\ respectively}, and we assume a gas-to-dust ratio of 133 \citep{2011A&A...525A.103C}. {The 3D cube model of more than one million cells is divided in two regions : an envelope and a core. The envelope, of constant density (1000 \cc), is always filled with a standard diffuse grain size distribution since NIR studies seem to be able to reproduce scattering in the outer parts of the investigated clouds with such distributions \citep{2012A&A...544A.141J}. The core itself is divided in two parts of approximately equal thickness (Fig. \ref{fig:cloudprofile}) in which different grain characteristics are tested. The threshold {between the two core layers} is situated at density  2~\pdix5 \cc\ (for M$_\mathrm{high}$), and  5~\pdix4 \cc\ (for M$_\mathrm{low}$). Grain property gradients can thus be investigated from the envelope standard distribution to the inner core evolved grains.}
%The different layers can be filled with different grain characteristics. 

%%%----%%%

\subsubsection{Dust properties}\label{sect:dust}

\begin{table}
\caption{\label{tab:grain_size}The different types of modeled grains and their properties.}
\centering
\begin{tabular}{ccccc}
\hline\hline
Grain model name & a$_\mathrm{min}$ & a$_\mathrm{max}$ & a$_\mathrm{cut}$ & $\langle$a$\rangle$  \\
  & nm &\,\mic &\,\mic & nm \\
\hline  
\bf{DustEm extended \tablefootmark{1} }\\
\hline
aSil / CBx2 & 4 & 2 & 0.2/0.150 & 15.37 / 25.2 \\ 
Cx.2 & 4 & 2 & 0.2 & 28.25\\
aS25 / Cx25 & 4 & 2 & 0.25 & 16.12 / 30.9\\ 
aS.3 / Cx.3 & 4 & 2 & 0.3 & 16.76 / 33.24 \\
aS.4 / Cx.4 & 4 & 2 & 0.4 & 17.8 / 37.3\\
aS.5 / Cx.5 & 4 & 2 & 0.5 & 18.65 / 40.8\\
aS.6 / Cx.6 & 4 & 2 & 0.6 & 19.37 / 43.87\\
aS.7 / Cx.7 & 4 & 2 & 0.7 & 19.99 / 46.59\\
aS.8 / Cx.8 & 4 & 2 & 0.8 & 20.55 / 48.9\\ 
aS.9 / Cx.9 & 4 & 2 & 0.9 & 21.06 / 50.75\\
aS1m / Cx1m & 4 & 2 & 1.0 & 21.52 / 52.14\\
aS2m / Cx2m & 4 & 5 & 2.0 & 24.72 / 70.56\\
aS5m / Cx5m & 4 & 9 & 5.0 & 29.12 / 97.18\\
S10 / C10 & 10 & 2 & 0.15 & 31.44 / 43.47\\
S20 / C20 & 20 & 2 & 0.5 & 66.1 / 106.88\\
S50 / C50 & 50 & 2 & 1.0 & 154.72 / 236.4\\
\hline  
\bf{Other bare grains} \\
\hline
WD31\tablefootmark{2}  (Si/Gra) & 0.35 & 0.3/1.0 & 0.25/0.4 \\
WD55\tablefootmark{2}  (Si/Gra) & 0.35 & 0.3/1.5 & 0.25/0.6 &\\
WD55B\tablefootmark{2}  (Si/Gra) & 0.35 & 0.3/6.0 & 0.25/3. &\\
ORNI2\tablefootmark{3}  & 0.1 & 0.6 & 0.3\\
\hline
\bf{With ices}\tablefootmark{3}\\
\hline
ORI2 & 0.1 & 0.6 & 0.3 \\
ORI3 & 0.1 & 2.5 & 1.2\\
\hline 
\bf{Porous}\tablefootmark{4}& $\rho$ & $\alpha$ & a$_\mathrm{cut}$ \\
& g\,\cc & &\,\mic \\
\hline
YSA 0\% & 2.87 & -2.4  & 0.234 \\
YSA 10\% & 2.59  & -2.4 & 0.242 \\
YSA 25\% & 2.16  & -2.4 & 0.256 \\
YSA 40\% &1.72  &-2.4 & {0.276} \\
\hline
\bf{Fractal}\tablefootmark{5}& & & & a\\
&  & & &\mic \\
\hline
MIN0.2& & & & 0.2 \\
MIN0.8& & & & 0.8 \\
MIN1.2& & & & 1.2 \\
MIN2.4& & & & 2.4 \\
MIN4.0& & & & 4.0 \\
\end{tabular}
\tablefoot{
Except for DustEm grains, only an approximation of the real law is given for the sake of comparison - see references for the exact dust law. For the fractal aggregates a single size is used, for the others we used a size distribution.
\tablebib{ (1)~\cite{2011A&A...525A.103C}; (2)~\cite{2001ApJ...548..296W}; (3)~\cite{2009A&A...502..845O}; (4)~\cite{2012A&A...542A..21Y}; (5)~Min et al., in prep.}}

\end{table}%

{The main dust models used in this paper are based on an extrapolation of standard grains able to fit the observations in the diffuse medium (comprised of a mixture of silicates and carbonaceous grains, in the classical proportion of 3/4 to 1/4, respectively) from DustEm. DustEm is a software package implemented by \citet{2011A&A...525A.103C} which is able to compute the dust emission from a spherical grain size distribution and grain optical properties (obtained internally by using Mie theory and the IDL code DustProp). Since we focused our study on scattering and absorption and not emission, DustEm is used as a tool to average the grain properties on their size distribution wavelength by wavelength. In each cloud layer (see Sect. \ref{sect:cloud}), we mixed the two dust species and adjusted their size distributions independently. This hypothesis is justified by the fact that the two species are supposed to be able to coagulate in separate ways (deduced from the extinction curves, \citealt{2014MNRAS.437.1636H}).}

{Our extrapolation starts from the biggest {grain size distributions} of DustEm (hereafter aSil and CBx2 according to \citealt{2011A&A...525A.103C} notations, Table \ref{tab:grain_size}). The goal is to investigate to what extent grain growth inside molecular clouds, especially coagulation, is able to explain the coreshine observations.} For both species, 1\,\mic\ grains are known to be sufficiently efficient in scattering and producing coreshine (\citealt{2010Sci...329.1622P} and Fig. \ref{fig:efficiencies}) but some questions remain: is the suppression of the smallest grains required to observe coreshine? Is there a size limitation for the distribution? What is the influence of the change of the slope of the power--law or of the complete dust distribution itself? What is the impact of taking into account ice mantles and fluffy grains?

\begin{figure*}[hdtp]
\centering
\sidecaption
\includegraphics[width=15cm]{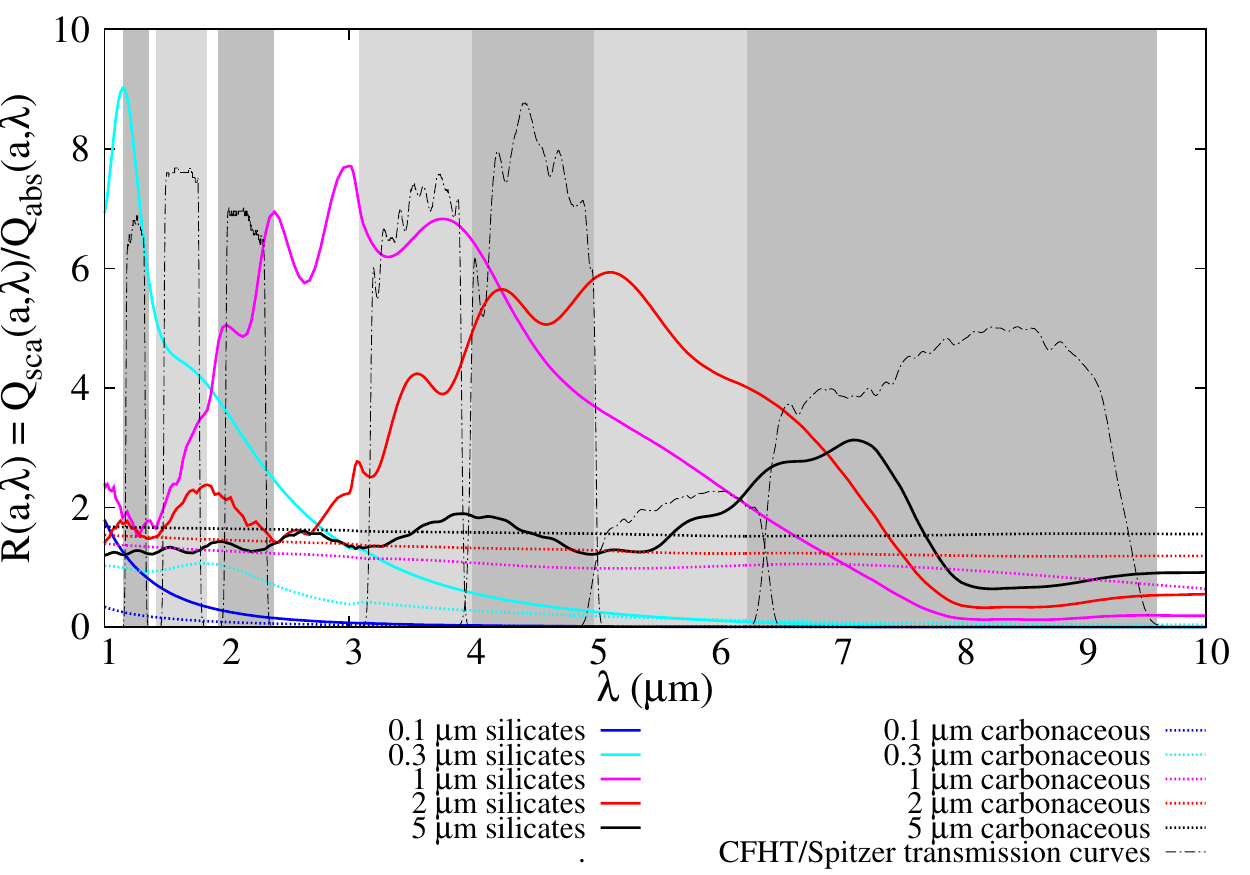}
\caption{Ratio of the scattering and absorption efficiencies for individual grain types and grain sizes. The grain properties come from \cite{2011A&A...525A.103C}. The grey bands delineate the different filter bandpasses (CFHT J, H, Ks, Spitzer 3.6, 4.5, 5.8, 8.0\,\mic).}
\label{fig:efficiencies}
\end{figure*}

In this perspective, we tried different short cutoffs (minimal size for {the grain size distribution}: a$_\mathrm{min}$) and long exponential cutoffs ($e^{-(a/a_\mathrm{cut}-1)^2}$, where a is the grain radius, a$_\mathrm{cut}$ the cutoff radius) {which applied} on the original size--distribution power--law \citep[Table \ref{tab:grain_size}]{2011A&A...525A.103C}.  {In particular, the index of the original power--law is $-$3.4 for the silicates and $-$2.8 for the carbonates. The a$_\mathrm{min}$ value evolves from 4 nm (aSil, CBx2) up to 50 nm (C50, S50) {while also the high cut--off value changes}. Nevertheless, we also chose to  compare {grain size distributions} with the same high cut-offs and different short cut-offs to test the influence of small grains (e.g. from Cx1m to C50). All {the grain size distributions} between CBx2 and Cx1m, as well as aSil and aS1m, have also been used to reproduce a smooth increase in size through the cloud (GRAD model hereafter). The change of the whole distribution itself has been investigated thanks to the use of different grain types from the litterature (\citealt{2001ApJ...548..296W}, \citealt{2009A&A...502..845O}) as well as the porosity  \citep{2012A&A...542A..21Y} and the fractal dimension (Min et al. in prep).}

For all the grain models, the optical properties are averaged in each cell of the cloud.  {DustEm provides us with averaged parameters for each grain size distribution taking into account the grain distribution law and the dust mass for each species.}
%{Since we considered averaged properties for {the whole grain size distribution} deduced from DustEm}, 
{Therefore, we have to consider three averaged parameters:}  the mean scattering efficiency  $\langle$$\mathrm{Q_{sca}}$$\rangle$, the mean absorption efficiency $\langle$$\mathrm{Q_{abs}}$$\rangle$, and the mean phase asymmetry factor  $\langle$$\mathrm{g}$$\rangle$. This phase function, the probability for a photon packet to be scattered in a given direction, is well--approximated by the Henyey--Greenstein expression 
\begin{equation}\label{eq:HG}
{\mathrm{P_{\textrm{HG}}(\mu) = \frac{1}{2} \frac{1 - \langle g \rangle^{2}}{(1 - 2\langle g \rangle \mu + \langle g \rangle^{2})^{3/2}}}}, 
\end{equation}
with $$\mathrm{\mu\ = cos (\theta),}$$
and $$ \mathrm{\int_{-1}^{1} \! P_\mathrm{HG}(\mu) \, \mathrm{d}\mu = 1,}$$
 which is accurate enough for spherical grains up to a form factor $x=2\pi$$\langle$a$\rangle$$/ \lambda$ equal to 10, with $\langle$a$\rangle$ the mean size of the {grain size distribution} and $\lambda$ the wavelength. In this paper, the form factors of the size averaged dust models vary from 0.04 to 0.4 at 3.6\,\mic\ (Table \ref{tab:grain_size}). Only the axial backward scattering might not be taken sufficiently into account. {The full radiative transfer equations are described in Steinacker et al. (in prep.).} Nevertheless, the averaging process on the whole size distribution has been tested by comparing a full distribution and the same distribution cut in three parts and put together in the same cell. No difference could be found providing the size distribution discretization had been increased up to 600 steps in DustEm.

To understand the competition between scattering and absorption at NIR and MIR wavelengths, it is important to look at the individual properties for the different sizes and grain components (Fig. \ref{fig:efficiencies}). One can notice from Fig. \ref{fig:efficiencies} that for a given total dust mass, $\langle$Q$_\mathrm{sca}$$\rangle$ increases faster than $\langle$Q$_\mathrm{abs}$$\rangle$ when $\langle$a$\rangle$ increases. This is undoubtedly linked to the emergence of coreshine in dark clouds \citep{2010A&A...511A...9S}. However, real clouds present a mixture of different grain types and grain sizes which imply some degeneracy. They are partly removed by modeling the cloud at several discrete wavelengths (from J band to 8\,\mic).

Because \cite{2011A&A...525A.103C} grains are not the only ones able to explain the observations in the diffuse medium, and in order to explore beyond spherical grains, we added a sample of dust grain varieties. In our grid of models, silicate and {graphite} mixtures (\citealt{2001ApJ...548..296W}, WD31, WD55, WD55B), {porous grains without ices} (\citealt{2012A&A...542A..21Y}, YSA 0\%, YSA 10\%, YSA 20\%, YSA 40\%), {monomer fractal aggregrates} {with different sizes} (Min et al. in prep.\footnote{see http://events.asiaa.sinica.edu.tw/meeting/20131118/talk/\\2013112111\_Talk\_MichielMin.pdf}, MIN0.2, MIN0.8, MIN1.2, MIN2.4, MIN4.0) or {compact agglomerates} with ices (\citealt{2009A&A...502..845O}, ORI2, ORI3) and {their counterpart without ices (\citealt{2009A&A...502..845O}, ORNI2)} were included with different maximal sizes (see Table \ref{tab:grain_size} and references for more details). We do not attempt to provide a review of all the grain types available in the literature, among which a-C:H grains \citep{Jones:2013ci}, and iron inclusions in silicates (Jaeger et al. in prep\footnote{see http://events.asiaa.sinica.edu.tw/meeting/20131118/talk/\\2013112117\_Talk\_CorneliaJaeger.ppt}) are other possibilities. The goal is to open the scope of what kind of grains will be suitable to explain our observations.

%-------------------------RESULTS----------------------------------
%__________________________________________________________________

\section{Results}\label{sect:res}

%%%----%%%

\begin{figure*}[t]
\flushleft
\sidecaption
\includegraphics[width = 14cm]{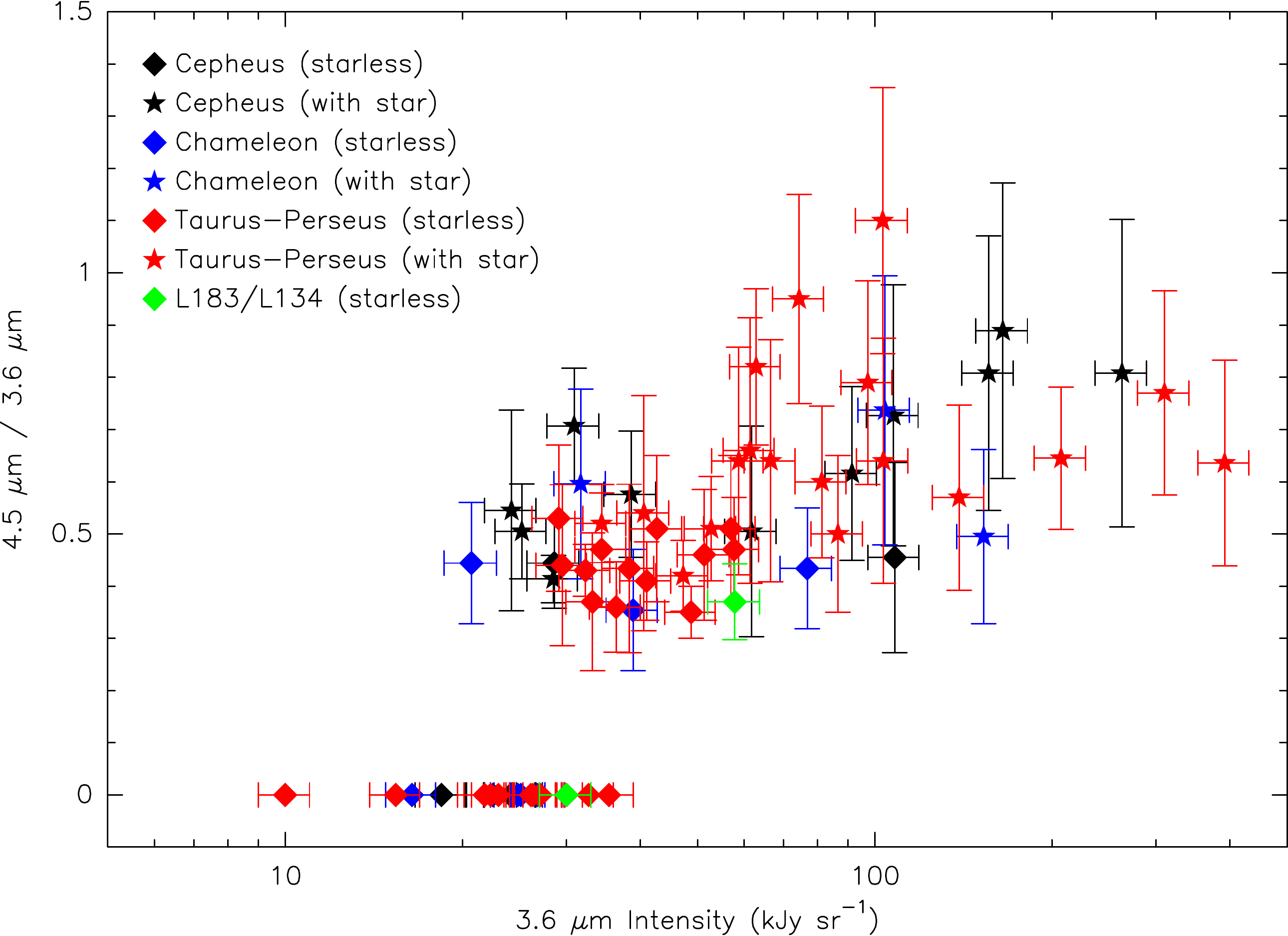}
\caption{Coreshine ratio for the selected data sample of 72 sources as a function of the 3.6\,\mic\ intensity. Diamonds indicate starless core, stars indicate clouds {with a known local protostar}.}
\label{fig:Observations}
\end{figure*}

{The coreshine strength is linked to the grain properties and to the environmental conditions. The modeling has to deal with the absolute coreshine intensity which is a contrast problem and a competition between scattering and absorption. These  extinction processes are linked to the dust properties, and in particular to the grain size distribution. However, these properties are degenerate when considering  a single wavelength. Therefore, we adopted a multi--wavelength approach in which the albedo variation between wavelengths, dependent on the dust population itself, is well--characterized by {intensity ratio} measurements. 

In Sect. \ref{sect:result_obs}, we present the measurements of the two key quantities, the 3.6\,\mic\ intensity and {the coreshine ratio (between 4.5 and 3.6\,\mic\ wavelengths)} for the observational sample of the four regions. In Sect. \ref{sect:contrast}, we expose the conditions to make coreshine appear and retrieve the correct absolute intensity from modeling while, in Sect.  \ref{sect:ratio}, we detail the use of the different {intensity ratios} as a tool.}}

\subsection{{Surface brightness and coreshine {ratio} in observations}}\label{sect:result_obs}

\begin{figure*}[t]
\flushleft
\sidecaption
\includegraphics[width=14cm]{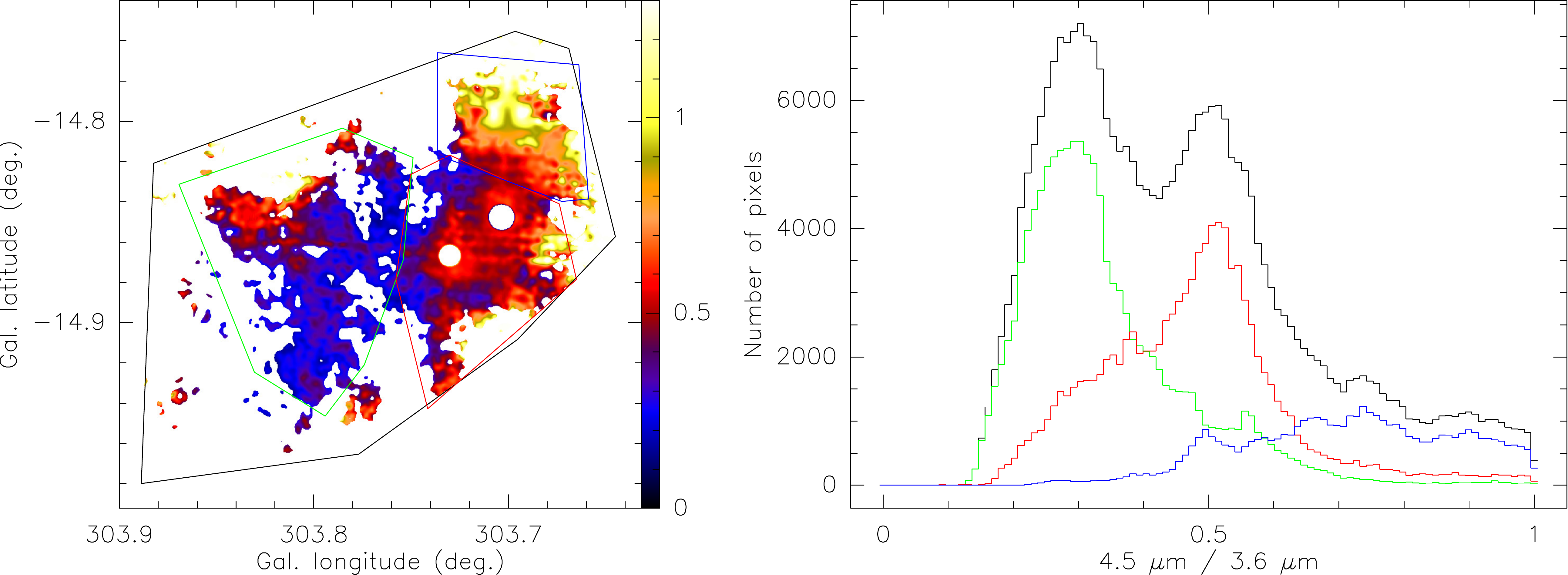}
\caption{\object{G303.72-14.86} coreshine (4.5\,\mic\ / 3.6\,\mic) ratio map and ratio histogram. Each histogram is calculated for the region delimited by the polygon of the same color.}
\label{fig:ecc815}
\end{figure*}

\begin{figure*}
\flushleft
\sidecaption
\includegraphics[height=7.3cm]{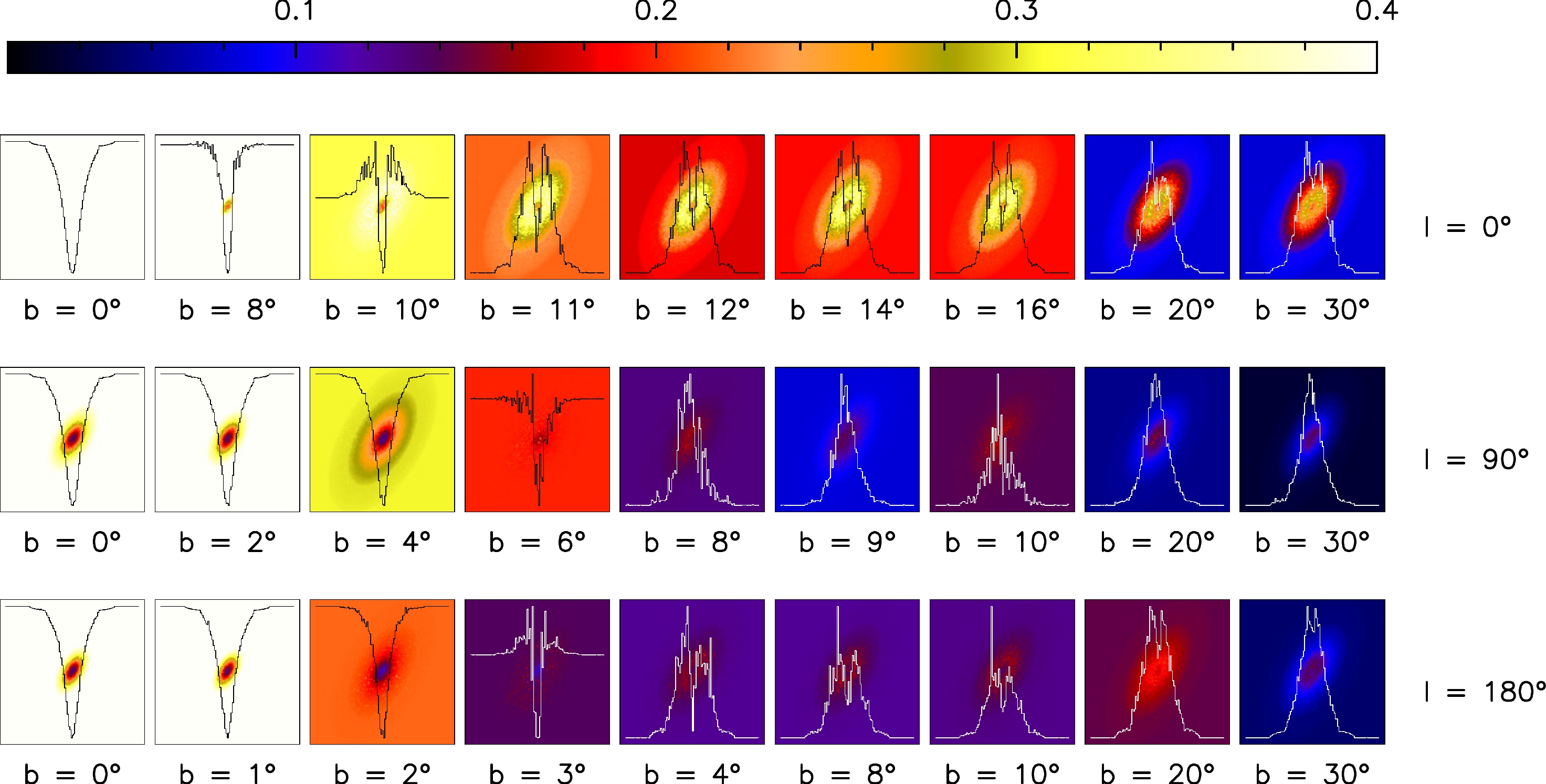}
\caption{Emergence of coreshine for three galactic longitudes and nine galactic latitudes (given under each plot in degrees). The images show the cloud with its background. The central profile (cut along x axis) helps to visualize the coreshine emergence. Fixed color scale images (MJy\,sr$^{-1}$) display in particular the background field variation (I$_\mathrm{back}$).}
\label{fig:GC}
\end{figure*}

{We investigated  two thirds of the detected coreshine cases, located in four different directions of the Galaxy, to look for regional variations, either {spatial} or physical. The two main regions we ignored (Orion/Monoceros and $\rho$ Oph) present an equal number of detections and non-detections, which indicates that these regions might be more sensitive to {specific local conditions} and are beyond the scope of this work. Figure \ref{fig:Observations} gives the 4.5\,\mic\,/\,3.6\,\mic\ ratio as a function of the 3.6\,\mic\ intensity for all the sources (values are given in Table 1). The horizontal bars represent a $\pm$\,10\,\% uncertainty {that is a conservative value of the error budget, which is dominated by the background removal}, while the vertical bars represent the range of ratios around the peak ratio value (Sect. \ref{sect:analysis} and Fig. \ref{fig:L183_sex}). 
Some of the weakest  sources  (21) have a signal--to--noise ratio at 4.5\,\mic\ which is  too low  to  safely estimate  the coreshine ratio. Their 3.6\,\mic\ intensity ranges from 10 to 40 kJy\,sr$^{-1}$. For the others, the 3.6\,\mic\ intensity ranges from 21 to 390  kJy\,sr$^{-1}$, and their ratio from 30 to 110\,\%. Starless cores are in a narrower range, 10 to 100 kJy\,sr$^{-1}$ and ratios from 35 to 51\,\% (the L1517C case -- 53\,\% -- is weak and noisy and probably not different from its neighbors, L1517 A and B -- 51\,\%). This upper limit of $\sim$ 50\,\% is well--explained by  the fact that the incoming DIRBE illumination ratio is only 70\,\% at 4.5 \mic\ with respect to 3.6 \mic\ and that the scattering efficiency of grains up to 1\,\mic\  is always lower at 4.5\,\mic\ than at 3.6\,\mic\ (Fig. \ref{fig:efficiencies}).} {On the contrary, the stronger and redder coreshine flux of many sources with local or embedded {Young Stellar Objects (YSOs)} is clearly linked to the YSOs themselves which provide more photons than the ISRF locally, with a much redder color due to their dust cocoon.}

{{However, some of the sources with nearby or embedded protostars may remain comparable to the starless cores.} There are three possible reasons: the source is i) too weak (like the VeLLO -- Very Low Luminosity Object -- in L1521F), ii) too deeply buried (like the strong jet driver L1157--mm/IRAS 20386+6751), or iii) too far outside the cloud though this is difficult to estimate since the relative position of the cloud and the YSO are not known with enough precision along the line of sight {(see Table 1 and references therein)}. {This is the case for the L1157 region which shows a low ratio (41\,\%) near the embedded driving source of its strong outflow, and partly along the southern outflow itself, while {it displays} a high ratio (71\,\%) just beyond the tip of this southern outflow where no YSO can be seen. A possible explanation is that the outflow provokes a shock region (B1 seen in H$_{2}$) which emits in the MIR with a 4.5 to 3.6 \mic\ ratio of 2.5 that might change {the coreshine ratio} of that part of the cloud. }

{For several sources the embedded object has a limited range of influence and if the core is extended, a multimodal {ratio} distribution appears (Fig. \ref{fig:ecc815}). In the cloud (\object{G303.72-14.86}) presented in Fig. \ref{fig:ecc815}, the strongest source is at the top end of the cloud, inside the blue polygon, and {two other interacting but weaker stars} are situated just below, inside the red polygon. Sources themselves are masked, hence the white spots. The situation for YSOs is quite complex, because other physical effects (e.g dust alignment due to polarization - \citealt{2013ApJ...770..151C} - and/or dust destruction) could have a profound impact on the coreshine effect. Indeed, for L1152 and L1228 there is the suggestion from NIR (JHKs) and MIR (Spitzer) that there are regions where shocks have destroyed grains  \citep{2009ApJ...690..496C}, thus potentially affecting the coreshine effect. This is probed by the fact that even though the radiation is enhanced by the local source, the 5.8 \mic\ band remains in absorption which would not be possible with too large grains.}

{Though the scattering effect does not seem to change much across the Galaxy as shown in \citet[their Fig. 4]{2010Sci...329.1622P}, we expected to find some difference between the four regions when taking into account the background and the contrast problem (see next Section). Of course, \cite{2010Sci...329.1622P} indicate scattering effects towards the Galactic center while we will see hereunder it is impossible to see any but Fig. \ref{fig:Observations} clearly shows that away from the Galactic plane and bulge, there is no difference between the regions either in absolute flux or {in ratio values} while the background values I$_\mathrm{back}$ vary by a factor up to around four between Cepheus and {Taurus--Perseus} (Table \ref{tab:Iback_table}). Moreover, the main incoming radiation field, from the Galactic center, is either behind the cloud, on its side or behind the observer (in the {Taurus--Perseus direction}, which implies some efficient backward scattering for anti-center cores, \citealt{Steinacker_detection}). The dispersion seems to be dominated by individual cloud properties instead, with little sensitivity on {the coreshine ratio}. }

{Finally, the very high proportion of coreshine cases in these regions is an interesting question. Either the grown grains are already present in the local diffuse medium of these regions and become apparent when the dust is concentrated enough or the grains coagulate relatively fast once the clouds are formed and before the turbulence dissipates in the cores {\citep{SteinackerL1506}}. In the first case, the presence of the grown grains in the diffuse medium might explain the excess of emission at 3.6 and 4.5\,\mic\ seen by \citet{2006A&A...453..969F}. This could be checked by examining the differential MIR spectrum on lines of sight sampling the diffuse medium inside and outside the Gum/Vela region for which the low number of cores with coreshine {and the high presence of PAHs} may indicate the {erosion} of large grains by the blast wave \citep{Pagani2012}. The second hypothesis would require the modeling of grain growth together with cloud contraction and turbulence dissipation}. 

\subsection{Coreshine emergence}\label{sect:contrast}

To study the coreshine {emergence} as a function of the galactic position, we use the all-sky background map obtained after stellar subtraction (following first method, see Sect. \ref{sect:back}) and we have done the modeling for several elevations towards the Galactic center direction, 90\degr\ from it (in longitude) and towards the Galactic anticenter direction (Fig. \ref{fig:GC}) for a given cloud (M$_\mathrm{low}$) and {grain size distribution} (aS1m/Cx1m) models. We easily explain that it is not possible to observe coreshine in the galactic plane because of the strong background field, even with favorable grain properties ({grain size distribution} up to 1\,\mic\ grains and  $\langle$Q$_\mathrm{sca}$$\rangle$ greater than  $\langle$Q$_\mathrm{abs}$$\rangle$). For a galactic longitude ($\mathrm{l}$) of 0\degr, the bulge dominates the background field so strongly  that coreshine is not able to appear until a galactic latitude ($\mathrm{b}$)  around 10\degr. On the contrary, in the anti-center direction coreshine is able to appear rapidly with elevation, as low as $\mathrm{b}$  $\ge$ 3\degr. {When the main illumination field comes from behind, at small angles, coreshine appears for higher background values than for other illumination directions ($\sim$300 \kjy instead of $\sim$120 \kjy, Fig. \ref{fig:GC}).}
{The values are compatible to what was found by \cite{Steinacker_detection}, here we also confront these criteria directly to the detections.} This modeling correctly explains the observations (Fig. \ref{fig:planck}) but the $\mathrm{b}$ values given in Fig.~\ref{fig:GC} have to be taken qualitatively since the values can change with the presence of local sources, the grain properties and the background proportion estimates. Finally, of all the clouds that contain enough micron-size grains  to efficiently scatter the MIR light, only those outside the Galactic plane are detectable via their coreshine emission and the 50$\%$ detection of positive coreshine cases (\citealt{2010Sci...329.1622P}, Paladini et al. in prep.) is only a lower limit in terms of grown grains in clouds\footnote{for the same contrast  issue the investigation of coreshine in other galaxies could be a challenge.}.

Beyond the galactic plane contrast problem, clouds will also appear in absorption if  $\langle$Q$_\mathrm{sca}$$\rangle$ is lower than $\langle$Q$_\mathrm{abs}$$\rangle$. Nevertheless, there are not a lot of cases in real absorption outside of the Galactic plane in our data collection (Table \ref{tab:sources}).  Scattering is always present in dark clouds, even when the cloud is seen in absorption. 
The coreshine phenomenon starts to appear when the scattering signal is able  to exactly compensate the extinction of the background field. Incidentally, clouds would totally disappear if this equilibrium was reached {at MIR wavelengths}. 
The coreshine phenomenon could be understood either by the presence of large grains ($\sim$ 0.5 -- 1\,\mic) or a stronger local radiation field which would enhance only I$_\mathrm{sca}$ (eq. \ref{eq:trans}). The enhancement of the local radiation field has to be considered cautiously since it has to be consistent with far--IR emission of the clouds. Here, we are not refering to embedded YSOs but to the large scale local radiation field surrounding starless cores. Incidentally, \citet{2001ApJ...557..193E} show that for a selection of prestellar cores in Taurus and $\rho$ Oph regions, far-IR observations seem to require a lower ISRF than standard to be fitted. {On the contrary} for the $\rho$ Oph region, \citet{2014A&A...562A.138R} argue that the ISRF is one order of magnitude higher than standard.}

At 5.8 and 8\,\mic, all the clouds of the cold Spitzer survey \citep{2010Sci...329.1622P} appear in absorption. While at 8\,\mic, this is partly due to the 9.7\,\mic\ silicate absorption feature wing that intercepts half of the 8\,\mic\ filter width, at 5.8\,\mic, it is {only} due to 
the background field strength. Indeed, the diffuse part I$_\mathrm{diff}$  increases globally more and more with the wavelength (Table \ref{tab:Iback_table}) in the IRAC bands range as can be seen in \cite{2006A&A...453..969F}. In parallel to this increase of the diffuse background field, the scattering efficiencies drop. {This provides additional constraints, in particular our modeling can eliminate the grain models for which the 5.8\,\mic\ map shows emission. This is especially true for distributions which include grains in the range 2--5~\mic\ for both silicates and carbonaceous grains (aS5m/Cx5m) and puts an upper limit for the grain size based on the 5.8 \mic\ diagnostic deep inside the core. Nevertheless, this limit in size has to be taken cautiously since it is dependent on the grain size distribution law and on the grain composition. It has to be understood as un upper limit on the abundance of grains larger than 2\mic, and for example, the WD55B distribution stays compatible with absorption at 5.8 \mic.} The 8\,\mic\ map always appears in absorption in our simulations and is not, as expected, sensitive to the grain properties but could be used to constrain the silicate column density.

\subsection{{Intensity ratios as {discriminants}}}\label{sect:ratio}

{We aim to separate the grain models according to their ability to explain the observed {intensity ratios between bands}. We start from the coreshine ratio (4.5/3.6) and extend the method to a comparison between NIR and MIR bands (K/3.6) and to NIR ratio (J/K ratio). The four modeled directions cover a range of local properties like the Galactic Center direction, and the differential local background field $\mathrm R_{\lambda1/\lambda2}(\mathrm I_{back})$ = I$_\mathrm{back}$($\lambda_1$) / I$_\mathrm{back}$($\lambda_2$) (Table \ref{tab:Iback_table}).} 

{To measure the ratios, we took the median value inside a polygon on the simulated map ratios. We obtain the model uncertainty by comparing these ratios computed with different I$\mathrm{_{back}}$ estimates (obtained with the second method, Sect \ref{sect:back}), namely the center pixel values versus the nine pixel average values, and the two bg estimates (bg1, bg2) (Table \ref{tab:Iback_table}). {The suitable grain models are the ones which satisfy all the criteria taking into account the uncertainties on the background variation.} For the M$_\mathrm{low}$ cloud model (see Section \ref{sect:cloud}), the chosen polygon is very central and corresponds to a mean value of \Av\ = 23 mag (assuming Rv = 5.5) while the mean polygon extinction for the M$_\mathrm{high}$ model is about \Av\ = 21 mag chosen to be between the two layers and outside the internal depression. }

\subsubsection{Grain properties deduced from the coreshine ratio}

{We plot the coreshine ratio derived from the model calculations as a function of the 3.6 \mic\ coreshine intensity for all the grain models and separately for the four modelled directions corresponding to the observed regions (Fig. \ref{fig:compar} to \ref{fig:Cepheus}). All the models that do not show coreshine at both 3.6 and 4.5\,\mic\ are not displayed on the ratio plots, especially WD31, ORI2, ORNI2, YSA models. Figure \ref{fig:compar} confirms that the two different cloud models that investigate different zones with {close averaged visual extinction} give coherent results {and that we correctly represent the observational range}. Moreover, the plots are not really sensitive to the background proportion itself {concerning the coreshine ratio range} (bg = 95 or 75\,\% for {the Taurus--Perseus region}, Fig. \ref{fig:compar},  and bg = 75 or 50\,\% for the L183 direction,  Fig. \ref{fig:L183}). {However, this implies some degeneracy in the grain property solutions for some regions (Fig. \ref{fig:compar}). Conversely, regions where the I$_\mathrm{sca}$ term is dominant (eq. \ref{eq:trans}, \ref{fig:L183}) give  more reliable results. This regime is reached for the L183 and Cepheus regions.} {For the following,} we keep the estimates by \citet{2014A&A...561A..91L} as a reference (bg1, Table \ref{tab:Iback_table}).}

\begin{figure*}
\centering
\includegraphics[trim = 0.85cm 0.3cm 1.3cm 0.5cm, clip,width=14cm]{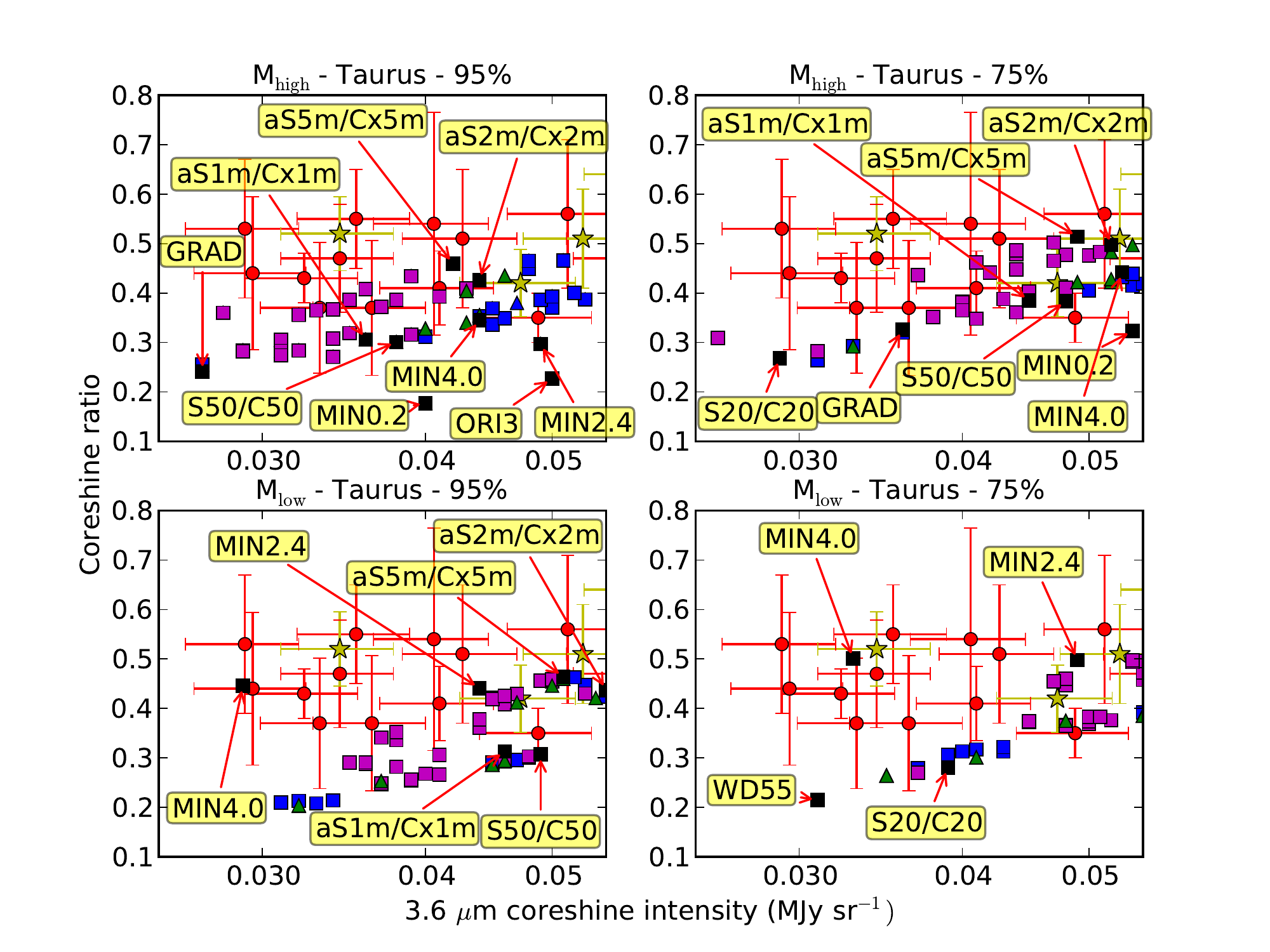}
\caption{{Coreshine ratio} versus 3.6\,\mic\ intensity {in the Taurus--Perseus direction. Zoom on the starless core cases, full range is displayed in Fig. \ref{fig:Taurus} }- Difference between M$_\mathrm{high}$ {(upper row)} and M$_\mathrm{low}$ {(lower row)} modeling without internal source for the two background fractions {(95\%, bg1, left, 75\%, bg2, right)}. OBSERVATIONS - red circles: starless cores; yellow stars: cores with YSOs. MODELS - black squares: single grain population; blue squares: carbonates bigger than silicates in at least one of the two layers; purple squares: silicates bigger than carbonates in at least one of the two layers; green triangles: everything else. {Grain model names refer to Table \ref{tab:grain_size} for spatially constant grain size distribution.}}
\label{fig:compar}
\end{figure*}

\begin{figure*}
\centering
\includegraphics[trim = 0.85cm 0.3cm 1.3cm 0.5cm, clip,width=14cm]{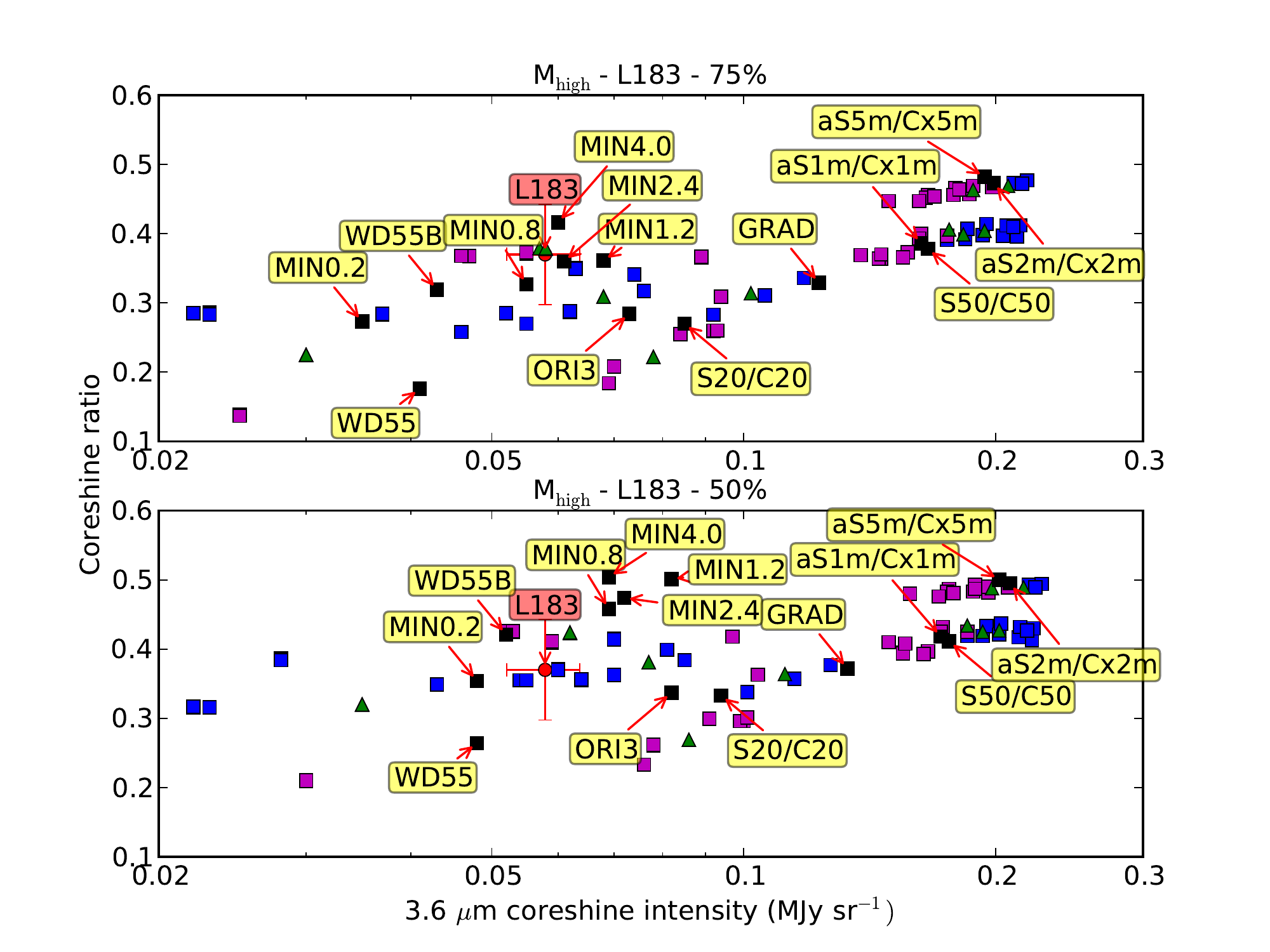}
\caption{Same as Fig. \ref{fig:compar} in L183 direction {with the two background proportion estimates: 75\% (bg1, top), and 50\%  (bg2, bottom).}}
\label{fig:L183}
\end{figure*}

The first question is to determine whether small grains have any impact on the coreshine modeling. We found that no differences appear between models which contain small grains between 4 and 10 nm and the ones without grains below 10 nm (from aSil to S10 grains and from CBx2 to C10, Table \ref{tab:grain_size}, \ref{tab:solutions}). Removing the small grains  at constant dust mass has been tested up to 50 nm (S50, C50) and the results are very close to the ones with the same high cut--off (a$_\mathrm{cut}$ = 1\,\mic) but with a starting size of 4 nm (aS1m, Cx1m), especially when taking into account the uncertainties {(see e. g. Fig. \ref{fig:L183})}. The suppression of  the smaller grains is therefore not mandatory in explaining the coreshine phenomenon because they do not contribute nor attenuate the signal significantly.

\begin{figure*}
\centering
\includegraphics[trim = 0.85cm 0.3cm 1.3cm 0.5cm, clip,width=14cm]{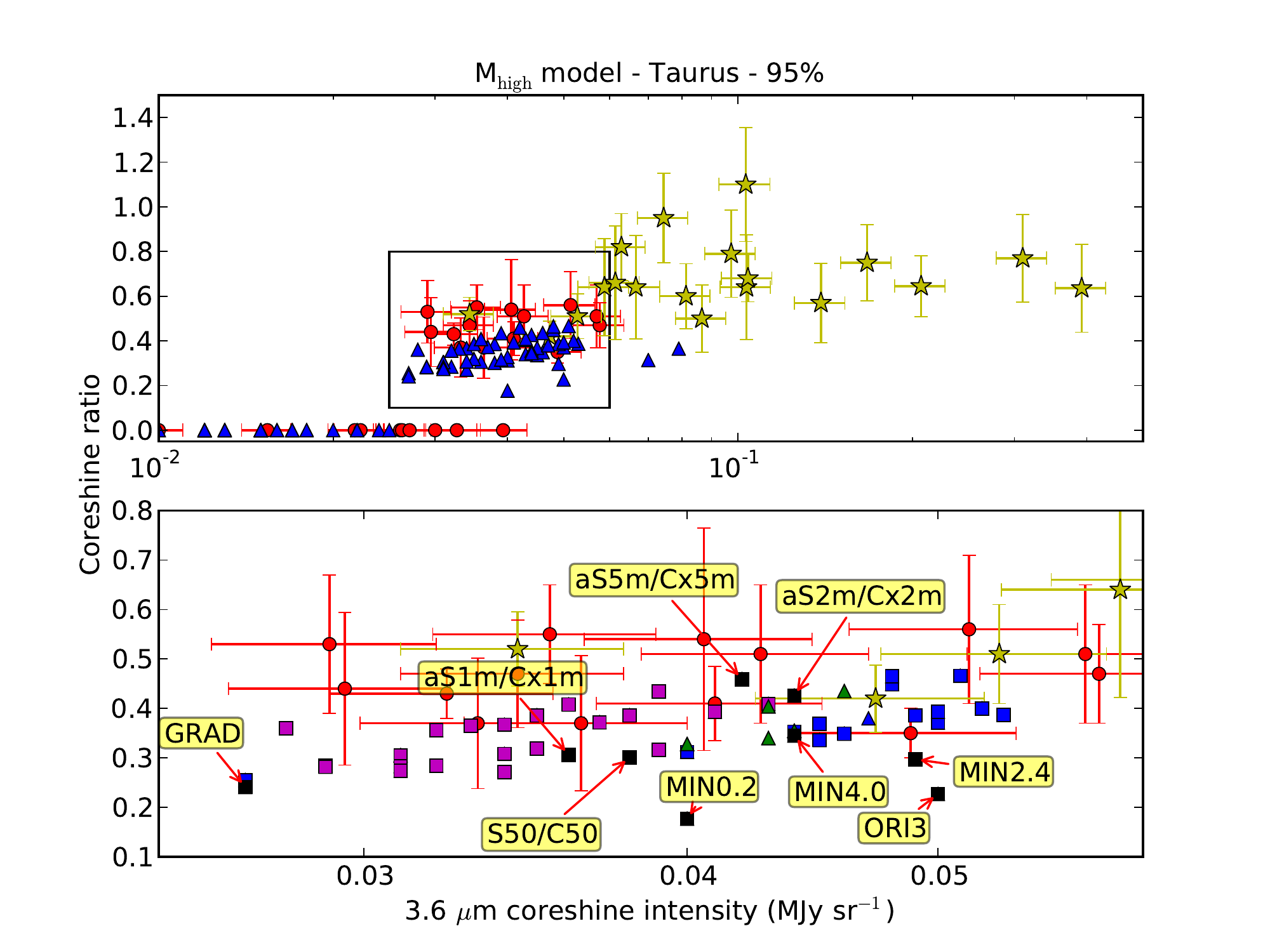}
\caption{{\bf Top:} observations, same as Fig.  \ref{fig:compar} in {Taurus--Perseus direction}. Blue triangles: all (undifferentiated) starless models. {\bf Bottom}: zoom of the rectangle in the top image, symbols identical as in Fig. \ref{fig:compar}}
\label{fig:Taurus}
\end{figure*}

\begin{figure*}
\centering
\includegraphics[trim = 0.85cm 0.3cm 1.3cm 0.5cm, clip,width=14cm]{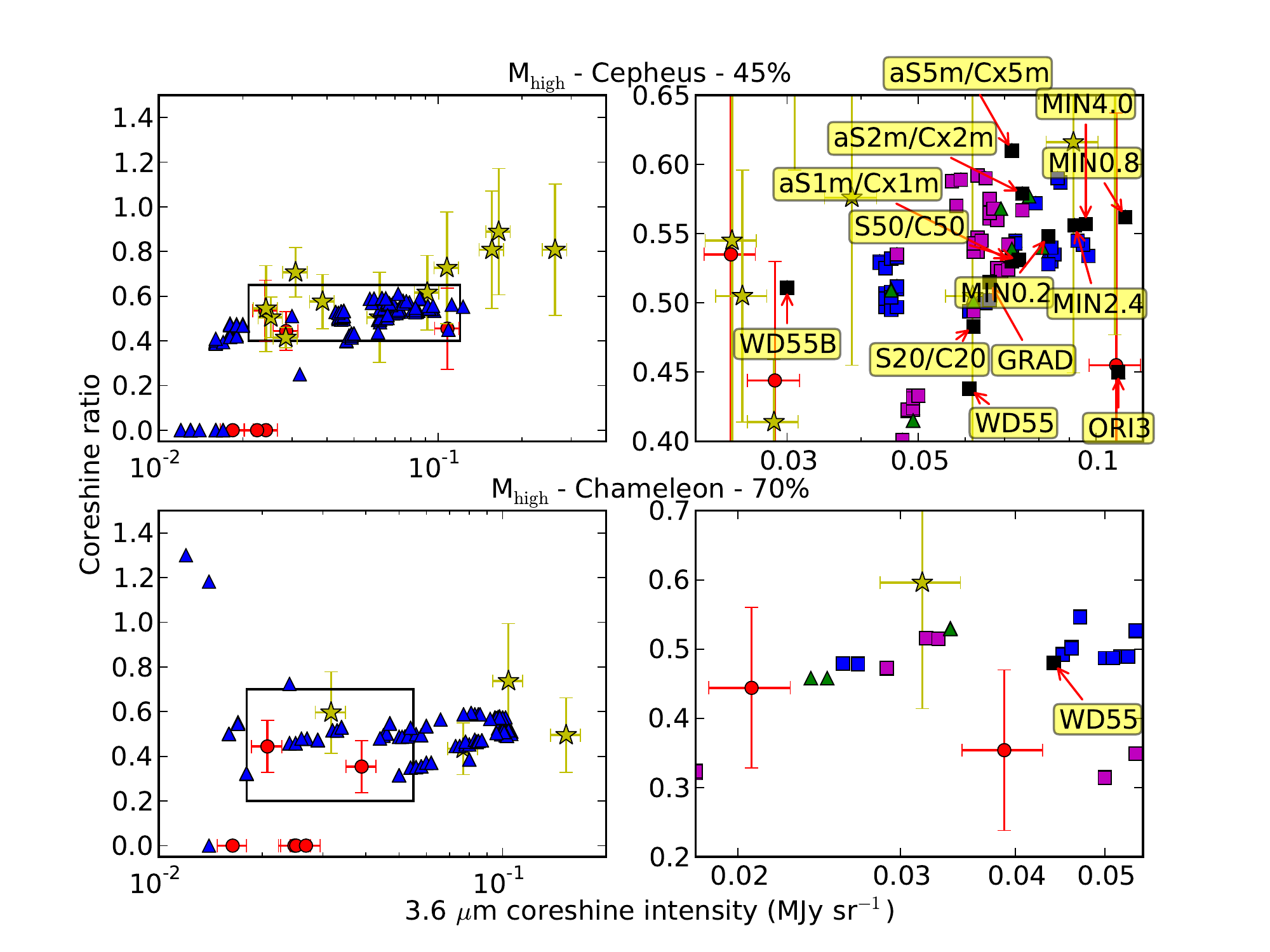}
\caption{Same as Fig. \ref{fig:compar} in the Chameleon and Cepheus directions: {top Cepheus region, top right: zoom on the starless observational zone, bottom : Cepheus region with its zoom (bottom right)}.}
\label{fig:Cepheus}
\end{figure*}

On the other hand, we tested the large grain size increase beyond the expected limits \citep{2013A&A...559A..60A}. The more the mean size of the distribution $\langle$a$\rangle$ increases, the more the 3.6\,\mic\ flux increases as well. The grain models on the right side of the ratio plots (Fig. \ref{fig:compar} to \ref{fig:Cepheus}) are those with the highest number of big grains, with an exponential cut-off at respectively 1\,\mic\ (aS1m, Cx1m){\bf ,} 2\,\mic\ (aS2m, Cx2m){\bf ,}  or 5\,\mic\ (aS5m, Cx5m). {However, when the high cut--off of the grain size distribution increases, the coreshine ratio starts to saturate and consequently grain growth alone is not a pertinent answer to increase the ratio beyond 50\,\% for {the Taurus--Perseus} and L183 regions (Fig. \ref{fig:L183}} and \ref{fig:Taurus}) and 65\,\% for Cepheus and Chameleon regions (Fig. \ref{fig:Cepheus}). This is well--explained by the combination of three effects. Firstly, the integrated illumination field intensity at 4.5\,\mic\ deduced from the scaled DIRBE map is only 70~\% of the 3.6\,\mic\ intensity. Secondly, an I$_\mathrm{back}$ lower at 4.5\,\mic\ than at 3.6\,\mic\  helps to increase the ratio as explained in Sect. \ref{sect:contrast}. Since the ratio values of I$_\mathrm{back}$ at 4.5 and 3.6\,\mic\, $\mathrm R_{4.5/3.6}(\mathrm I_\mathrm{back})$, stay around 1 for the {Taurus--Perseus region} and L183 direction, they have a minor impact on the coreshine ratio, while the lower I$_\mathrm{back}$(4.5\,\mic)  towards Chameleon and Cepheus provides an explanation of why ratios near 65\,\% are reached (Fig. \ref{fig:Cepheus}). 
In the third place, the ratio of the scattering and absorption efficiencies (Fig. \ref{fig:efficiencies}) becomes flatter with wavelength for both grain types when the grain size is increased up to 5\,\mic.

We also confirm what has been found previously (\citealt{2010Sci...329.1622P}, \citealt{2010A&A...511A...9S}). Classical diffuse medium grains (WD31 \citealt{2001ApJ...548..296W}, aSil and CBx2) are not efficient enough to scatter in the MIR range. Indeed, it is well known that the ratio of visual extinction to reddening, R$_\mathrm{V}$ = 3.1, is not valid in dense, cold environment and R$_\mathrm{V}$ = 5.5 has been advocated  \citep{2001ApJ...548..296W}. This change of slope has been explained by grain growth. While the WD55 model gives  a 3.6\,\mic\ coreshine flux between 20 and 50 kJy\,sr$^{-1}$ and a coreshine ratio about 20\,\%, depending on the local conditions (direction and I$_\mathrm{back}$ values), the WD55B model, which includes grains up to 10\,\mic\ in size,  has been found promising by other observations in the MIR \citep{2012arXiv1211.6556A}, and allows the coreshine ratio to increase up to 40\%.

The previous results were obtained for compact spherical grains, and we also wanted to investigate  the consequences of the fluffiness and of the coagulation, which has to be understood here as an agglomeration of smaller grains. We tested the evolved grains of \cite{2009A&A...502..845O},   with or without ices,   with different evolution time scales. We aimed to compare three different populations: i) a reference population, without ices, and an evolution time scale of 1 \pdix 5 years (ORNI2)  which happened to produce no coreshine in the modeling, ii) another one starting from the same distribution and which evolved in the presence of ice mantles (ORI2), and iii) one which has grown during a longer timescale of 3 \pdix 5 years with ices (ORI3). Despite the presence of ices which favor coagulation, the ORI2 {grain size distribution} has not had enough time to reach a sufficient size (Fig. 1 in \citealt{2009A&A...502..845O}) to make coreshine appear as our models confirm. On the contrary, the ORI3 model traces some evolution like the WD55B model and shows a ratio compatible with the observations. For these two models, even though the ratio is compatible with the observations, the 3.6\,\mic\ flux is higher than observed  which is acceptable since  we could always adjust the cloud model in order to obtain the right flux with an identical ratio.

What appears  from these comparisons is the fact that the first key role of ices is to favor growth but the {3.05\,\mic\ H$_2$O ice feature included in the edge of the IRAC filter could also play a role} in increasing the coreshine ratio {by changing the dust optical properties in the same way} as the silicate feature enhances the absorption in the 8\,\mic\ filter. Since, without ices, the maximal size reachable in Ormel bare grain models is about 1\,\mic, comparable to the size obtained with the ORI2 model which produces no coreshine, it is not possible to compare the ORI3 model to the same size distribution without ices to disentangle the ice effect from the pure growth effect. 

{Scattering optical properties are supposed to be more sensitive to grain surface variations than absorption efficiencies. Consequently, we expect to see a direct impact of the flufiness on the coreshine. To calculate optical properties for fluffy grains, one can focus on the porosity with a simple approach where the proportion of silicates, carbonaceous grains, and vacuum in the grains do not depend on their sizes (\citealt{2012A&A...542A..21Y}) or one can explore the fractal dimension which means to focus on the dissymetry of the grains themselves build by agglomeration of monomers \citep{2006A&A...445.1005M}.} 
%To investigate the impact of fluffiness alone, 
%We compared our set of DustEm modified models to bare aggregates taken from the literature \citep{2006A&A...445.1005M,2012A&A...542A..21Y}. 
First, we studied the influence of the degree of porosity. The grains from \cite{2012A&A...542A..21Y} which could explain the far-IR emission in L1506C \citep{2013A&A...559A.133Y}, a condensation in a Taurus filament,  are not efficient enough to produce coreshine at 4.5\,\mic\ while the 3.6\,\mic\ coreshine flux, about 13 kJy\,sr$^{-1}$ (compared to 33 kJy\,sr$^{-1}$, Table 1), does not vary from 0\,\% porosity to 40\,\% porosity (YSA models). Because porosity does not seem to be an answer to change the 3.6\,\mic\ coreshine intensity, and correspondingly the coreshine ratio, we tried to test the influence of fractal structure. As preliminary results, we tested fractal aggregates which show optical properties quite different from the compact spherical grains \citep{2006A&A...445.1005M}. These fluffy monomer aggregates behave like small spherical silicate particles for the 9.7 \,\mic\ silicate feature (Min et al., in prep.).{ Their sizes vary from 0.2\,\mic\ to 4\,\mic\footnote{private communication, Min et al. (in prep)}}.

We had to approximate the real phase function provided by Min et al. by an equivalent asymmetry coefficient ($\langle$g$\rangle$, eq. \ref{eq:HG}).
As it will be discussed below, the phase function is not expected to modify the coreshine ratio by much, especially because this approximation applies to both wavelengths. Indeed, the two coreshine wavelengths are close enough so that the respective variation in the phase function is weak. These grains are really promising since they do not necessarily imply a significant grain growth to produce a ratio about 40 to 50\,\% and could be an interesting answer for clouds which would not show much evolution from other tracers like depletion (e.g. L1521E \citealt{2004A&A...414L..53T}, \citealt{2010Sci...329.1622P}). \emph{Furthermore, we find the same saturation of the ratio for the agglomerates as for pure spherical grains with grain growth, which seems to confirm a maximal size efficiency as can also be deduced from Fig. \ref{fig:efficiencies}. Fractal aggregates behave like big spherical grains to produce a high coreshine ratio.}

To limit the number of models, we restrained ourselves to two layers for the part dedicated to the core (Sect. \ref{sect:cloud}), that is why we do not aim to fit exactly the observations. We introduced a finer slicing in 10 layers inside the core, filled with grains presenting a high cut-off increasing from 0.2\,\mic\ to 1\,\mic\ (GRAD model). We obtained a ratio {close to} a  two-layer core model with a$_\mathrm{cut}$ of 1\,\mic\ (aS1m and Cx1m mixed) but with {less flux} {(see e.g. Fig. \ref{fig:Taurus})}, which is expected for the observations. \emph{This emphasizes again the fact that the cloud model and the way we fill it with several layers would act only on the absolute coreshine fluxes without modifying the ratio}. 

The phase function can {slightly} change the ratio especially if backward scattering becomes a non--negligible quantity. This is the case for {the Taurus--Perseus region} for which the main anisotropic radiation source (the Galactic Center) comes from behind us. Nevertheless, even if we consider an extreme case with $\langle$g$\rangle$ = -0.99 which corresponds to a probability of 75\% to have axial backward scattering, the ratio in {the Taurus--Perseus} region, with the same $\langle$ Q$_\mathrm{sca}$ $\rangle$ and $\langle$ Q$_\mathrm{abs}$ $\rangle$ absolute values, increases from 35 to 40\,\% for grains up to 1\,\mic\ (aS1m, Cx1m). The variation is even smaller for less extreme probability changes. In the most extreme case, even if the ratio is not really sensitive to the phase function modification, the absolute intensity at 3.6\,\mic\ has been enhanced by almost a factor 30 to reach a few hundred kJy\,sr$^{-1}$. {Incidentally, the emergence of coreshine for some dust models will be dependent on this parameter.} We estimate that the Henyey-Greenstein approximation is satisfactory
 {for most of the directions} since the ratio is only slightly sensitive to phase function variations. 
{{We will investigate the impact on the absolute flux by considering the true phase function when we perform a full cloud model, especially for regions in the regime where the influence of I$_\mathrm{back}$ is dominant (Sect. \ref{sect:contrast}).}}

\begin{figure*}
\centering
\includegraphics[width=13cm]{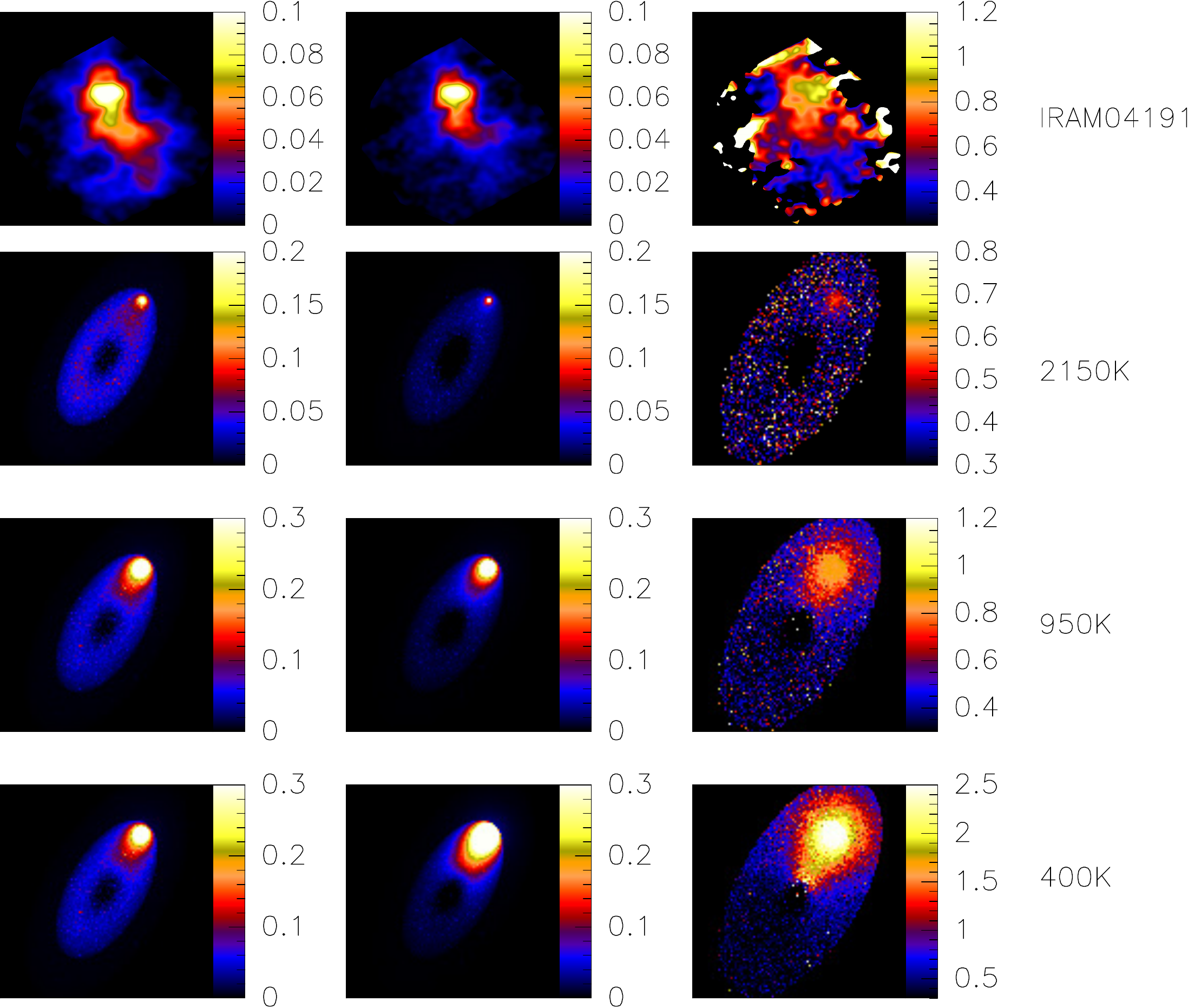}
\caption{From left to right: IRAC 1 (3.6\,\mic), IRAC2 (4.5\,\mic), coreshine ratio. First row: analysis of IRAM04191 Spitzer observations. Next rows: Modeling in the dense case (M$_\mathrm{high}$) for the three internal sources. {Color scales are given in \mjy.}}
\label{fig:sources}
\end{figure*}

\subsubsection{Impact of a local source}\label{sect:local_source}

Since the presence of an embedded object is often found to enhance the coreshine ratio and the coreshine intensity in its vicinity (Table 1, Fig. \ref{fig:Observations} and {\ref{fig:ecc815}}), we explored the influence of a Class 0 or a Class I object in our model.
{The full treatment of a protostar reddened by its compact and dense dust cocoon is beyond the scope of this work. To investigate its possible effects qualitatively, we inserted blackbodies (stars) in the model  with temperatures and fluxes typical of Class 0 or I objects at coreshine wavelengths.} We ran the test with three different temperatures, 400, 950 and 2150 K and adjusted the flux accordingly. {At 400 K, the 4.5/3.6 ratio is 3.8 and is representative of a Class 0 case as it could be observed for \object{IRAM04191} (\citealt{Chen:2012dr}). At 950 K, it is almost flat (1.2)  and we obtain a flux and {a coreshine ratio} comparable to what could be deduced for a Class 0/I object like \object{IRAS 04016+2610} or \object{IRAS 04361+2547}  \citep{2007ApJS..169..328R}. At 2150 K, it is typical of a solar type Class I protostar and its 4.5/3.6 ratio is 0.8, like \object{IRAS 04325+2402} \citep{2007ApJS..169..328R}, close to the DIRBE all-sky ratio of 0.7. The Class distinction is not very important here since it depends very much on the geometry of the source and other factors that are beyond the scope of this paper. Several papers discuss these classification problems  \citep[e.g.,][]{2007ApJS..169..328R,0004-637X-692-2-973,2009ApJS..185..198K}}.
 
For the densest cloud model (M$_\mathrm{high}$), we introduced, at the edge of the cloud, these three different stars. { The test with the Class I object (2150K)  shows a small enhancement of the coreshine ratio in the vicinity of the object (Fig. \ref{fig:sources}). With the Class 0/I object at 950K, the ratio is increased to $\sim$0.8. Finally, with the Class 0 object (400K), depending on its actual flux, the ratio can vary from below 0.5 for a negligible contribution (as in L1157 near the embedded object) to {$≥$ 2} if the local source flux dominates.} We therefore can clearly reproduce the trend of the observations. However, we are not trying to make exact fits, and to go further in the modeling one has to constrain the local radiation field emanating from the embedded source to avoid the degeneracy between the grain properties and the local source flux and color.

{Indeed, the question of whether the presence of an embedded protostar can be helpful or not in modeling the coreshine effect arises. If the protostar is weak and deeply embedded, its properties will be difficult to assess and its impact on the cloud scattered light will only add another degree of freedom. Similarly, if the source is clearly outside of the core, its distance to the cloud along the line of sight will remain a free parameter.
If the YSO is clearly embedded in the cloud, but not too deeply,  or in contact with it, so that its contribution can be measured directly, little uncertainty is added. This has  {three more advantages:} Firstly, the ISRF becomes dominated by the local source and is therefore better constrained, secondly, as the scattered flux will be higher, the error due to the background uncertainty in the modeling becomes smaller and, thirdly the 5.8 \mic\ criterium becomes even more constraining in enhanced ISRF conditions. Indeed, a stronger ISRF increases the capability to produce coreshine (either smaller grains can shine or big grains shine more). In a local stronger field, the non-detection of emission at 5.8 \mic\ puts therefore a more stringent constraint on the abundance of grains above 1 \mic\ as presented in Sect. \ref{sect:contrast}. However, the presence of a Class 0 or Class I object indicates a more evolved cloud and, presumably, more evolved grains. Therefore studying starless clouds or protostellar clouds are both important and have both caveats.

\subsubsection{Extension to the NIR}

{The information obtained from the coreshine ratio}, which is more sensitive to the grain models than to the other free parameters, can be extended to NIR wavelengths. {Specific studies on  NIR scattering have been done before (\citealt{2014A&A...563A.125M} and references therein).} Our approach assumes that the different wavelengths are close enough to investigate the same volume of the cloud and far enough to see a variation in the slope for the different grain types (Fig. \ref{fig:efficiencies}). Because our NIR observations are limited to a few cases, we only set a range for the expected {ratio values} which we admit does not provide much constraint.
 Nevertheless we attempted to see if the multi-wavelength approach could lead to a sample of suitable grain models.

{First, we studied the K \,/\,3.6\,\mic\ (= NIR/MIR) ratio as a function of the coreshine ratio towards L183 and {the Taurus--Perseus complex}. The observational range obtained for this ratio is between 3.5 and 6 for L183 and 3.2 and 7 for {Taurus--Perseus region} (Fig. \ref{fig:nirmir_L183} and \ref{fig:nirmir_Taurus}).} This eliminates models with a valid coreshine ratio but a NIR/MIR ratio which is too low (ellipse, Fig. \ref{fig:nirmir_L183}). In both directions, the selected models correspond to relatively {large grain size distributions}, which is what we expected since the K band is supposed to sample a region close to the core (see Fig. \ref{fig:model16} and \citealt{2013A&A...559A..60A}). Nevertheless, when {the grain size distributions} contain too many large grains (aS2m/Cx2m, aS5m/Cx5m) the NIR/MIR ratio becomes too low. Actually, grain growth on spherical grains lowers the NIR/MIR ratio because when $\langle$a$\rangle$ increases, the K intensity increases but more slowly than the 3.6\,\mic\ coreshine intensity. {This applies mainly to clumps with A$_\mathrm{V}$ $>$ 10 mag, when the K band already suffers significant saturation.} In the same way, the fractal aggregates, even the 0.2\,\mic\ monomer aggregates, also seem to predict a  NIR/MIR ratio that is too low. On the contrary, ice mantles (ORI3) give a higher NIR/MIR ratio and are one of the more suitable grains for {the Taurus--Perseus region}. In both directions, {grain size distributions} which contain bigger silicates than carbonates are more likely to be the ones able to explain the observations (purple squares - Fig. \ref{fig:nirmir_L183} and \ref{fig:nirmir_Taurus}). 
{Some spatially constant grain size distributions} are also able to explain them, especially the ones up to 1\,\mic\ with or without ices (aS1m/Cx1m, ORI3). Finally, the L183 direction is well explained by silicates up to 1~\mic\ and with different grain size distributions for the carbonates (from a high cut-off of 0.15\,\mic, CBx2, to 1\,\mic, C50) while {the Taurus--Perseus direction} can tolerate a larger variety of grains as a solution, for example model A is one with standard grains in the external core layer and grains up to 1\,\mic\ (S50/C50) in the inner layer {(Table \ref{tab:solutions})}. In this case only the second layer is visible in coreshine (Fig. \ref{fig:model16}) and almost the same region is emitting in K band while the J signal is more extended.

\begin{figure*}
\centering
\includegraphics[trim = 1.5cm 0.5cm 1.3cm 0.5cm, clip,width=14cm]{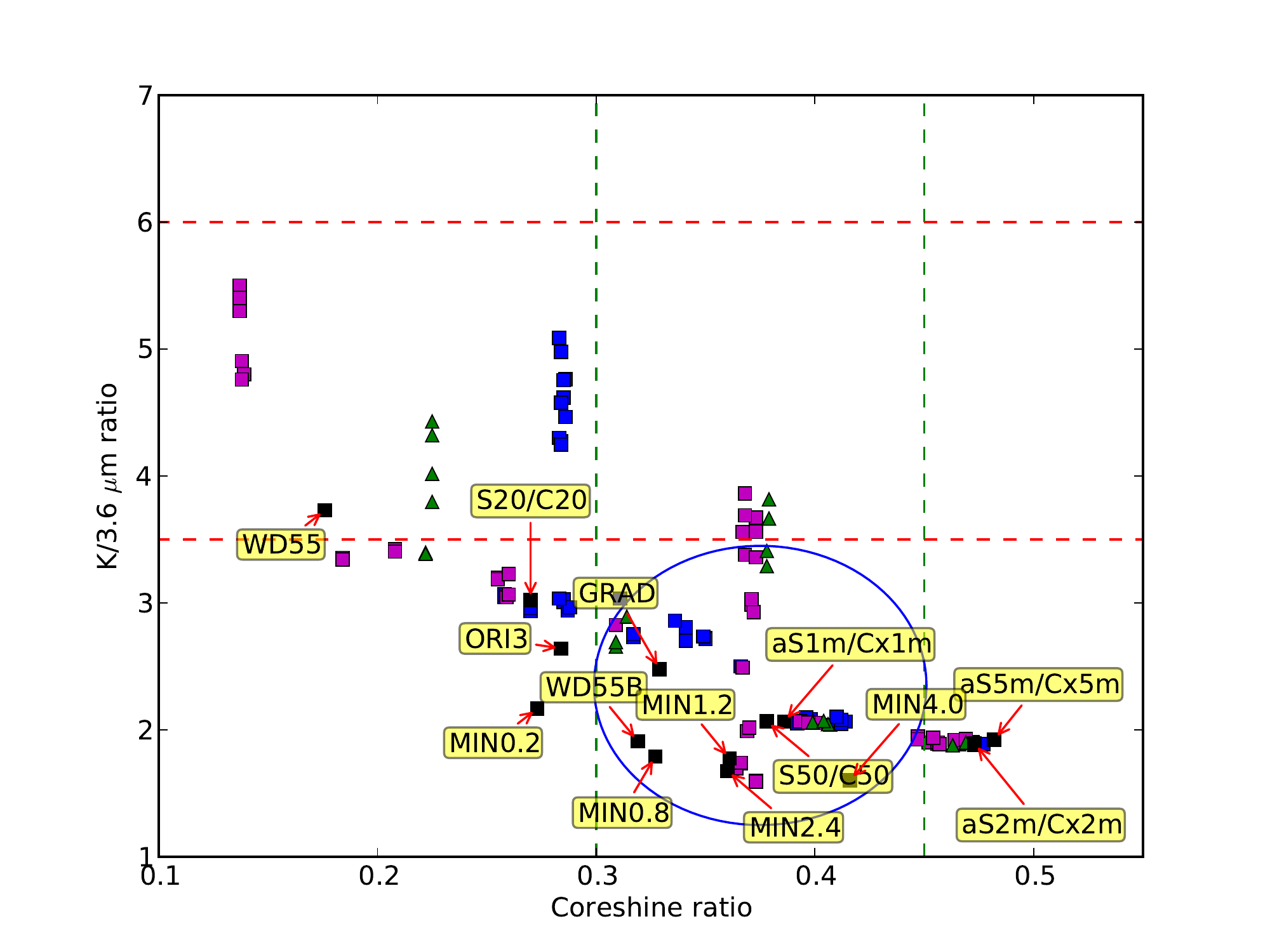}
\caption{{NIR/MIR ratio versus coreshine ratio} plot for L183 for the M$_\mathrm{high}$ model. OBSERVATIONS:  red dashed lines: observational range of the NIR/MIR ratio, green dashed lines: observational range of the coreshine ratio. MODELS: same as Fig. \ref{fig:L183}. Blue ellipse delineates the coreshine compatible grains which are eliminated from the NIR/MIR ratio.}
\label{fig:nirmir_L183}
\end{figure*}

\begin{figure*}
\centering
\includegraphics[trim = 1.5cm 0.5cm 2.cm 1cm, clip,width=14cm]{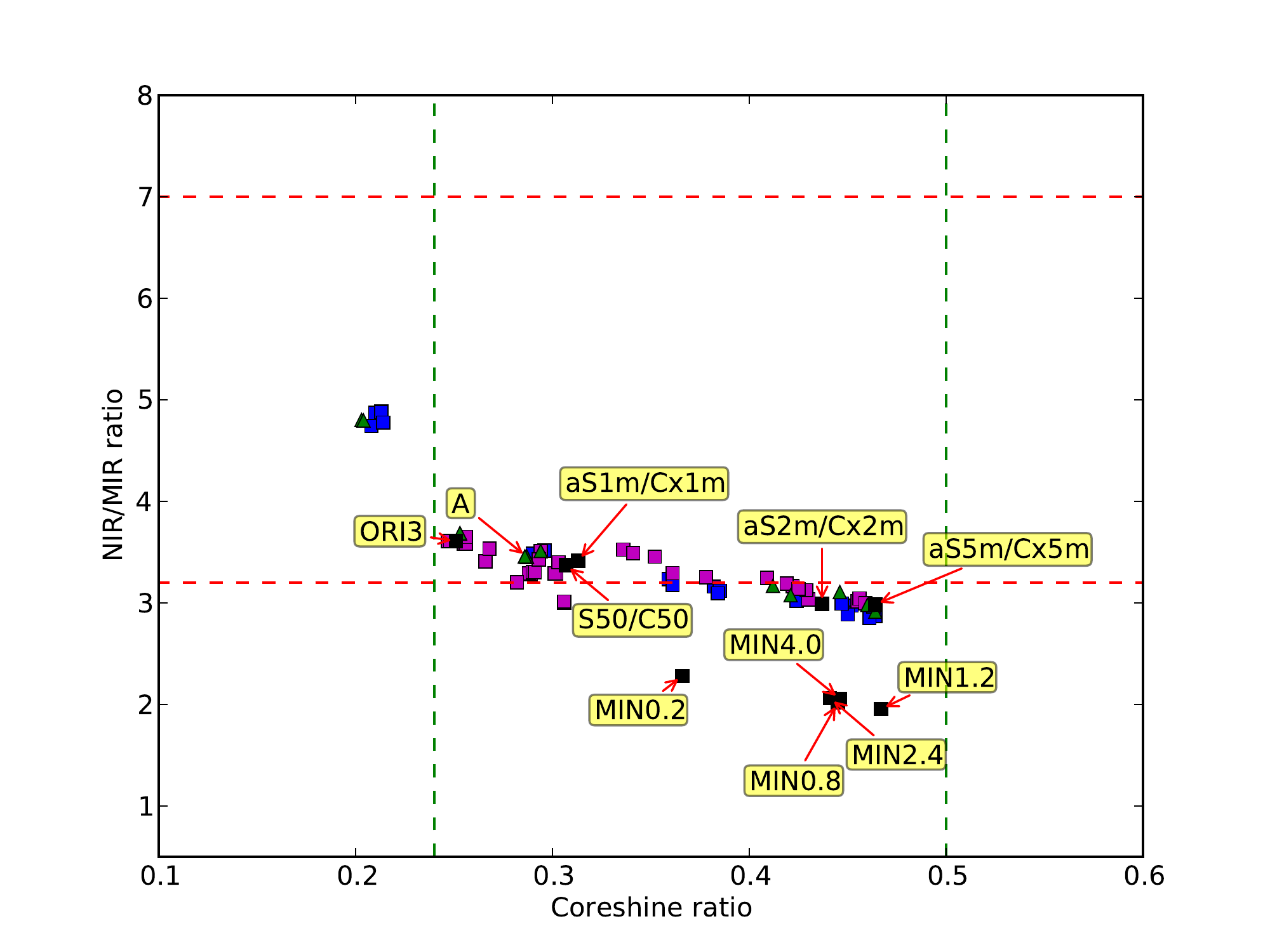}
\caption{Same as Fig. \ref{fig:nirmir_L183} for {the Taurus--Perseus} region with the M$_\mathrm{low}$ model.}
\label{fig:nirmir_Taurus}
\end{figure*}

The simple cloud model we are using could be a limitation to investigating shorter wavelengths which are supposed to be more sensitive to the smaller grain part of the distribution contained in the extended envelope. Nevertheless we obtained interesting results about the J/K ratio (NIR ratio). The typical observational range obtained for the J/K ratio is from 0.3 to 3 and this ratio is highly position dependent which is well--reproduced in the modeling (model A, Fig. \ref{fig:model16}). Globally, it shows a bimodal distribution: one around 3 for the part dominated by the J region of emission and one below 1 for the central part dominant at the K wavelength. In particular, the model output median values for this ratio, which averaged this bimodal behavior, varies from 0.3 to 0.9 only (Table \ref{tab:solutions}), for the remaining suitable grains. Therefore it could not be used as an indicator to differentiate the models and is supposed to be more sensitive to the cloud structure than to the grain properties.

{In conclusion, the preliminary constraints put by the multi-wavelength approach give us suitable grain models as solution for {the Taurus--Perseus and L183 regions} in standard ISRF illumination conditions. L183 ratios are well reproduced by grain models which contain bigger silicates or a mix between bigger carbonates in one layer and bigger silicates in the other layer while the {Taurus--Perseus region} admit also {spatially constant grain size distribution models} as solutions (Table \ref{tab:solutions}). In particular, grains which could explain the observations in the {Taurus--Perseus region} are globally bigger than the ones able to explain L183 observations. The addition of water ices lower the coreshine ratio but increases the NIR/MIR ratio and could be an answer for some cores in {Taurus--Perseus region} but has to be considered carefully. Bigger silicates have also tendency to increase the NIR/MIR ratio while decreasing the coreshine ratio, and carbonates increase the later and yield higher flux at 3.6 \mic. In any case, grain composition have an important role to play and the classical spherical silicates/carbonates approach has to be modulated, for example {the fractal dimension} is a major actor even if it is not able to explain the NIR/MIR ratio alone for both regions.}

\begin{table}
\flushleft
\caption{\label{tab:solutions}Suitable grain types for L183 and {Taurus--Perseus} regions.}
\begin{center}
\begin{tabular}{llcccc}
\hline 
\hline 
Outer layer & Inner Layer & 3.6/4.5 & NIR/MIR & NIR/NIR \\
\hline
\hline
\multicolumn{2}{l}{{\bf L183} (M$_\mathrm{high}$)}\\
\hline
\multicolumn{2}{l}{\textit{bigger silicates}}\\
\hline
aSil C10 & S50 C20 & 0.367 & 3.559 & 0.934 \\
S10 CBx2 & S50 CBx2 & 0.373 & 3.672 & 0.881 \\
S10 CBx2 & S50 C10 & 0.373 & 3.560 & 0.888 \\
S10 CBx2 & S50 C20 & 0.368 & 3.863 & 0.923 \\
S10 C10 &S50 C20 & 0.368 & 3.690 & 0.926 \\
\hline
\multicolumn{2}{l}{\textit{others}}\\
\hline
S10 CBx2 & S50 C50 & 0.379 & 3.665 & 0.933 \\ 
S10 C10 & S50 C50 & 0.379 & 3.816 & 0.936 \\
\hline
\multicolumn{2}{l}{{\bf Taurus--Perseus} (M$_\mathrm{low}$)}\\
\hline
\hline
\multicolumn{2}{l}{\textit{bigger silicates}}\\
\hline
aSil CBx2 & S50 CBx2 & 0.289 & 3.284 & 0.337 \\
aSil CBx2 & S50 C10 & 0.288 & 3.288 & 0.336 \\
aSil CBx2 & S50 C20 & 0.247 & 3.608 & 0.534 \\
aSil C10 & S50 C10 & 0.288 & 3.289 & 0.336 \\
aSil C10 & S50 C20 & 0.248 & 3.613 & 0.533 \\
aS1m Cx1m & aS2m Cx1m & 0.361 & 3.294 & 0.697 \\
aS1m Cx1m & aS5m Cx1m& 0.336 & 3.525 & 0.736 \\
aS1m Cx1m & aS5m Cx2m & 0.409 & 3.246 & 0.833 \\
aS2m Cx1m & aS2m Cx1m & 0.378 & 3.254 & 0.694 \\
aS2m Cx1m & aS5m Cx1m & 0.352 & 3.455 & 0.726 \\
aS5m Cx1m & aS5m Cx1m & 0.341 & 3.490 & 0.726 \\
S10 CBx2 & S50 CBx2 & 0.291 & 3.303 & 0.337 \\
S10 CBx2 & S50 C10 & 0.290 & 3.304 & 0.336 \\
S10 CBx2 & S50 C20 & 0.248 & 3.613 & 0.535 \\
S10 C10 & S50 C10 & 0.291 & 3.304 & 0.337 \\
S10 C10 & S50 C20 & 0.248 & 3.611 & 0.535 \\
S20 CBx2 & S50 CBx2 & 0.282 & 3.207 & 0.346 \\
S20 CBx2 & S50 C10 & 0.282 & 3.206 & 0.346 \\
S20 CBx2 & S50 C20 & 0.255 & 3.583 & 0.534 \\
S20 CBx2 & S50 C50 & 0.293 & 3.428 & 0.611 \\
S20 C10 & S50 C10 & 0.282 & 3.204 & 0.346 \\
S20 C10 & S50 C20 & 0.256 & 3.584 & 0.533 \\
S20 C10 & S50 C50 & 0.293 & 3.426 & 0.611 \\
S20 C20 & S50 C20 & 0.256 & 3.650 & 0.582 \\
S50 CBx2 & S50 C20 & 0.266 & 3.407 & 0.510 \\
S50 CBx2 & S50 C50 & 0.302 & 3.291 & 0.593 \\
S50 C10 & S50 C20 & 0.266 & 3.411 & 0.509 \\
S50 C10 & S50 C50 & 0.301 & 3.291 & 0.592 \\
S50 C20 & S50 C20 & 0.268 & 3.533 & 0.562 \\
S50 C20 & S50 C50 & 0.303 & 3.401 & 0.651 \\
\hline
\multicolumn{2}{l}{\textit{bigger carbonates}}\\
\hline
aS1m Cx1m & aS1m Cx2m& 0.359 & 3.236 & 0.750 \\
aSil C50 & S50 C50 & 0.294 & 3.509 & 0.703 \\
S10 C20 & S50 C50  & 0.290 & 3.486 & 0.683 \\
S10 C50 & S50 C50 & 0.294 & 3.505 & 0.706 \\
S20 C50 & S50 C50 & 0.296 & 3.516 & 0.688 \\
\hline
\multicolumn{2}{l}{\textit{others}}\\
\hline
\tablefootmark{a}aSil CBx2 & S50 C50& 0.287 & 3.453 & 0.620 \\
aSil C10 & S50 C50 & 0.287 & 3.455 & 0.620 \\
aSil C20 & S50 C20 & 0.253 & 3.683 & 0.589 \\
aSil C20 & S50 C50 & 0.291 & 3.487 & 0.682 \\
S10 CBx2 & S50 C50 & 0.286 & 3.456 & 0.623 \\
S10 C10 & S50 C50 & 0.286 & 3.457 & 0.623 \\
S10 C20 & S50 C20& 0.253 & 3.687 & 0.589 \\
S20 C20 & S50 C50 & 0.294 & 3.512 & 0.666 \\
\hline
\multicolumn{2}{l}{\textit{constant}}\\
\hline
S50 C50 & S50 C50 & 0.307 & 3.371 & 0.672 \\
aS1m Cx1m & aS1m Cx1m & 0.313 & 3.415 & 0.686 \\
ORI3 & & 0.251 & 3.608 & 0.709 \\
\hline 
\end{tabular}
\tablefoot{
\tablefoottext{a}{model A}}
\end{center}
\end{table}

\begin{figure}
\centering
\includegraphics[width=8cm]{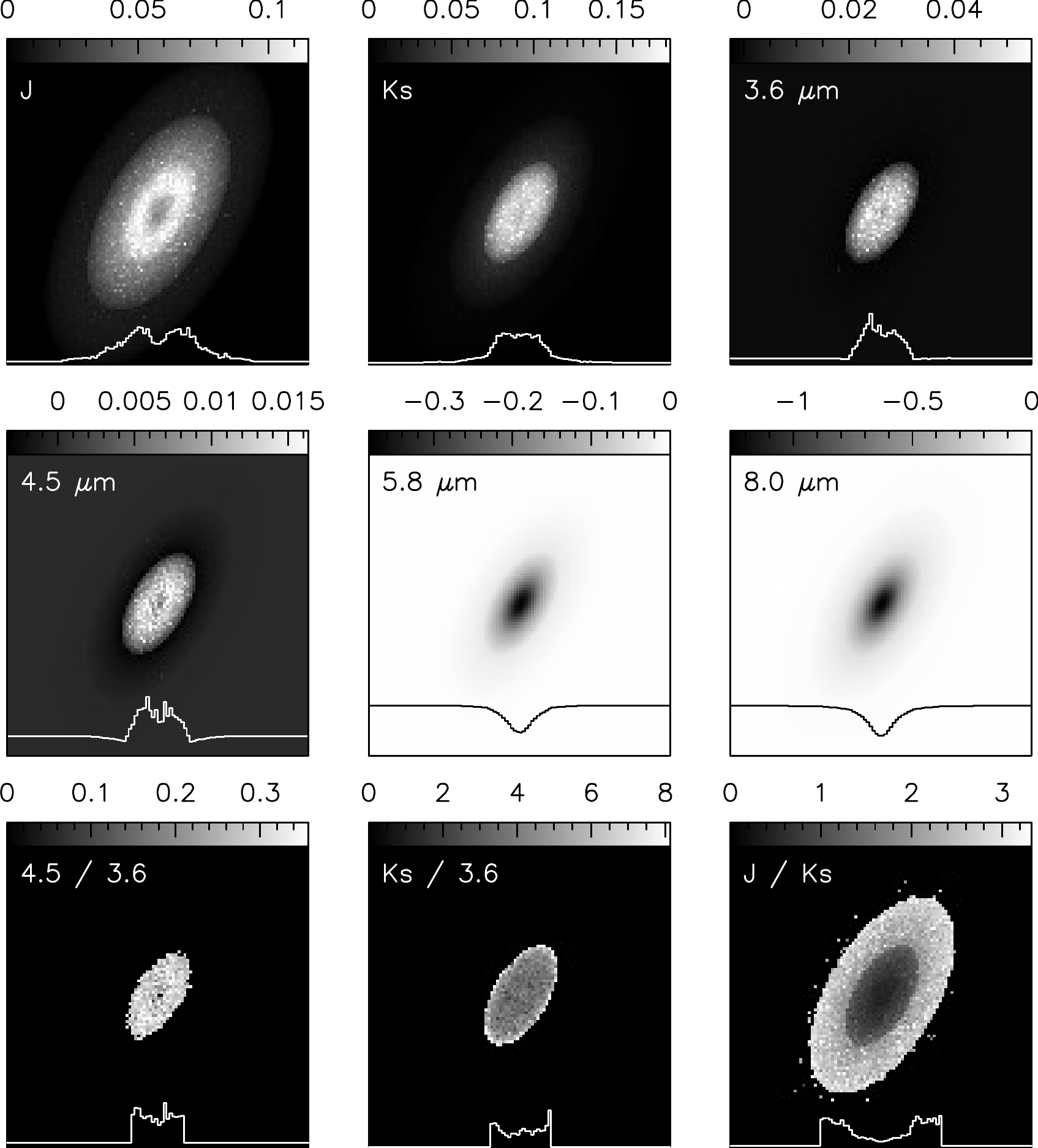}
\caption{Model A simulation outputs {for Taurus--Perseus region}. Last row is ratio maps. The histogram displays the horizontal cut along the x axis through the center. Grey scale given in \mjy.}
\label{fig:model16}
\end{figure}

\section{Conclusion}\label{sect:conclu}
We explored a radiative transfer model in order to study the scattering of NIR and MIR light in dark clouds by varying the grain properties and evaluating the impact of other parameters such as the ISRF and the background field intensities. We confront our large sample of coreshine observations to the modeling and we selected a handful of cases to explore the capability of coreshine to disentangle dust properties. Our results are the following:
\begin{itemize}
\item{The emergence of coreshine is a contrast issue which has to be treated carefully. The cloud background field estimation is a key point to the modeling. We adapted the Levenson method with careful flux conversion to the pertinent wavelengths. Moreover, the choice of keeping the standard ISRF is the safest approach and also a real challenge which will benefit from emission modeling in future studies. }
\item{Merging previous samples of coreshine observations to build sufficient statistics, some Galactic regions appear to be favored. This could be considered in a big picture as the presence of previously grown grains in the initial diffuse medium of these individual regions. It opens the scope of modeling coreshine with regards to individual regions or for clouds in close environmental conditions (like L1517A,B,C,D or L134 and L183).}
\item{The use of 3.6 and 4.5\,\mic\ coreshine bands and especially their ratio{, referred to as coreshine ratio,} brings additional constraints on the grain properties in the core.} 
\item{{The highest coreshine ratios} and fluxes are obtained for clouds which contain an embedded object, probing the influence of local change in the radiation field which has been tested qualitatively in our modeling.}\\
\end{itemize}
Concerning the grain properties, the main conclusions are:
 \begin{itemize}
\item{Small grains have no influence in the modeling on the coreshine ratio.} 
\item{ For any given coreshine ratio, the absolute 3.6\,\mic\ flux value is somewhat adjustable from both the cloud model and the phase function.}
\item{The size increase is mandatory but not sufficient to explain coreshine. The coreshine ratio quantity tends to saturate with the grain size increase,  both for spherical compact grains and for agglomerates of monomers. While bare spherical grains show a saturation of this ratio above a high exponential cut-off of 1\,\mic, fluffiness helps to raise the saturation limit. }
\item{Since dust grains inside molecular clouds are expected to be icy, the role of ice mantles needs to be further investigated. Indeed, we expect ices to favor fluffiness and growth but in the meantime the only model we tested shows that it decreases the coreshine ratio. New models of icy grains are strongly needed.}
{\item{Both the NIR/MIR ratio for the core outer layer and the absence of emission at 5.8 \mic\ for any layers eliminate a mix of silicates and carbonates that both include grains above $\sim$2 \mic\ in meaningful quantity.}}
\item{Porosity has no impact at the studied wavelengths.}
\item{The comparison with other wavelengths for a valid coreshine ratio can help to disentangle between the cloud models and the grain properties.}
\item{In the case where the NIR and MIR do not sample the same zone, we might be able to peel the cloud layer by layer thanks to the multi-wavelength approach. If the same region is sampled then the {grain size distribution} properties have to be customized to be compatible with the observations at the different wavelengths.}
\end{itemize}

In this paper, we constituted a database for the grain behavior covering grain growth by coagulation (up to 5\,\mic\ in size). We extended our study to aggregates and the presence of ice mantles and found several promising grain types which are able to reproduce the observations. We constrained the ISRF and the background value for different lines of sight and deduced the impact of the other free parameters on the modeling. The complementarity with NIR observations introduced to highlight the promising perspective of a 3D multi--wavelength cloud modeling has to be investigated on a real cloud following all the key points above and extending the study to far--IR emission.

\begin{acknowledgements}
\thanks{This research has made use of observations from Spitzer Space Telescope and data from the NASA/IPAC Infrared Science Archive, which are operated by the Jet Propulsion Laboratory (JPL) and the California Institute of Technology under contract with NASA.\\
Based on observations obtained with WIRCam, a joint project of CFHT, Taiwan, Korea, Canada, France, and the Canada-France-Hawaii Telescope (CFHT) which is operated by the National Research Council (NRC) of Canada, the Institute National des Sciences de l'Univers of the Centre National de la Recherche Scientifique of France, and the University of Hawaii.\\
  This publication makes use of data products from the Wide-field Infrared Survey Explorer, which is a joint project of the University of California, Los Angeles, and the Jet Propulsion Laboratory/California Institute of Technology, funded by the National Aeronautics and Space Administration.\\
  This publication makes use of data products from the Two Micron All Sky Survey, which is a joint project of the University of Massachusetts and the Infrared Processing and Analysis Center/California Institute of Technology, funded by the National Aeronautics and Space Administration and the National Science Foundation.\\
  This research makes a large use of the CDS (Strasbourg, France) services, especially Aladin \citep{2000A&AS..143...33B}, Simbad and Vizier \citep{2000A&AS..143...23O} and of the NASA Lambda data base at GSFC.}\\
We thank J-Ph Bernard, Nathalie Ysard and Vincent Guillet for fruitful discussions, Michiel Min for providing us his agglomerates of monomers in advance of publication and the DIM ACAV and "Région Ile de France" for financial support. MJ and and V-MP acknowledge the support of Academy of Finland grant 250741. MA, JS acknowledge support from the ANR (SEED ANR-11-CHEX-0007-01). We thank the referee for the careful reading of our paper and the valuable comments.\\

\end{acknowledgements}

\bibliographystyle{aa}

\bibliography{modelisation}

\longtab{1}{
%\flushleft
\begin{longtable}{lccccccc}
\caption{{3.6\,\mic\ coreshine intensity and 4.5/3.6 coreshine ratio sorted by increasing ratio value for each detected cloud of the four regions.}}\\
\hline\hline
Name & 3.6\,\mic\ intensity  & \multicolumn{2}{c}{4.5/3.6 }& Protostar& Class\tablefootmark{a}&L$_\mathrm{Bol}$\tablefootmark{b}  &refs\\ % 
& kJy\,sr$^{-1}$&median & FWHM&&&L\sun&\\ 
\hline
\endfirsthead
\caption{continued.}\\
\hline\hline
Name & 3.6\,\mic\ intensity  & \multicolumn{2}{c}{4.5/3.6 }& Protostar& Class\tablefootmark{a}&L$_\mathrm{Bol}$\tablefootmark{b}&refs\\
& kJy\,sr$^{-1}$&median &FWHM&&&L\sun&\\ 
\hline
\endhead
\hline
\endfoot
Taurus--Perseus\\
\hline
\object{G171.80$-$09.78} & 10 & -- & -- \\
\object{CB24} & 15 & -- & -- \\
\object{L1503} & 22 & -- & -- & \\
\object{G179.18$-$19.62} & 22 & -- & -- \\
\object{G182.19$-$17.71} & 23& -- & -- & \\
\object{G170.81$-$18.34} & 26 & -- & -- &  \\
\object{L1552} & 26 & -- & --&  \\
\object{G169.82$-$19.39}& 27 & -- & -- & \\
\object{CB20} & 27 & -- & --  &  \\
\object{G173.45$-$13.34} & 30  & -- & -- & \\%2 coeurs pas planck + fort 
\object{G177.89$-$20.16} & 33  & -- & -- \\
\object{B18--3 = G174.39$-$13.43	} &35  & -- &\\% --  & à revoir le ratio\\
\object{G154.68$-$15.34} & 49 & 0.35 & 0.10 & \\%très petite zone bof\\
\object{G170.99$-$15.81} & 36 & 0.36 & 0.17 &  \\
\object{L1506C }& 33 & 0.37 & 0.26 \\
\object{L1507A (G171.51$-$10.59)} & 41 & 0.41 & 0.15 & \\
\object{IRAS03282+3035} & 47 & 0.42 & 0.14 & IRAS 03282+3035 &Class 0& 1.2&1\\
\object{G173.69$-$15.55} & 32 & 0.43 & 0.10 &\\
\object{L1544} & 39 &  0.43 & 0.32 \\
\object{L1512}& 30 & 0.44 & 0.31\\
\object{L1521E}& 51 & 0.46 & 0.25& \\
\object{L1498}& 34 & 0.47 & 0.22 \\
\object{G171.34$-$10.67}&58&0.47 & 0.20 &  \\
\object{G170.26$-$16.02} & 87 & 0.50 & 0.30 & IRAS 04181+2654AB &Class I&--&2\\
\object{L1521F}  & 53 & 0.51 & 0.20 & VeLLO\tablefootmark{c} & Class 0 & 0.36 &3\\
\object{L1517A} & 57 & 0.51 & 0.28 &\\
\object{L1517B} & 43 & 0.51 & 0.28 & \\
\object{G157.10$-$08.70} & 34 & 0.52 & 0.15 & IRAS 03586+4112 (?)\tablefootmark{d}\\
\object{L1517C} & 29 & 0.53 & 0.28 & \\
\object{L1507} & 41 & 0.54 & 0.35 &2MASS J04432023+2940060&Class II &--&4 \\
\object{IRAM04191} & 140 & 0.57 & 0.36 & IRAM04191--IRS &Class 0&0.28& 5 \\
\object{L1439} & 81 & 0.60 & 0.29 & IRAS 04559+5200 &Class I & $\geq$ 0.5&6,7\\%  3.6\arcmin\ southwest of the starless core\\
\object{TMC2}\tablefootmark{e}&100 & 0.64 & 0.47 & IRAS 04294+2413& Class 0 ? & 0.78 (L$_\mathrm{IR}$)& 8 \\
\object{L1448mm}&59 & 0.64 & 0.44 & L1448-mm&Class 0&8.6&9\\%\object{L1448mm} & 0.936\\
\object{G163.21$-$08.40} & 67 & 0.64 &0.46&  IRAS 04218+3708 (?)\tablefootmark{d}\\
\object{G157.12$-$11.56} & 390 & 0.64 & 0.39&IRAS 03484+3845 (?)\tablefootmark{d}&  \\
\object{G155.45$-$14.59} & 210 & 0.64 & 0.27 &  IRAS 03330+3727 (?)\tablefootmark{d}\\
\object{G160.51$-$16.84} &61 & 0.66 & 0.51 &  B5 IRS1 (IRAS 03445+3242) &Class I&3.8&10\\ %plusieurs objets
\object{G171.91$-$15.65}& 310 & 0.77 & 0.39& DG Tau B & Class I & 0.86& 2,11\\
\object{Barnard18--1}&97&0.79&0.39&IRAS 04292+2422(E+W)& Class I&0.6&10\\
\object{G163.32$-$08.42} & 63 & 0.82 & 0.30 &IRAS 04223+3700&Class 1&2.7&10\\
\object{G158.86$-$21.60} & 74 & 0.95 & 0.40 &IRAS 03249+2957&Class I&0.3 &1\\
\object{Barnard1}	& 100 & 1.1 & 0.51 & IRAS 03301+3057 (cluster)& Class I& 2.7 (L$_\mathrm{IR}$)&12 \\
\hline
L183 / L134 \\
%\hline
\hline
\object{L183} & 58 & 0.37 & 0.14\\
\object{L134}\tablefootmark{f} & $>$ 30&-- & -- \\
\hline
Chameleon\\
%\hline
\hline
\object{G302.89$-$14.05} & 16 & -- & -- & \\
\object{G298.34$-$13.03}&25&-- & -- &\\
\object{G303.28$-$13.32}& 25 &-- & -- \\
\object{Mu8}&27	&-- & --  \\
\object{G297.09$-$16.02} &39	&0.35&0.23 \\
\object{G303.09$-$16.04} &77&	0.43&	0.23 &\\%!sp9 \\
\object{G303.68$-$15.32}&21&	0.44&	0.23\\
\object{G303.39$-$14.26}&150&	0.50&	0.33&IRAS 12553$-$7651 &Class I &1.2 &13\\ %and 5.6 R$_{\sun}$) embedded (K4.5 of 4500 K, 90 sec\\
\object{G303.15$-$17.34}\tablefootmark{g}&32	&0.60&0.36&\\
\object{G303.72$-$14.86}&100 &	0.74&	0.52&IRAS 13014$-$7723& Class II & 1.6 &14,15 \\
\hline
Cepheus\\
%\hline
\hline
\object{G093.20+09.53}&18 & -- & --   \\
\object{L1155E}&	 	22 & -- & --  	\\
\object{G093.16+09.61}&		24   & -- &  -- 	& \\
\object{G130.56+11.51}	&	24   & -- & --      \\
\object{L1157}&		28 &  0.41 &  0.09	&IRAS 20386+6751 &Class 0 & 11 & 16\\%!très enfouie, pas d'influence 
\object{L1155C}	& 	29  & 0.44 &  0.17&	\\ %n
\object{L1247}&		110 &  0.46 &  0.36	&\\ %n
\object{L1251A}&	 	62 &  0.50 &  0.40&	L1251A--IRS1--4& Class I, I, 0 \& 0& IRS3 = 0.8 & 17 \\
\object{L1333}\tablefootmark{h}&		25 &   0.50 &  0.18&	\\ %source submm en fait
\object{L1148}& 	 	24	& 0.54&	0.38&	L1148--IRS&Class I&0.44&18\\
\object{L1152}&		39  &    0.58  & 0.24	& L1152 1--3& Class 0, I, \& I&1.5, 0.6, 2.1&18 \\
\object{L1262}	 	&92 &0.62  & 0.33	&  IRAS 23238+7401& Class 0 & 1.5 & 19\\ %$\sim$ 90\arcsec\ east of the starless core\\
\object{L1157--outflow}&	31  & 0.71 &  0.22& outflow shock region ?\\%	&!zone de choc outflow ?
\object{L1228}&		110  &  0.73  & 0.50&IRAS 20582+7724 (cluster)&Class I &2.3 &18\\%Star cluster G111.67+20.22 Includes  a object  L\sun \\
\object{L1221}&		160  &  0.81 &  0.52&	L1221--IRS1 \& 3 & Class 0 \& 0 & 3.0 \& 1.4&18\\
\object{L1251C}\tablefootmark{i} &		260  &  0.81 &  0.59&	IRAS 22343+7501 (cluster)&Class I& 33&18\\
\object{L1251B}\tablefootmark{j}&		160  &    0.89 &  0.57& IRAS 22376+7455 (cluster) & Class 0& 15&18\\

\hline
\end{longtable}
\tablefoot
{
\tablefoottext{a}{Protostar classes depend on the criterion (spectral index or T$_\mathrm{Bol}$). Whenever possible we use the second one. Geometry effects also count. See \citet{2007ApJS..169..328R,2009ApJS..185..198K} for further details.}\\
\tablefoottext{b}{if (L$_\mathrm{IR}$) is indicated, it is the integrated IR luminosity, L$_\mathrm{Bol}$ being not available.}\\
\tablefoottext{c}{Very Low Luminosity Object}\\
\tablefoottext{d}{A bright IRAS source is observed in the vicinity of the cloud but no study of the source has been found in the literature. Its YSO status is therefore not proved but probable.}\\
\tablefoottext{e}{TMC2 is usually considered to be starless \citep[e.g.][]{BradyFord:2011ke}. The IRAC coreshine images and ratio show that the nearby YSO at $\sim$4\arcmin\ is illuminating it.}\\
\tablefoottext{f}{The Spitzer L134 3.6\,\mic\ map is too narrow compared to the coreshine extent in the WISE image and a part of the flux is missed.}\\
\tablefoottext{g}{A bright star at position 12h57m20s $-$80\degr15\arcmin42\arcsec\ (J2000) seems to illuminate the cloud. It is unknown in SIMBAD.}\\
\tablefoottext{h}{L1333 contains a submm source: JCMTSF J022611.7+752732 \citep{2008ApJS..175..277D}.}\\
\tablefoottext{i}{\citet{2009ApJS..185..198K} have renamed L1251A in L1251W, and L1251C in L1251A. We keep SIMBAD definition of the source parts.}\\
\tablefoottext{j}{This part of the L1251 source is named either L1251B or L1251E depending on the authors, SIMBAD separates them by 3.5\arcmin.}\\
\tablebib{(1)~\cite{0004-637X-692-2-973}; (2)~\cite{2010ApJS..186..259R}; (3)~\cite{Dunham:2008ks}; (4)~\cite{2010ApJS..186..111L}; (5)~\cite{2006ApJ...651..945D}; (6)~\cite{2004ApJ...617..418S}; (7)~\cite{2013AA...560A..41L};(8)~\cite{1998ApJ...502..296O}; (9)~\cite{2013ApJ...770..123G}; (10)~\cite{2010AJ....140.1214C};(11)~\cite{1986ApJ...311L..23J}; (12)~\cite{2005AJ....130.1795W}; (13)~\cite{2008ApJ...680.1295S}; (14)~\cite{2008ApJ...676..427A}; (15)~\cite{2009ApJS..181..321E}; (16)~\cite{2013AA...558A..94G}; (17)~\cite{2010ApJ...709L..74L}; (18)~ \cite{2009ApJS..185..198K}; (19)~\cite{Stutz:2010kc}}
}
\label{tab:ratio_table}
}

\newpage

%------------------APPENDIX--------------------

\Online
\begin{appendix}

\newpage
\section{Isotropy versus anisotropy for the incident radiation field}\label{sect:aniso}
\begin{figure*}[t!]
\centering
\sidecaption
\includegraphics[width=12cm]{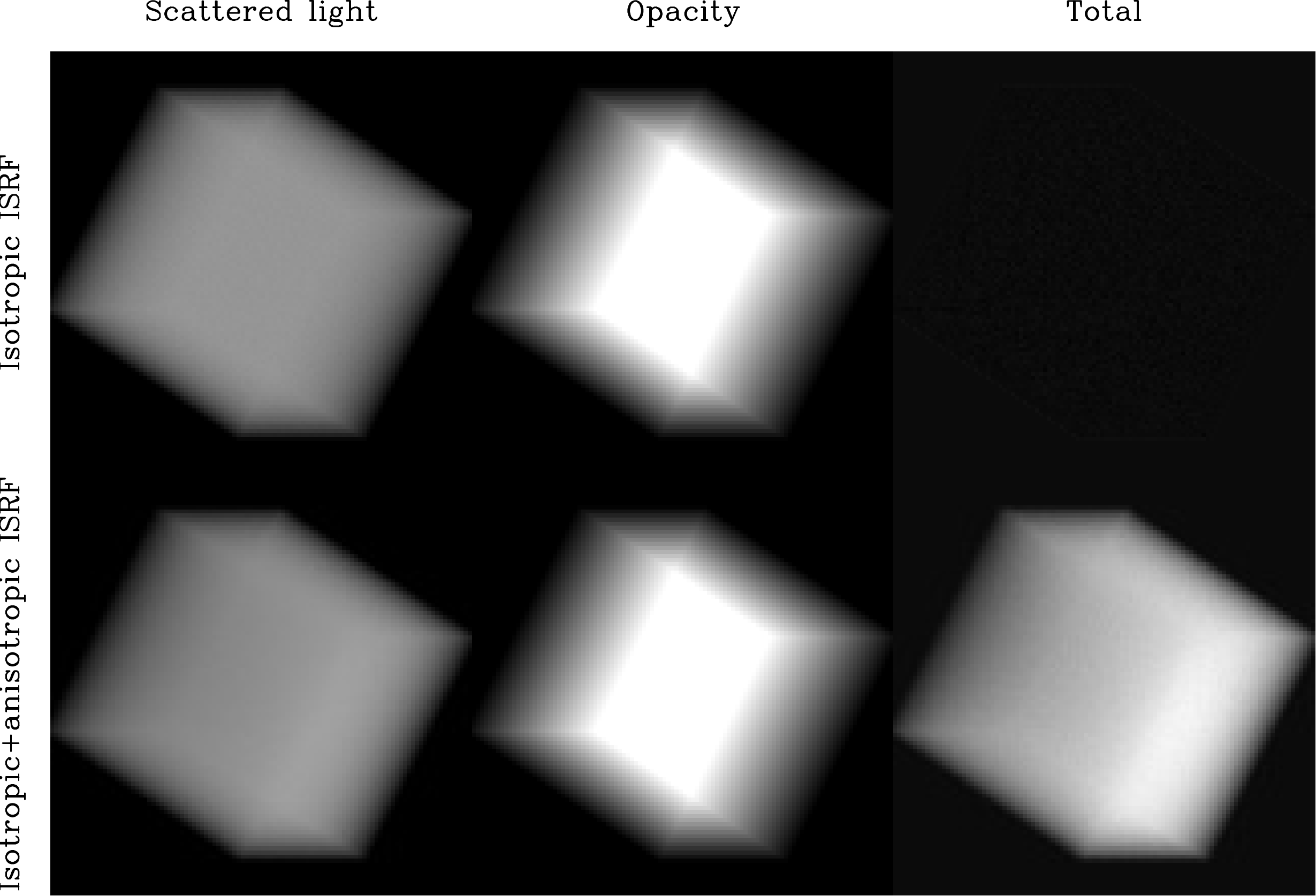}
\caption{A cubic cloud filled with scattering (absorptionless) dust at constant density in an isotropic (upper row) or isotropic+anisotropic (lower row) field. The left column shows the scattered photons only, the middle column the opacity along the line of sight (identical for both cases) and the right column the net surface brightness.}
\label{appfig:iso}
\end{figure*}

Coreshine (and any kind of scattering processes) can be seen in emission only if it follows a number of conditions as discussed in Sect. \ref{sect:contrast}. Another compulsory condition is the anisotropy of the incident ISRF. This anisotropy can come from large scales (like galactic structure) or small scales (like stars, near or far). To illustrate the importance of anisotropy we consider two simple cases: only isotropic illumination or the combination of one type of anisotropic source with the isotropic illumination.

If the ISRF is isotropic, there is no preferred direction for photons to travel. To see the cloud in emission, we need to introduce some anisotropy to concentrate photons towards a privileged direction, i.e. the observer. The cloud would therefore have to produce this anisotropy. To increase the number of photons towards the observer, the cloud would have to act like a telescope mirror pointed at the Earth to collect photons from a large number of directions to redirect them in a single direction  (this would of course lower the number of photons scattered in some other directions). Since the cloud has no such focusing capabilities and the observer no privileged position, an isotropic ISRF cannot make a cloud glow. In fact, for any direction across the cloud there are as many deviated photons away from that line than photons from other directions being deviated into that propagation line and all the scatterings cancel out. This is in an ideal case without absorption. In presence of absorption, a part of the photons are lost and the cloud can only appear as a darker object against the background sky, never in emission. The field must therefore be anisotropic. 

In the anisotropic case, there is one noticeable direction which is the line going from the localized light source to the cloud (we consider a single source of photons, like a nearby star, superimposed on an isotropic ISRF). This is the path with the highest number of photons traveling towards the cloud. Like in the isotropic case, photons enter from all directions and are partly deviated, partly untouched (and partly absorbed but we keep ignoring absorption here). The difference with the isotropic case is that along this particular line, across the cloud, there will be a larger number of deflected photons away from it than photons brought back onto it. The contrary can happen for some or all of the other directions (depending on details such as the phase function of the grain scattering properties). Therefore, away from this path, the number of deviated photons leaving the cloud is increasing (while they are decreasing along the path). Eventually, there can be more photons deviated towards the observer from the anisotropic source than photons coming from behind the cloud and deviated away from the observer. The net effect is to show the cloud in emission.

%If the ISRF is isotropic, there is no preferred direction for photons to travel. Thus an excess of photons from the cloud towards the observer (the cloud would be seen in emission) is possible only if the observer is in a privileged direction. This would be the case if the cloud could concentrate the light towards the observer like a telescope mirror. Since the cloud has no such capabilities and the observer no privileged position, a net excess of photons must come from an anisotropic field. In that case, the light coming from the source is scattered in other directions when reaching the cloud and the cloud can transmit photons and be seen in excess above the background if the latter is low enough. If the anisotropic source of light is behind the cloud for the observer, the cloud can only be seen in absorption as the total number of photons (scattered and unscattered) is constant if one neglects the absorption contribution to the opacity, or smaller if not. In the isotropic field case, the cloud will always appear in absorption as long as the absorption coefficient (Q$_\mathrm{sca}$) is not zero or simply disappear from the sky if there is no photon loss by absorption (the cloud would not be transparent but translucid). 
To illustrate this effect we ran a model based on a cube of constant density, tilted at 30\degr\ angles on two axes to see the edges and three sides. The cube is either in an isotropic field or a composite of isotropic and anisotropic fields. It is filled with dust with scattering capability only (absorption coefficient is set to zero). Figure \ref{appfig:iso} shows the results. 
The left column shows slightly different images of scattered photons but taking into account the background illumination absorption due to the cloud opacity (displayed in the central figure), following eq. \ref{eq:trans}, the cloud completely disappears in the isotropic field. This is explained by the fact that all the photons scattered towards the observer (as seen in the left panel) are exactly compensated by the photons scattered away from the line of sight for the radiation field coming from behind the cloud. A close inspection of the image reveals the cube by the numerical noise of the {Monte Carlo radiative transfer code} only. If the cloud opacity had not been set to zero, the cloud would have been seen in absorption against the background, while for the isotropic+anisotropic case, the cloud would have appeared in emission or in absorption depending on the balance (Sect. \ref{sect:contrast}).

\section{Global method for stellar contribution subtraction.}\label{sect:JPB}

We assume that the intensity in the DIRBE{1} (1.25\,\mic) and DIRBE2 (2.2\,\mic) bands is only due to the sum of individual stellar contributions (PSC):\\
\begin{equation}\label{eq:hyp1}
I_K = I^*_K = \sum I_{PSC,K} = DIRBE{2} 
\end{equation}
\begin{equation}\label{eq:hyp2}
I_J = I^*_J = \sum I_{PSC,J} = DIRBE{1}.
\end{equation}

The intrinsic color I$_{int}$ of the stellar component has been measured from high galactic latitude and low dust emission regions and is also equal to: 
\begin{equation}\label{eq:intcolor}
I_{int}(\lambda_{1}/\lambda_{2}) = \frac{I^*_{\lambda2} \times exp(\tau_{\lambda2})}{I^*_{\lambda1} \times exp(\tau_{\lambda1})}
\end{equation}

We deduce the extinction on each line of sight by using the measured value of the intrinsic color between J and K ($ I_\mathrm{int}(J/K) = 1. $ ), its definition (eq. \ref{eq:intcolor}) and the previous assumptions (eq. \ref{eq:hyp1} and \ref{eq:hyp2}):
\begin{equation}
I_\mathrm{int}(J/K) = 1. =  \frac{I^*_{K} \times exp(\tau_{K})}{I^*_{J} \times exp(\tau_{J})} = \frac{ DIRBE{2}}{ DIRBE{1}} \times exp (\tau_K - \tau_J)
\end{equation}

Finally we use eq. \ref{eq:intcolor} to yield the stellar contribution in each band:

\begin{equation}
I^*_{3.6} =I_{int}(J/3.6) \times I^*_J \times exp(\tau_{J})/exp(\tau_{3.6})
\end{equation}

Taking into account our assumption (eq. \ref{eq:hyp2}) and the conversion coefficient deduced from the extinction curve of \cite{1985ApJ...288..618R} with R$_V$ = 3.1 we obtain:

\begin{equation}
I^*_{3.6} =I_{int}(J/3.6) \times DIRBE{1} \times exp(\tau_{J})/exp(0.5\tau_{K})
\end{equation}

\begin{equation}
I^*_{3.6} =I_{int}(J/3.6) \times DIRBE{1} \times \frac{DIRBE2}{DIRBE1} \times \frac{1.}{exp(0.5)}
\end{equation}

\begin{equation}
I^*_{3.6} =I_{int}(J/K) \times I_{int}(K/3.4)  \times DIRBE2 \times \frac{1.}{exp(0.5)}
\end{equation}
\noindent with I$_{int}(J/K) $ = 1. and $I_{int}(3.4/K) = 1.7$ \citep{1994A&A...291L...5B}.

We obtain the diffuse emission map:

\begin{equation}
I_\mathrm{diff}(3.4) = DIRBE{3} - I^*_{3.4}
\end{equation}

In the same way using $I_\mathrm{int}(4.9/3.4) = 2.1$ we compute the other diffuse emission map:

\begin{equation}
I_\mathrm{diff}(4.9) = DIRBE{4} - I^*_{4.9}
\end{equation}

\newpage
\section{Source sample}\label{sect:sources}

\longtab{1}{
\centering
\begin{longtable}{lcccc}
\caption{\label{tab:sources}{Source table classified region by region}}\\
\hline
\hline
Name&Gal. Longitude&Gal. Latitude&Status\tablefootmark{a}&Region\\ %Comments\\
\hline
\endfirsthead
\caption{continued.}\\
\hline\hline
Name&Gal. Longitude&Gal. Latitude&Status\tablefootmark{a}&Region\\ %Comments\\
\hline
\endhead
\hline
\endfoot
\object{G190.15$-$14.34} & $-$169.85 & $-$14.34 &  P? &  Orion\\
\object{L1570} & $-$169.32 &  $-$0.45 &  N &  Orion\\
\object{CB42} & $-$167.42 &  $-$2.80 &  N &  Orion\\
\object{CB41} & $-$167.30 &  $-$2.93 &  N &  Orion\\
\object{G195.09$-$16.41} & $-$164.90 & $-$16.41 &  C &  Orion\\
\object{G196.21$-$15.50} & $-$163.78 & $-$15.50 &  P &  Orion\\
\object{B35A} & $-$163.07 & $-$10.36 &  P &  Orion\\
\object{G198.03$-$15.24} & $-$161.96 & $-$15.25 &  P &  Orion\\
\object{CB46} & $-$156.98 &  $-$3.73 &  C &  Orion\\
\object{G209.28$-$19.62} & $-$150.71 & $-$19.63 &  N &  Orion\\
\object{G214.69$-$19.94} & $-$145.31 & $-$19.95 &  C? &  Orion\\
\object{G203.57$-$30.08} & $-$156.42 & $-$30.09 &  N &  Eridanus\\
\object{G202.21$-$09.17} & $-$157.79 &  $-$9.18 &  N &  Monoceros\\
\object{L1633} & $-$153.13 &  $-$4.39 &  A &  Monoceros\\
\object{G211.70$-$12.17} & $-$148.29 & $-$12.18 &  N &  Monoceros\\
\object{G215.41$-$16.39} & $-$144.58 & $-$16.39 &  C &  Monoceros\\
\object{G216.69$-$13.88} & $-$143.31 & $-$13.88 &  C &  Monoceros\\
\object{G216.76$-$16.06} & $-$143.24 & $-$16.06 &  C &  Monoceros\\
\object{G219.28$-$09.27} & $-$140.71 &  $-$9.27 &  U &  Monoceros\\
\object{G219.35$-$09.70} & $-$140.65 &  $-$9.71 &  N &  Monoceros\\
\object{G219.37$-$07.68} & $-$140.63 &  $-$7.69 &  N? &  Monoceros\\
\object{G219.26$-$17.89} & $-$140.73 & $-$17.90 &  C &  Lepus\\
\object{G227.30$-$03.77} & $-$132.69 &  $-$3.77 &  N &  Canis major\\
\object{BHR7} & $-$107.47 &   0.07 &  C &  Gum/Vela\\
\object{CG30/31} & $-$106.82 &  $-$1.66 &  P &  Gum/Vela\\
\object{BHR13} & $-$106.41 &   2.95 &  A &  Gum/Vela\\
\object{BHR14} & $-$106.18 & $-$10.91 &  P+C? &  Gum/Vela\\
\object{BHR16} & $-$104.56 &  $-$3.95 &  C? &  Gum/Vela\\
\object{DC257.3$-$2.5} & $-$102.72 &  $-$2.45 &  C?+A &  Gum/Vela\\
\object{BHR21} & $-$100.56 & $-$12.73 &  C+P &  Gum/Vela\\
\object{BHR22} & $-$100.48 & $-$16.45 &  P &  Gum/Vela\\
\object{BHR30} &  $-$94.73 &  $-$0.01 &  A &  Gum/Vela\\
\object{BHR31} &  $-$94.35 &  $-$7.69 &  A &  Gum/Vela\\
\object{DC266.0$-$7.5} &  $-$94.00 &  $-$7.41 &  A &  Gum/Vela\\
\object{BHR36} &  $-$92.64 &  $-$7.51 &  A &  Gum/Vela\\
\object{BHR37} &  $-$92.54 &  $-$7.41 &  A+P &  Gum/Vela\\
\object{BHR34} &  $-$92.42 &  $-$6.47 &  C?+P? &  Gum/Vela\\
\object{BHR40} &  $-$92.42 &  $-$6.44 &  P &  Gum/Vela\\
\object{BHR38/39} &  $-$92.36 &  $-$6.01 &  P &  Gum/Vela\\
\object{BHR41} &  $-$92.35 &  $-$7.36 &  A &  Gum/Vela\\
\object{BHR42} &  $-$92.04 &  $-$7.78 &  P &  Gum/Vela\\
\object{BHR44} &  $-$90.53 &   3.95 &  A &  Gum/Vela\\
\object{BHR43} &  $-$90.50 &   2.95 &  A &  Gum/Vela\\
\object{BHR47} &  $-$87.55 &   2.01 &  A &  Gum/Vela\\
\object{BHR53} &  $-$85.78 &  $-$0.39 &  A &  Gum/Vela\\
\object{BHR55} &  $-$84.00 &   1.85 &  A &  Gum/Vela\\
\object{BHR56} &  $-$83.81 & $-$10.59 &  C &  Gum/Vela\\
\object{BHR59} &  $-$68.94 &  $-$1.66 &  A &  Carina\\
\object{BHR71} &  $-$62.28 &  $-$2.78 &  A &  Musca\\
\object{DC298.3$-$131 (G298.34$-$13.03)} &  $-$61.66 & $-$13.04 &  C &  Musca\\
\object{BHR76} &  $-$59.42 &  $-$3.13 &  N &  Musca\\
\object{Mu8} &  $-$58.78 &  $-$8.28 &  C &  Musca\\
\object{BHR78} &  $-$58.78 &  $-$0.37 &  A &  Crux\\
\object{BHR83} &  $-$57.90 &   7.44 &  C &  Centaurus\\
\object{G295.13$-$17.56} &  $-$64.86 & $-$17.56 &  N &  Chamaleon\\
\object{G297.09$-$16.02} &  $-$62.91 & $-$16.02 &  C &  Chamaleon\\
\object{G302.89$-$14.05} &  $-$57.11 & $-$14.05 &  C? &  Chamaleon\\
\object{G303.09$-$16.04} &  $-$56.91 & $-$16.04 &  C &  Chamaleon\\
\object{G303.15$-$17.34} &  $-$56.84 & $-$17.35 &  C &  Chamaleon\\
\object{G303.28$-$13.32} &  $-$56.71 & $-$13.32 &  C? &  Chamaleon\\
\object{G303.39$-$14.26} &  $-$56.60 & $-$14.27 &  C &  Chamaleon\\
\object{G303.68$-$15.32} &  $-$56.32 & $-$15.33 &  C &  Chamaleon\\
\object{G303.72$-$14.86} &  $-$56.27 & $-$14.86 &  C &  Chamaleon\\
\object{BHR86} &  $-$56.13 & $-$14.16 &  U &  Chamaleon\\
\object{DC338.2+16.4} &  $-$21.84 &  16.38 &  C &  Lupus\\
\object{DC338.8+16.5$-$2} &  $-$20.97 &  16.73 &  C &  Lupus\\
\object{L1681 ($\rho$ Oph E)} &   $-$7.00 &  16.65 &  C &  $\rho$ Oph\\
\object{$\rho$ Oph D} &   $-$6.35 &  17.71 &  N &  $\rho$ Oph\\
\object{G354.19+16.27} &   $-$5.80 &  16.28 &  C &  $\rho$ Oph\\
\object{$\rho$ Oph 9} &   $-$5.63 &  16.17 &  C+P+A &  $\rho$ Oph\\
\object{G356.96+07.27} &   $-$3.03 &   7.27 &  N &  $\rho$ Oph\\
\object{B59} &   $-$2.89 &   7.12 &  C &  $\rho$ Oph\\
\object{L1772} &   $-$1.30 &   6.03 &  U &  $\rho$ Oph\\
\object{L4} &    0.24 &  11.71 &  C &  $\rho$ Oph\\
\object{L43} &    1.35 &  20.98 &  C &  $\rho$ Oph\\
\object{B68} &    1.52 &   7.08 &  N &  $\rho$ Oph\\
\object{{Fest 1-457}} &  1.71 & 3.651 & N & $\rho$ Oph\\
\object{B72} &    1.78 &   6.95 &  N &  $\rho$ Oph\\
\object{L63} &    1.84 &  16.59 &  C &  $\rho$ Oph\\
\object{L100 (G003.07+09.95)} &    3.08 &   9.97 &  U &  $\rho$ Oph\\
\object{L111} &    3.30 &  10.43 &  C &  $\rho$ Oph\\
\object{CB68} &    4.51 &  16.34 &  C &  $\rho$ Oph\\
\object{L158} &    4.86 &  19.62 &  N &  $\rho$ Oph\\
\object{G004.92+17.95} &    4.92 &  17.95 &  ? &  $\rho$ Oph\\
\object{L162  (G005.03+19.07)} &    5.03 &  19.08 &  N &  $\rho$ Oph\\
\object{L173} &    5.30 &  11.08 &  C &  $\rho$ Oph\\
\object{L191 (G006.08+20.26)} &    6.09 &  20.26 &  N &  $\rho$ Oph\\
\object{L204C$-$2 (in G006.41+20.56)} &    6.34 &  20.46 &  P + C? &  $\rho$ Oph\\
\object{G006.41+20.56 (core$-$s4)} &    6.42 &  20.56 &  C &  $\rho$ Oph\\
\object{L234E} &    7.65 &  21.18 &  P &  $\rho$ Oph\\
\object{L260 (G008.67+22.14	)} &    8.68 &  22.14 &  C &  $\rho$ Oph\\
\object{L328} &   13.03 &  $-$0.83 &  A & $\rho$ Oph?\\
\object{CB103} &   23.89 &  11.12 &  C &  $\rho$ Oph\\
\object{L723} &   52.98 &   3.05 &  A/N &  $\rho$ Oph\\
\object{L1780} &   $-$1.10 &  36.88 &  P &  Serpens\\
\object{L134A} &    4.24 &  35.81 &  C &  Serpens\\
\object{L183 (G006.04+36.74)} &    6.00 &  36.74 &  C &  Serpens\\
\object{G011.40+36.19} &   11.40 &  36.19 &  N &  Serpens\\
\object{L429$-$C} &   21.62 &   3.75 &  A &  Serpens\\
\object{L438} &   22.29 &   4.97 &  N &  Serpens\\
\object{L462$-$1} &   23.69 &   7.56 &  N &  Serpens\\
\object{L483} &   24.89 &   5.40 &  A &  Serpens\\
\object{L492} &   25.50 &   6.18 &  C &  Serpens\\
\object{L507} &   26.72 &   6.71 &  N &  Serpens\\
\object{L648$-$1(G043.02+08.36)} &   43.02 &   8.37 &  C &  Hercules\\
\object{L531} &   28.46 &  $-$6.41 &  C &  Aquila\\
\object{G032.93+02.68} &   32.94 &   2.69 &  A &  Aquila\\
\object{B335} &   44.94 &  $-$6.56 &  C &  Aquila\\
\object{L673} &   46.28 &  $-$1.25 &  A &  Aquila\\
\object{L673$-$7} &   46.46 &  $-$1.46 &  A &  Aquila\\
\object{L675} &   46.52 &  $-$2.02 &  A &  Aquila\\
\object{CB188} &   46.53 &  $-$1.01 &  A &  Aquila\\
\object{L694$-$2} &   48.41 &  $-$5.83 &  C &  Aquila\\
\object{L771(G057.08+04.46)} &   57.10 &   4.45 &  N &  Vulpecula\\
\object{G089.03$-$41.28} &   89.03 & $-$41.29 &  N &  Pegasus\\
\object{B158} &   89.64 &  $-$6.63 &  C &  Cygnus\\
\object{G092.26+03.80} &   92.26 &   3.81 &  A &  Cygnus\\
\object{L1014} &   92.57 &  $-$0.25 &  N &  Cygnus\\
\object{L1021} &   93.00 &   0.71 &  N &  Cygnus\\
\object{G093.16+09.61} &   93.16 &   9.61 &  C &  Cygnus\\
\object{G093.20+09.53} &   93.21 &   9.54 &  C &  Cygnus\\
\object{G093.22$-$04.59} &   93.23 &  $-$4.59 &  P? &  Cygnus\\
\object{G093.31$-$11.68} &   93.32 & $-$11.68 &  N &  Lacerta\\
\object{CB228} &   93.89 &   7.60 &  C &  Cepheus\\
\object{B148} &   96.31 &  10.02 &  C &  Cepheus\\
\object{L1148} &  102.18 &  15.26 &  C &  Cepheus\\
\object{L1155E} &  102.61 &  15.20 &  C &  Cepheus\\
\object{L1155C$-$2} &  102.70 &  15.37 &  C &  Cepheus\\
\object{L1165} &  103.17 &   2.68 &  N &  Cepheus\\
\object{L1166} &  103.29 &   3.18 &  N &  Cepheus\\
\object{G105.55+10.40} &  105.56 &  10.41 &  N &  Cepheus\\
\object{L1197} &  106.35 &   0.48 &  N &  Cepheus\\
\object{G108.23+15.61} &  108.24 &  15.62 &  N? &  Cepheus\\
\object{L1221} &  110.65 &   9.64 &  C &  Cepheus\\
\object{L1228} &  111.67 &  20.22 &  C &  Cepheus\\
\object{Bern48} &  112.40 &  20.59 &  U &  Cepheus\\
\object{L1251A$-$2} &  113.99 &  14.92 &  C &  Cepheus\\
\object{L1251A} &  114.19 &  14.81 &  C &  Cepheus\\
\object{L1251C} &  114.48 &  14.69 &  C &  Cepheus\\
\object{L1251B} &  114.68 &  14.48 &  C &  Cepheus\\
\object{L1262} &  117.12 &  12.41 &  C &  Cepheus\\
\object{L1247} &  125.42 &  12.41 &  C &  Cepheus\\
\object{G128.25+20.78} &  128.25 &  20.78 &  U &  Cepheus\\
\object{L1152} &  102.36 &  15.98 &  C &  Draco\\
\object{L1157} &  102.65 &  15.80 &  C &  Draco\\
\object{G108.85$-$00.80} &  108.85 &  $-$0.80 & P & Cas\\
\object{L1253} &  115.84 &  $-$3.54 & N & Cas\\
\object{L1301} &  122.09 &  $-$0.36 & A & Cas\\
\object{CB6} &  122.60 &   5.00 & P? & Cas\\
\object{L1325} &  127.27 &   0.55 & A & Cas\\
\object{G127.88+02.66} &  127.88 &   2.67 & ? & Cas\\ %inner depression?{LP : bof changed 01$-$20$-$2014}\\
\object{L1333} &  128.89 &  13.69 & C & Cas\\ %IRAS 023687453?\\
\object{G128.95$-$00.18} &  128.96 &  $-$0.19 & A+P & Cas\\
\object{L1345} &  130.36 &   0.77 & N & Cas\\
\object{G130.56+11.51} &  130.56 &  11.51 & C & Cas\\ %{changed 01$-$20$-$2014}\\
\object{L1355} &  133.55 &   8.61 & N & Cas\\
\object{G131.35$-$45.73} &  131.36 & $-$45.73 & N & Pisces\\
\object{G145.87+17.77} &  145.88 &  17.78 &  C &  Cameleopardalis\\
\object{L1389} &  147.02 &   3.39 &  C &  Cameleopardalis\\
\object{G149.41+03.37} &  149.41 &   3.38 &  A? &  Cameleopardalis\\
\object{G149.58+03.45} &  149.59 &   3.45 &  P &  Cameleopardalis\\
\object{G150.22+03.91} &  150.23 &   3.92 &  C &  Cameleopardalis\\
\object{G151.45+03.95} &  151.46 &   3.96 &  P? &  Cameleopardalis\\
\object{CB24} &  155.76 &   5.91 &  C &  Auriga\\
\object{L1439} &  156.05 &   6.02 &  C &  Auriga\\
\object{G159.65+11.39} &  159.65 &  11.40 &  N? &  Auriga\\
\object{G170.77$-$08.51} &  170.77 &  $-$8.52 &  U &  Auriga\\
\object{L1512} &  171.86 &  $-$5.24 &  C &  Auriga\\
\object{L1517} &  172.38 &  $-$8.09 &  C &  Auriga\\
\object{L1448} &  158.06 & $-$21.42 & C & Aries\\%Class 0\\
\object{L1455 (G158.86$-$21.60)} &  158.86 & $-$21.60 & C & Aries\\ %{LP : CS (2 coeurs)}\\
\object{L1457 (G158.88$-$34.18) } &  158.88 & $-$34.18 & C & Aries\\ %{LP : weak}\\
\object{G158.97$-$33.01} &  158.97 & $-$33.02 & C & Aries\\ %{LP : CS}\\\
\object{IRAS03282+3035} &  159.09 & $-$20.66 & C & Aries\\ %Class 0\\
\object{G159.67$-$34.31} &  159.68 & $-$34.32 & N? & Aries\\ %{changed 01$-$20$-$2014 LP : petit+weak au centre d'un grand trou à 12\,\mic!}\\
\object{G154.68$-$15.34} &  154.69 & $-$15.35 & C & Perseus\\ %{changed 01$-$20$-$2014}\\
\object{G155.45$-$14.59} &  155.46 & $-$14.59 & C? & Perseus\\ %{changed 01$-$20$-$2014 a droite}\\
\object{G157.10$-$08.70} &  157.10 &  $-$8.71 & C & Perseus\\ %{changed 01$-$20$-$2014}\\
\object{G157.12$-$11.56} &  157.13 & $-$11.57 & C & Perseus\\ %{LP : CS}\\
\object{Barnard1} &  159.20 & $-$20.12 & C & Perseus\\ %added 09/13\\
\object{B5 (G160.51$-$16.84)} &  160.51 & $-$16.84 & C & Perseus\\%{LP : CS}\\
\object{G162.90$-$08.63} &  162.91 &  $-$8.63 & N & Perseus\\ %changed 09/13\\
\object{G163.21$-$08.40} &  163.21 &  $-$8.40 & C? & Perseus\\%{changed 01$-$20$-$2014}\\\
\object{G163.32$-$08.42} &  163.32 &  $-$8.42 & C & Perseus\\%{changed 01$-$20$-$2014}\\
\object{G169.82$-$19.39} &  169.83 & $-$19.39 &  C &  Taurus\\
\object{L1498} &  169.97 & $-$19.00 &  C &  Taurus\\
\object{G170.26$-$16.02} &  170.27 & $-$16.02 &  C &  Taurus\\
\object{G170.81$-$18.34} &  170.82 & $-$18.35 &  C? &  Taurus\\
\object{L1521B$-$2(G170.99$-$15.81)} &  170.87 & $-$15.87 &  C &  Taurus\\
\object{L1503} &  170.92 & $-$10.93 &  C &  Taurus\\
\object{L1506C} &  171.15 & $-$17.57 &  C &  Taurus\\
\object{L1507A (G171.51$-$10.59)} &  171.34 & $-$10.70 &  C &  Taurus\\
\object{G171.34$-$10.67} &  171.34 & $-$10.67 &  C &  Taurus\\
\object{L1521F (G171.49$-$14.908)} &  171.49 & $-$14.90 &  C &  Taurus\\
\object{CB23} &  171.50 & $-$10.60 &  C &  Taurus\\
\object{L1521$-$2} &  171.55 & $-$14.67 &  U &  Taurus\\
\object{G171.80$-$09.78} &  171.80 &  $-$9.78 &  C &  Taurus\\
\object{G171.91$-$15.65} &  171.91 & $-$15.66 &  C? &  Taurus\\
\object{L1521E} &  172.09 & $-$15.20 &  C &  Taurus\\
\object{G173.45$-$13.34} &  173.45 & $-$13.34 &  C &  Taurus\\
\object{L1524$-$4} &  173.62 & $-$16.26 &  C &  Taurus\\
\object{B18$-$2 (G173.69$-$15.55)} &  173.66 & $-$15.54 &  C &  Taurus\\
\object{B18$-$1} &  173.82 & $-$15.87 &  C &  Taurus\\
\object{TMC2} &  174.06 & $-$15.81 &  C &  Taurus\\
\object{G174.08$-$13.24} &  174.09 & $-$13.25 &  C? &  Taurus\\
\object{CB20} &  174.31 & $-$15.01 &  C &  Taurus\\
\object{B18$-$3 (G174.44$-$15.75)} &  174.45 & $-$15.74 &  C &  Taurus\\
\object{G174.50$-$19.88} &  174.51 & $-$19.89 &  P &  Taurus\\
\object{B18$-$5} &  174.72 & $-$15.44 &  U &  Taurus\\
\object{G177.89$-$20.16} &  177.89 & $-$20.16 &  C &  Taurus\\
\object{L1544} &  177.98 &  $-$9.71 &  C &  Taurus\\
\object{L1552} &  179.02 &  $-$6.75 &  C &  Taurus\\
\object{G179.18$-$19.62} &  179.19 & $-$19.63 &  C? &  Taurus\\
\object{IRAM04191} &  179.56 & $-$23.50 &  C &  Taurus\\
\object{G182.19$-$17.71} &  182.20 & $-$17.72 &  C &  Taurus\\
\end{longtable}
\tablefoot{
\tablefoottext{a}{A = Absorption, C = Coreshine, N = nothing, P = PAHs (or bright-rimmed cores or emission in all 4 bands) U = useless. Can be combined when two or three parts of the same cloud display different statuses.}
}
}

\end{appendix}

\end{document}